\documentclass[onecolumn]{aastex631}

\usepackage{amsmath}

\newcommand{\eq}[1]{Eq.~\eqref{#1}}
\newcommand{\fref}[1]{Fig.~\ref{#1}}
\newcommand{\sref}[1]{Sec.~\ref{#1}}

\newcommand{\dd}{\mathrm{d}}
\newcommand{\nn}{\nonumber}

\begin{document}

\title{Detection of astrophysical gravitational wave sources by TianQin and LISA}

\author[0000-0002-5467-3505]{Alejandro Torres-Orjuela}
\email{atorreso@hku.hk}
\affiliation{MOE Key Laboratory of TianQin Mission, TianQin Research Center for Gravitational Physics \& School of Physics and Astronomy, Frontiers Science Center for TianQin, Gravitational Wave Research Center of CNSA, Sun Yat-Sen University (Zhuhai Campus), Zhuhai 519082, China}
\affiliation{Department of Physics, The University of Hong Kong, Pokfulam Road, Hong Kong}

\author[0000-0002-7112-759X]{Shun-Jia Huang}
\thanks{Co-second author}
\affiliation{MOE Key Laboratory of TianQin Mission, TianQin Research Center for Gravitational Physics \& School of Physics and Astronomy, Frontiers Science Center for TianQin, Gravitational Wave Research Center of CNSA, Sun Yat-Sen University (Zhuhai Campus), Zhuhai 519082, China}

\author{Zheng-Cheng Liang}
\thanks{Co-second author}
\affiliation{MOE Key Laboratory of TianQin Mission, TianQin Research Center for Gravitational Physics \& School of Physics and Astronomy, Frontiers Science Center for TianQin, Gravitational Wave Research Center of CNSA, Sun Yat-Sen University (Zhuhai Campus), Zhuhai 519082, China}

\author[0000-0002-1197-2054]{Shuai Liu}
\thanks{Co-second author}
\affiliation{MOE Key Laboratory of TianQin Mission, TianQin Research Center for Gravitational Physics \& School of Physics and Astronomy, Frontiers Science Center for TianQin, Gravitational Wave Research Center of CNSA, Sun Yat-Sen University (Zhuhai Campus), Zhuhai 519082, China}

\author[0000-0002-7779-8239]{Hai-Tian Wang}
\thanks{Co-second author}
\affiliation{Kavli Institute for Astronomy and Astrophysics, Peking University, Beijing 100871, China}

\author{Chang-Qing Ye}
\thanks{Co-second author}
\affiliation{MOE Key Laboratory of TianQin Mission, TianQin Research Center for Gravitational Physics \& School of Physics and Astronomy, Frontiers Science Center for TianQin, Gravitational Wave Research Center of CNSA, Sun Yat-Sen University (Zhuhai Campus), Zhuhai 519082, China}

\author[0000-0002-7869-0174]{Yi-Ming Hu}
\email{huyiming@mail.sysu.edu.cn}
\affiliation{MOE Key Laboratory of TianQin Mission, TianQin Research Center for Gravitational Physics \& School of Physics and Astronomy, Frontiers Science Center for TianQin, Gravitational Wave Research Center of CNSA, Sun Yat-Sen University (Zhuhai Campus), Zhuhai 519082, China}

\author{Jianwei Mei}
\affiliation{MOE Key Laboratory of TianQin Mission, TianQin Research Center for Gravitational Physics \& School of Physics and Astronomy, Frontiers Science Center for TianQin, Gravitational Wave Research Center of CNSA, Sun Yat-Sen University (Zhuhai Campus), Zhuhai 519082, China}



\begin{abstract}
TianQin and LISA are space-based laser interferometer gravitational wave (GW) detectors planned to be launched in the mid-2030s. Both detectors will detect low-frequency GWs around $10^{-2}\,{\rm Hz}$, however, TianQin is more sensitive to frequencies above this common sweet-spot while LISA is more sensitive to frequencies below $10^{-2}\,{\rm Hz}$. Therefore, TianQin and LISA will be able to detect the same sources but with different accuracy depending on the source and its parameters. We consider some of the most important astrophysical sources -- massive black hole binaries, stellar-mass black hole binaries, double white dwarfs, extreme mass ratio inspirals, light and heavy intermediate mass ratio inspirals, as well as the stochastic gravitational background of astrophysical origin -- that TianQin and LISA will be able to detect. For each of these sources, we analyze how far they can be detected (detection distance) and how well their parameters can be measured (detection accuracy) using a Fisher Matrix analysis. We compare the results obtained by the three detection scenarios (TianQin alone, LISA alone, and joint detection by LISA and TianQin) highlighting the gains from joint detection as well as the contribution of TianQin and LISA to a combined study of astrophysical sources. In particular, we consider the different orientations, lifetimes, and duty cycles of the two detectors to explore how they can give a more complete picture when working together.
\end{abstract}



\section{Introduction}\label{sec:int}

The direct detection of gravitational waves (GWs) has opened a new window to observe and characterize compact objects throughout the Universe~\citep{ligo_virgo_2016a,GWTC1,GWTC2,GWTC3}. Currently, GW detection is restricted to the high-frequency band above $1\,{\rm Hz}$ covered by ground-based laser interferometry detectors like LIGO, Virgo, and KAGRA~\citep{ligo_2015,virgo_2012,kagra_2019} and the $\rm nHz$ band detected by pulsar timing arrays~\citep{ipta_2013,nanograv_2013,ppta_2013,cpta_2016,epta_2016,inpta_2018}. However, detection might be expanded in coming years to the intermediate band ($\rm dHz$) by space-based laser interferometry detectors such as DECIGO~\citep{decigo_2021} as well as ground-based and space-based atom interferometry detectors such as AION, ZAIGA, and AEDGE~\citep{aion_2020,zaiga_2020,aedge_2020}. Moreover, around the mid-2030s detection will be expanded to the low-frequency ($\rm mHz$) band by space-based laser interferometry detectors TianQin, LISA, and Taiji~\citep{tq_2016,lisa_2017,taiji_2015}. Detection of GW sources in the low-frequency band will allow us to study a variety of objects from the most massive black holes (BHs) at high cosmological redshifts to small compact objects in our galaxy like white dwarfs (WDs)~\citep{bayle_bonga_2022}. Moreover, different evolutionary stages of these sources will be detectable from the early inspiral to the merger and ring-down. Their detection is expected to revolutionize our understanding of compact objects and their population as well as to provide new insight into their interaction and evolution.

The wide range of masses that space-based laser interferometers will be able to detect results in a multitude of potential astrophysical sources. Some of the most important sources are massive BH binaries (MBHBs) residing in the center of galaxies as the result of galaxy mergers~\citep{komossa_2003,milosavljevic_merritt_2003a,milosavljevic_merritt_2003b}, stellar-mass BH binaries (SBHBs) formed either as the result of co-evolving massive stars~\citep{ligo_virgo_2016b,vanbeveren_2009,belczynski_dominik_2010,kruckow_tauris_2018,giacobbo_mapelli_2018,mapelli_giacobbo_2018,du-buisson_marchant_2020} or through dynamical assembling in dense stellar systems~\citep{ligo_virgo_2016b,portegies-zwart_mcmillan_2000,gultekin_miller_2004,gultekin_miller_2006,zevin_samsing_2019,tagawa_haiman_2020,samsing_dorazio_2020,rodriguez_chatterjee_2016,rodriguez_haster_2016}, double WDs (DWDs) that comprise the majority of compact stellar mass binaries in the Milky Way~\citep{nelemans_yungelson_2001a,yu_jeffery_2010,lamberts_garrison-kimmel_2018,breivik_coughlin_2020}, extreme mass ratio inspirals (EMRIs) that form when a small compact object is captured by a massive BH (MBH) in the center of nuclear clusters~\citep{amaro-seoane_2018a,mapelli_ripamonti_2012}, intermediate mass ratio inspirals (IMRIs) formed by either a stellar-mass BH (SBH) orbiting an intermediate-mass BH (IMBH)~\citep{will_2004,amaro-seoane_gair_2007,konstantinidis_amaro-seoane_2013,leigh_lutzgendorf_2014,haster_antonini_2016,macleod_trenti_2016,amaro-seoane_2018a,amaro-seoane_2018b,arca-sedda_amaro-seoane_2021,rizzuto_naab_2021} or an IMBH orbiting a MBH~\citep{basu_chakrabarti_2008,arca-sedda_gualandris_2018,arca-sedda_capuzzo-dolcetta_2019,derdzinski_dorazio_2019,bonetti_rasskazov_2020,derdzinski_dorazio_2021,rose_naoz_2022}, and the stochastic GW background produced by unresolved binaries~\citep{allen_romano_1999,romano_cornish_2017,christensen_2019}.

Despite TianQin and LISA having an overall similar design, they differ in several key aspects. Therefore, they will detect the same astrophysical sources but up to different distances and with different accuracy depending on their specific parameters. The distinct sensitivities open up the possibility to study a bigger parameter space thus increasing the information we can obtain about different sources as well as their respective population. This will allow us to get a better understanding of the formation and evolution of different sources, and the system in which they reside.

In this paper, we study and compare the detection of astrophysical GW sources by TianQin, LISA, and their joint detection. The goal of this study is to understand how well TianQin and LISA will detect the different sources as well as the gain we can obtain from joint detection. Therefore, we do not base the properties of the sources studied on any population models but instead explore the biggest possible parameter space for each source. Nevertheless, we focus on exploring regions of the parameter space that seem meaningful in an astrophysical context. For each source, we use state-of-the-art waveform models that allow exploring an extensive parameter space. However, for some sources, the waveforms available impose strict restrictions on the studies performed and the results must be understood as (the best possible) approximation to future detection.

This paper is organized as follows. In \sref{sec:gwd} we introduce the properties of TianQin and LISA that are most relevant for detection, including their noise curves, and basic methods to assess their detection capabilities. We analyze how far massive black hole binaries (MBHBs) can be detected (detection distance) and how well their parameters can be measured (detection accuracy) in \sref{sec:mbhb}. The detectability and parameter reconstruction for stellar-mass black hole binaries (SBHBs) and double white dwarfs (DWDs) are studied in \sref{sec:sbhb} and \sref{sec:dwd}, respectively. The detection distance and accuracy of extreme mass ratio inspirals (EMRIs) are studied in \sref{sec:emri} while we perform similar analyses for light and heavy intermediate mass ratio inspirals in \sref{sec:imri}. In \sref{sec:sgwb}, we study the detection of the galactic and extragalactic stochastical gravitational wave background (SGWB). We summarize our result in \sref{sec:res} and draw conclusions in \sref{sec:con}.

\section{Gravitational wave detection with TianQin and LISA}\label{sec:gwd}

TianQin and LISA are both space-based GW detectors consisting of three satellites in a triangular shape performing interferometry among them~\citep{tq_2016,lisa_2017}. Therefore, they have several features in common but others differ significantly leading to a difference in their detection band and sensitivity. In this section, we give a brief introduction to the realizations of TianQin and LISA as well as the resulting sensitivity curves. We further introduce later in this section, the data analysis methods used in the paper.

\begin{figure}[tpb] \centering \includegraphics[width=0.48\textwidth]{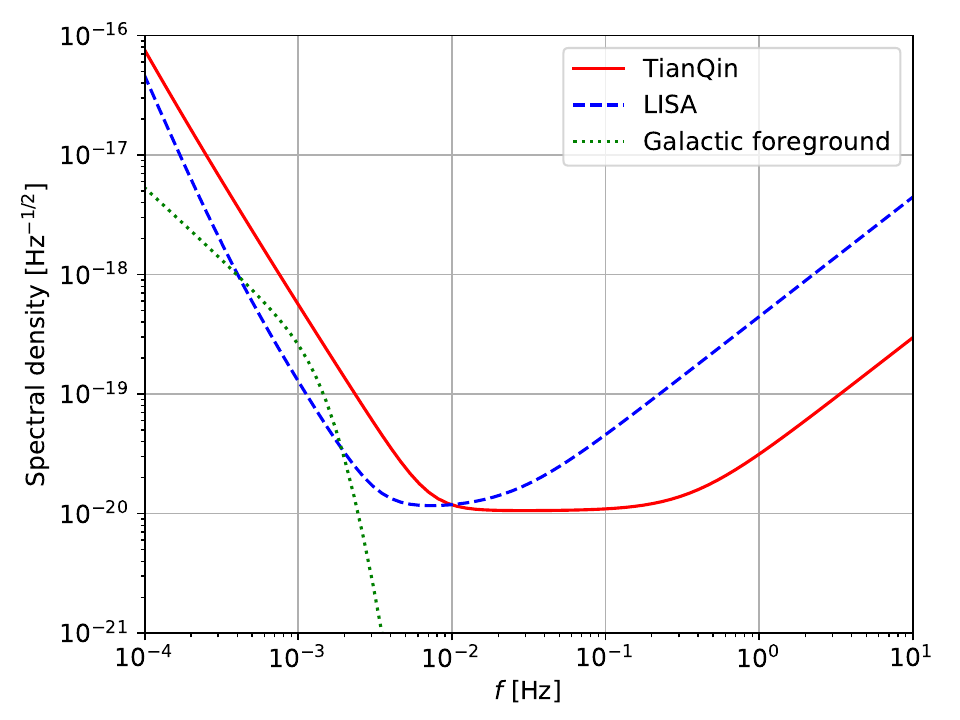}
\caption{
    The sensitivity curves for TianQin and LISA as functions of the frequency. The figure also shows the confusion noise from the galactic foreground for LISA and a 4-year mission. No confusion noise for TianQin is included since it lies below TianQin's sensitivity curve.
    }
\label{fig:sc}
\end{figure}

One of the most striking differences is that TianQin is going to fly on a geocentric orbit~\citep{tq_2021} while LISA is going to be on a heliocentric orbit~\citep{lisa_2017}. A direct consequence of the orbit is that TianQin will have an arm length $L_{\rm TQ} = \sqrt{3}\times10^5\,{\rm km}$ while LISA on its heliocentric orbit will have a bigger arm length $L_{\rm LISA} = 2.5\times10^6\,{\rm km}$. The difference in the arm length is one of the reasons why LISA is more sensitive to slightly lower frequencies while TianQin is more sensitive to slightly higher frequencies as shown in \fref{fig:sc}. The mission lifetime of TianQin is planned to be five years with a duty cycle of 50\,\% a schedule of three months on/three months off resulting in an effective observation time of 2.5 years. We include these on- and off-times by manipulating the data in a way where three months of unchanged signal (on-time) are followed by three months where the signal is set to zero (off-time). Moreover, we assume that the last three months of detection are always during an on-time. LISA is planned to have a mission lifetime of four years but with a higher duty cycle of at least 75\,\%~\citep{lisa_2022a}. Therefore, we assume that three consecutive years of data are collected. Another remarkable difference between TianQin and LISA is that the first has a fixed orientation towards RX J0806.3+1527~\citep{strohmayer_2005} whereas the second changes its orientation while orbiting around the sun completing one cycle in one year. Therefore, TianQin maintains a high sensitivity throughout the year but only for certain regions in the sky while LISA is sensitive to all regions in the sky but with varying sensitivity. This difference in the orientation and orbit between TianQin and LISA will affect their `response function' (also called `antenna pattern function') to the incoming GWs. We have adopted the antenna pattern functions introduced in \cite{tq_mbhb_2019b} and \cite{klein_barausse_2016} for TianQin and LISA, respectively. A summary of the difference in the parameters for TianQin and LISA is shown in TABLE~\ref{tab:td}.

\begin{table}[tpb]
    \centering
    \begin{tabular}{|c||c|c|} \hline
         & TianQin & LISA \\ \hline\hline
        Type of orbit & Geocentric & Heliocentric \\ \hline
        Arm length & $\sqrt{3}\times10^5\,{\rm km}$ & $2.5\times10^6\,{\rm km}$ \\ \hline
        Mission lifetime & 5 years & 4 years \\ \hline
        Duty cycle & 50\,\% & 75\,\% \\ \hline
        Orientation & RX J0806.3+1527 & cyclic \\ \hline
    \end{tabular}
    \caption{TianQin and LISA technical details}
    \label{tab:td}
\end{table}

From a data analysis point of view, which is the focus of this paper, the difference between TianQin and LISA is reflected by the difference in their sensitivity curves. TianQin's sensitivity curve as a function of the frequency $f$ can be parameterized as~\citep{tq_2021}
\begin{align}\label{eq:tqsc}
    \nn S_n^{\rm TQ}(f) =& \frac{10}{3L_{\rm TQ}^2}\left[P_x^{\rm TQ} + \frac{4P_a^{\rm TQ}}{(2\pi f)^4}\left(1 + \frac{10^{-4}\,{\rm Hz}}{f}\right)\right] \\
    &\times\left[1 + \frac{6}{10}\left(\frac{f}{f_\ast^{\rm TQ}}\right)^2\right],
\end{align}
where $(P_x^{\rm TQ})^{1/2} = 10^{-12}\,{\rm mHz^{-1/2}}$ is TianQin's displacement measurement noise, $(P_a^{\rm TQ})^{1/2} = 10^{-15}\,{\rm ms^{-2}\,Hz^{-1/2}}$ is TianQin's residual acceleration noise and $f_\ast^{\rm TQ} = c/(2\pi L_{\rm TQ}) \approx 0.28\,{\rm Hz}$ is TianQin's transfer frequency ($c$:~speed of light in vacuum). The sensitivity curve of LISA can be approximated as~\citep{robson_cornish_2019}
\begin{align}\label{eq:lsc}
    \nn S_n&^{\rm LISA}(f) =  \frac{10}{3L_{\rm LISA}^2}\bigg[P_x^{\rm LISA} + \frac{2P_a^{\rm LISA}}{(2\pi f)^4}\\
    &\times\left(1 + \cos^2\left(\frac{f}{f_\ast^{\rm LISA}}\right)\right)\bigg]\left[1 + \frac{6}{10}\left(\frac{f}{f_\ast^{\rm LISA}}\right)^2\right],
\end{align}
where $f_\ast^{\rm LISA} = c/(2\pi L_{\rm LISA}) \approx 0.019\,{\rm Hz}$ is LISA's transfer frequency,
\begin{equation}\label{eq:poms}
    (P_x^{\rm LISA})^{1/2} = 1.5\times10^{-11}\sqrt{1 + \left(\frac{2\times10^{-3}{\rm Hz}}{f}\right)^4}\,{\rm mHz^{-1/2}}
\end{equation}
is LISA's displacement measurement noise, and
\begin{align}\label{eq:pacc}
    \nn (P_a^{\rm LISA})^{1/2} = 3\times10^{-15}\sqrt{1 + \left(\frac{4\times10^{-4}\,{\rm Hz}}{f}\right)^2} \\
    \times\sqrt{1 + \left(\frac{f}{8\times10^{-3}\,{\rm Hz}}\right)^4}\,{\rm ms^{-2}\,Hz^{-1/2}}
\end{align}
is LISA's residual acceleration noise.

\fref{fig:sc} shows the square root of the sensitivity curves for TianQin $\sqrt{S_n^{\rm TQ}}$ and LISA $\sqrt{S_n^{\rm LISA}}$ for frequencies between $0.1\,{\rm mHz}$ and $10\,{\rm Hz}$. We see that TianQin is more sensitive in a slightly higher frequency band while LISA is most sensitive in a slightly lower frequency band. In particular, TianQin is most sensitive to frequencies between $10\,{\rm mHz}$ and $100\,{\rm mHz}$ while  LISA is most sensitive to frequencies around and below $10\,{\rm mHz}$. Nevertheless, both detectors have a strong sensitivity at frequencies around $10\,{\rm mHz}$ thus allowing a particularly accurate joint detection at this common ``sweet spot''. \fref{fig:sc} further shows the confusion noise from the galactic foreground for LISA as described in \cite{robson_cornish_2019} for a 4-year mission which will affect LISA detection for frequencies around $1\,{\rm mHz}$. TianQin detection, in contrast, is not affected by the galactic foreground and thus no curve for the confusion noise for TianQin is included in \fref{fig:sc}.

\subsection{Signal-to-noise ratio and Fisher Matrix analysis}\label{sec:snr}

The importance of a detector's sensitivity curve, $S_n(f)$, lies in the fact that it describes how well a signal can be detected. For a detected signal $h(t)$ and a model of the signal $h_m(t)$ in the time-domain, the noise-weighted inner product is defined as~\citep{sathyaprakash_schutz_2009}
\begin{equation}\label{eq:ipro}
    \langle h|h_m\rangle = 4\Re\left[\int_0^\infty\dd f\frac{\tilde{h}(f)^\ast\tilde{h}_m(f)}{S_n(f)}\right],
\end{equation}
where $\tilde{h}(f)$ and $\tilde{h}_m(f)$ are the Fourier transform of $h(t)$ and $h_m(t)$, respectively, an asterisk indicates the complex conjugate, and $\Re$ means to take the real part of the function. The optimal signal-to-noise ratio (SNR) $\rho$ is obtained when the model of the signal is identical to the detected signal, thus we define
\begin{equation}\label{eq:snr}
    \rho := \langle h|h\rangle.
\end{equation}
Moreover, in the case of joint detection, the combined SNR can be calculated from the two independent SNRs as $\rho_{\rm Joint} = \sqrt{(\rho_{\rm TianQin})^2 + (\rho_{\rm LISA})^2}$.

The SNR indicates how ``loud'' a signal or more specifically a detection is. Therefore, a higher SNR, in general, means that we can extract information from the detection with more accuracy. Detection accuracy does not only depend on the SNR but on the properties of the signal or rather its source. How precisely the parameters of the source $\boldsymbol{\lambda}$ can be extracted from the signal can be addressed using a Fisher Matrix analysis (FMA)~\citep{coe_2009}, which is a linear estimate of the measurement errors that asymptotes to the true error in the limit of high SNR. The Fisher Matrix can be defined as
\begin{equation}
    F_{ij} := \left\langle\frac{\partial h(\boldsymbol{\lambda})}{\partial \lambda_i},\frac{\partial h(\boldsymbol{\lambda})}{\partial \lambda_j}\right\rangle.
\end{equation}
The inverse of the Fisher Matrix $C = F^{-1}$ then approximates the sample covariance matrix of the Bayesian posterior distribution for the parameters. In particular, the diagonal elements $C$ indicate the error of the respective parameter, $\Delta\lambda_i = C_{ii}$, while the other elements indicate the correlation between different parameters. Note, throughout this paper we use $\Delta\lambda_i$ to denote absolute errors and $\delta\lambda_i := \Delta\lambda_i/\lambda_i$ to denote relative errors. We point out, that the detection accuracy of the joint detection can be obtained using a FMA by combining the results of the two independent detectors
\begin{equation}
    F_{\rm Joint} = F_{\rm TianQin} + F_{\rm LISA}
\end{equation}
and then inverting $F_{\rm Joint}$~\citep{isoyama_nakano_2018}.

We point out that in the case of a stochastic gravitational wave background (SGWB), as discussed in \sref{sec:sgwb}, the SNR and FMA need to be defined differently. However, their meaning is still the same -- the SNR indicates how loud a signal is while the FMA approximates the measurement errors in the high-SNR limit -- thus we do not introduce their calculation for the SGWB in detail. For an introduction to the measurement of the SGWB and the SNR using the cross-correlation method and the null channel method see, e.g., \cite{tq_sgwb_2022a} Sec. III.A and III.B, respectively. For a discussion of the FMA for the SGWB see, e.g., Sec. IV.A of \cite{wang_han_2021}.

\subsection{Performance ratio}\label{sec:per}

A major focus of this paper is to compare how well the three detection scenarios considered -- TianQin alone, LISA alone, and joint detection -- perform for the different sources. To quantify this difference for the entire range of a parameter $\lambda$, we define the performance ratio
\begin{equation}\label{eq:pr}
    Q_X[\lambda] := \int_{\lambda_{\rm min}}^{\lambda_{\rm max}}\dd\lambda\Big/\int_{\lambda_{\rm min}}^{\lambda_{\rm max}}\frac{\Delta\lambda_X}{\Delta\lambda_{\rm Joint}}\dd\lambda,
\end{equation}
where $X$ can be either TianQin or LISA, $\Delta\lambda_X$ ($\Delta\lambda_{\rm Joint}$) is the absolute error in detector $X$ (the joint detection), and $[\lambda_{\rm min},\lambda_{\rm max}]$ is the interval of the parameter considered. In case the parameter range considered spans several orders of magnitude, we define
\begin{equation}\label{eq:prl}
    Q_X[\lambda] := \int_{\lambda_{\rm min}}^{\lambda_{\rm max}}\dd(\ln\lambda)\Big/\int_{\lambda_{\rm min}}^{\lambda_{\rm max}}\frac{\Delta\lambda_X}{\Delta\lambda_{\rm Joint}}\dd(\ln\lambda).
\end{equation}

The performance ratio indicates how well a single detection scenario performs compared to the joint detection or how much a single detector contributes to the joint detection. Because the error in the joint detection is always smaller than in any of the two detectors alone, it is normalized to one and $Q_X[\lambda]$ close to one indicates that the parameter $\lambda$ is mainly constrained by detector $X$ while $Q_X[\lambda]$ close to zero indicates that parameter $\lambda$ is poorly constrained by detector $X$. We point out that $Q$ measures the performance for the entire parameter range $[\lambda_{\rm min},\lambda_{\rm max}]$, thus, even if a detector has a low performance ratio it still can constraint a parameter well in sub-intervals.

\section{Massive black holes binaries}\label{sec:mbhb}

Observations indicate that almost every galaxy has a massive black hole (MBH) in the mass range of $10^5 - 10^{10}\,{\rm M_\odot}$ at its center~\citep{kormendy_richstone_1995}. In the hierarchical merger scenario of galaxies, large galaxies form through the merger of smaller galaxies, which subsequently leads to pairs of MBHs in the center of the newly formed galaxy~\citep{graham_2023}. Due to the deep potential and other effects like dynamical friction or triple interaction, the MBHs are expected to come together until they form a MBH binary (MBHB) that emits GWs~\citep{komossa_2003,milosavljevic_merritt_2003a,milosavljevic_merritt_2003b,graham_2023}. Furthermore, models have been proposed on how MBHB can form directly inside the first galaxies under special conditions~\citep{bromm_loeb_2003}.

MBHBs are the most powerful astrophysical GW sources and among the loudest sources in the band of space-based detectors. Therefore, they can be used to perform detailed studies on the formation and growth mechanism of seed black holes (BHs)~\citep{bromm_loeb_2003,madau_rees_2001}, the co-evolution of MBHBs with their host galaxies~\citep{magorrian_tremaine_1998,kormendy_ho_2013}, and cosmography~\citep{hughes_holz_2003,tamanini_caprini_2016}. However, how well these studies can be performed depends on how accurately the parameters of the MBHB can be measured and how far we will be able to detect them. In this section, we use the waveform model IMRPhenomXHM which can describe MBHBs with aligned spins on non-eccentric orbits including several subdominant modes~\citep{garcia-quiros_colleoni_2020} to study the distance to which MBHBs will be detected by TianQin, LISA, and their joint detection. Furthermore, we estimate the detection accuracy for the most important parameters using Fisher Matrix analysis (FMA) where we use the conventional detection threshold of $\rho=8$~\citep{klein_barausse_2016}. For more details on the detection of MBHBs with TianQin and LISA see, e.g., \cite{tq_mbhb_2019a}, \cite{tq_mbhb_2019b}, \cite{tq_mbhb_2019c}, \cite{klein_barausse_2016}, and \cite{lisa_2022b}.

\subsection{Detectability}\label{sec:mbhbd}

\begin{figure}[tpb] \centering \includegraphics[width=0.48\textwidth]{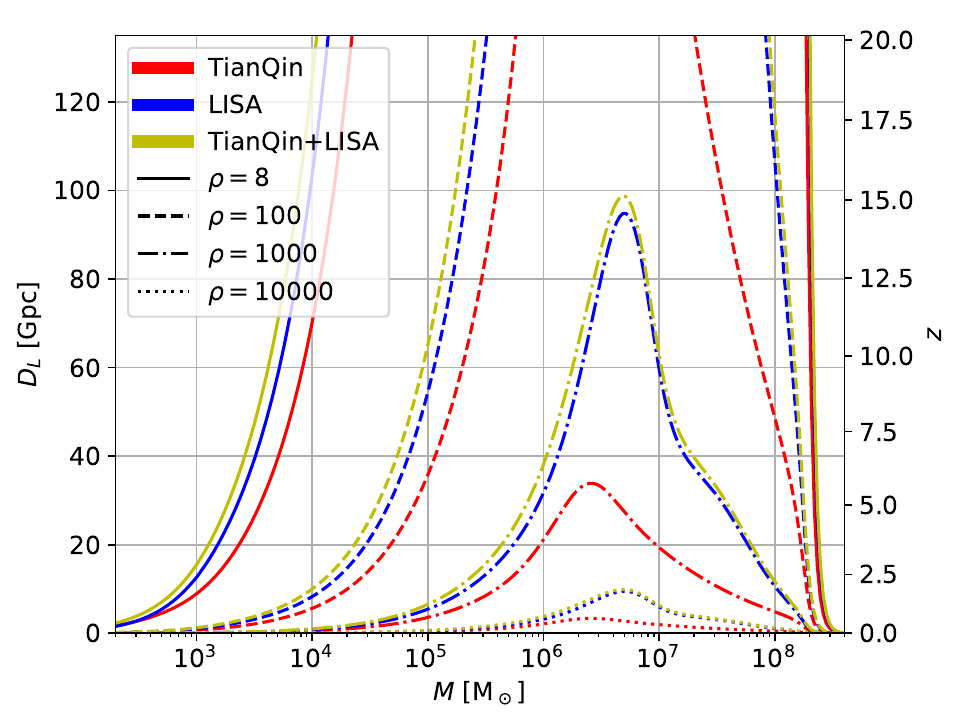}
\caption{
    The distance to which a MBHB can be detected with a fixed SNR $\rho$ for TianQin, LISA, and joint detection as a function of the binary's total mass $M$ in the observer frame. The left ordinate shows the luminosity distance $D_L$ while the right ordinate shows the corresponding redshift $z$.
    }
\label{fig:mbhb_hor_snr}
\end{figure}

In this subsection, we study how far a MBHB with equal-mass non-spinning components can be detected by TianQin, LISA, and joint detection. For all sources discussed in this paper, we have chosen the parameters so that TianQin and LISA have a similar SNR. We make this choice for two reasons: i) we want the results in the single detection scenarios to be comparable and similar SNR guarantees that the differences in detection are driven by the properties of the detectors instead of a particular choice of the source's parameters, and ii) by TianQin and LISA contributing similarly to the joint detection, it becomes easier to identify the gains we can get from said joint detection. However, for many sources, it is only possible to have a similar SNR in TianQin and LISA in small portions of the parameter space due to the different sensitivity of the two detectors. Moreover, the masses presented are always in the observer frame, we do not average over the sky localization, and we consider the source to be face-on if its inclination is not specified. \fref{fig:mbhb_hor_snr} shows the distance to which these sources will be detectable for a SNR of eight, 100, 1000, and 10000. We see that for a SNR of eight, MBHBs will be detectable to redshifts 20 and beyond if their mass lies between a few $10^4\,{\rm M_\odot}$ and some $10^8\,{\rm M_\odot}$ for all three detection scenarios. For a SNR of 100, LISA alone and a joint detection will still detect MBHBs to a redshift of 20 and beyond if their mass ranges between a few $10^5\,{\rm M_\odot}$ and around $10^8\,{\rm M_\odot}$. TianQin alone will still detect MBHBs to redshifts of 20 and higher for a SNR of 100 but only for masses between $5\times10^5\,{\rm M_\odot}$ and $2\times10^7\,{\rm M_\odot}$. For lower SNRs, the difference between LISA alone and TianQin alone becomes more prominent. If the SNR is 1000 LISA will be able to detect MBHBs out to a redshift of around 14 if their total mass is around $5\times10^6\,{\rm M_\odot}$ and to redshifts of five or higher if their mass ranges between $10^6\,{\rm M_\odot}$ and some $10^7\,{\rm M_\odot}$. TianQin in contrast only will detect MBHBs with a SNR of 1000 to a redshift of five if their mass is around $2\times10^6\,{\rm M_\odot}$. However, for total masses between around $5\times10^5\,{\rm M_\odot}$ and a few $10^7\,{\rm M_\odot}$ TianQin still will be able to detect MBHBs at redshifts of two or higher. Since LISA performs significantly better for MBHBs the detectability in the joint detection resembles the detectability for LISA alone very closely. Only for masses below $5\times10^6\,{\rm M_\odot}$ and for masses of a few $10^7\,{\rm M_\odot}$ the joint detection performs slightly better than LISA alone. This trend continues for a SNR of 10000 where LISA alone and the joint detection can detect MBHBs with a total mass between $10^6\,{\rm M_\odot}$ and a few $10^7\,{\rm M_\odot}$ to redshifts between 0.5 and 1.5. The same binaries will be detectable in TianQin for redshifts of around 0.5 and below.

\begin{figure}[tpb] \centering \includegraphics[width=0.48\textwidth]{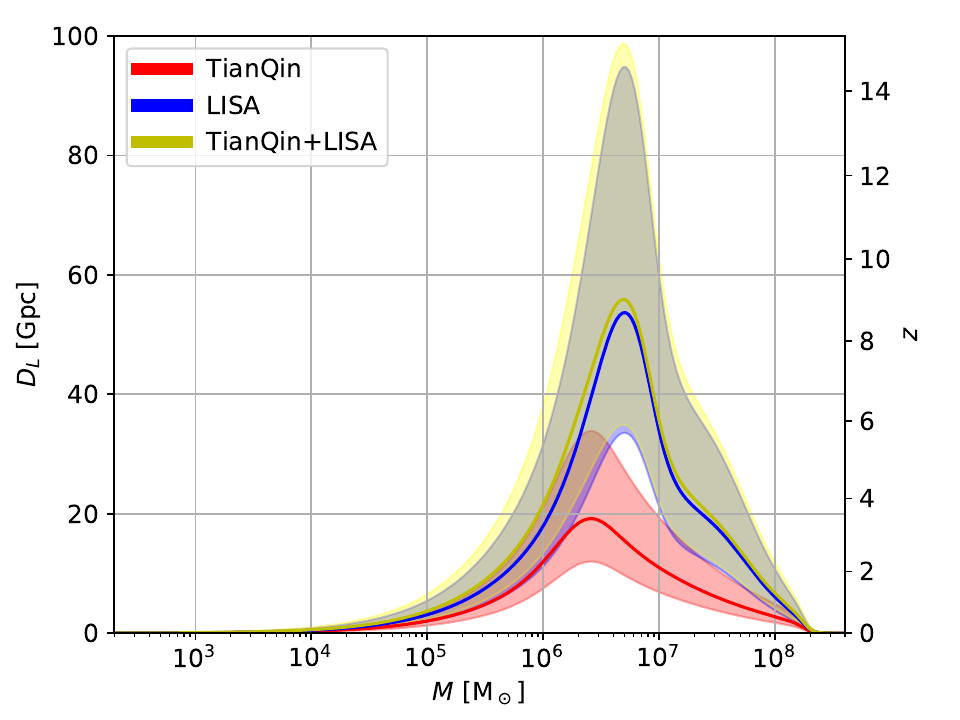}
\caption{
    The distance to which a MBHB with a SNR of 1000 can be detected for a varying inclination for TianQin, LISA, and joint detection as a function of the binary's total mass $M$. The upper edge of the shaded region corresponds to the face-on case while the lower edge corresponds to the edge-on case; the solid line corresponds to the average inclination of $60^\circ$. The left ordinate shows the luminosity distance $D_L$ while the right ordinate shows the corresponding redshift $z$.
    }
\label{fig:mbhb_hor_inc}
\end{figure}

The strain of a GW depends critically on the distance of the source but also its inclination. Face-on sources emit two polarizations while edge-on sources only emit one polarization and thus face-on sources have a significantly stronger signal than edge-on sources. Therefore, we show in \fref{fig:mbhb_hor_inc} the distances to which MBHBs will be detectable by TianQin alone, LISA alone, and joint detection for different inclinations and a SNR of 1000. As before the joint detection resembles the detectability of LISA closely with small contributions from TianQin for some specific mass ranges. We see that for a mass of around $5\times10^6\,{\rm M_\odot}$ LISA will be able to detect such a source to a redshift of around 14 if it is face-on. For an inclination of $60^\circ$ the same source will be detectable to a redshift of around 8.5 and if it is edge-on to a redshift of around 5.5. In general, we can see that face-on sources will be detectable at distances of roughly two times the distance for the average inclination and three times the distance of the edge-on case. For TianQin alone, we have a similar behavior where for the best case scenario of a MBHB with a mass of around $2\times10^6\,{\rm M_\odot}$ face-on sources will be detectable out to a redshift of around five, sources with an average inclination of $60^\circ$ to a redshift of around three, and edge-on source to a redshift of almost two.

\subsection{Parameter estimation}\label{sec:mbhbp}

In this subsection, we study how accurate the total mass in the observer frame $M$, the luminosity distance $D_L$, the inclination $\iota$, the effective spin along the angular momentum $\chi_z$, and the sky localization $\Delta\Omega$ of a MBHB can be detected in TianQin, LISA, and by joint detection. For our analysis, we consider an equal mass binary with spins aligned to the angular momentum and fix all parameters but one to the following values $M = 4\times10^6\,{\rm M_\odot}$ (in the observer frame), $z = 2$ ($D_L \approx 10.5\,{\rm Gpc}$), $\iota = 60^\circ$, $\chi_{z,1} = \chi_{z,2} = 0.5$, $\theta_{\rm bar} \approx 45^\circ$, and $\phi_{\rm bar} \approx 29^\circ$, where $\theta_{\rm bar}$ and $\phi_{\rm bar}$ are the latitude and azimuth angle in barycentric coordinates, respectively. Note that because we consider an equal mass binary, the choice of primary and secondary BHs is random; thus we only report the accuracy for the detection of $\chi_z$ for one BH.

\begin{figure}[tpb] \centering \includegraphics[width=0.48\textwidth]{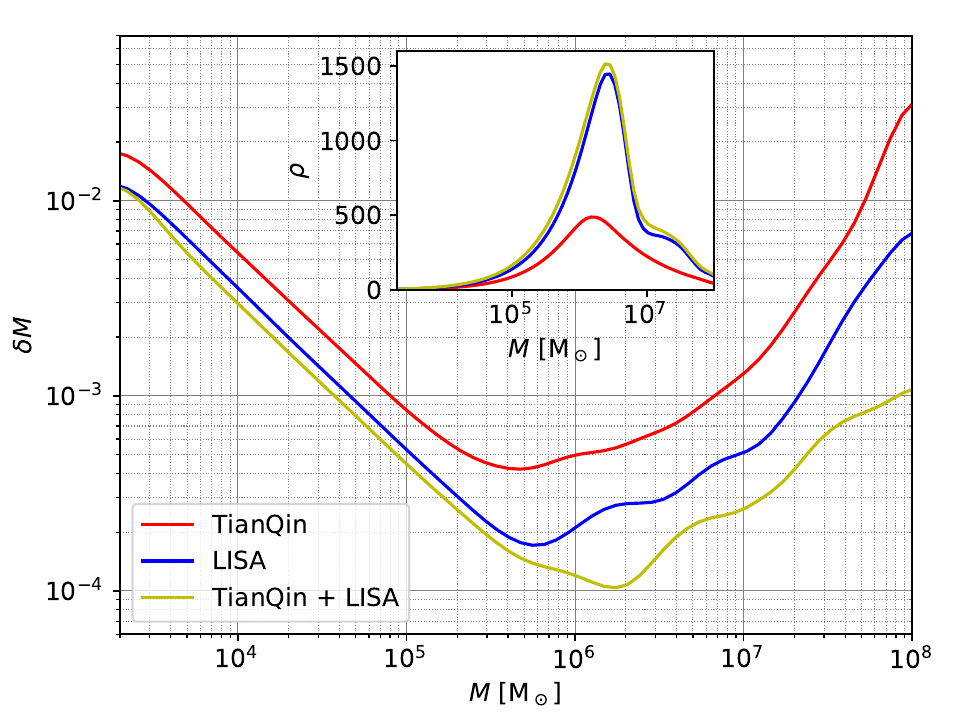}
\caption{
    The relative error in the total mass of a MBHB $\delta M$ as a function of the total mass in the observer frame $M$. The red line shows the detection accuracy for TianQin alone, the blue line for LISA alone, and the yellow line for joint detection. The inset shows the corresponding SNR of the source where the color coding is the same as in the main plot.
    }
\label{fig:mbhb_dm}
\end{figure}

\textit{Total mass:} \fref{fig:mbhb_dm} shows the relative error TianQin, LISA, and a joint detection will have when detecting the total mass of a MBHB. We see that the SNR in LISA is for most sources significantly higher than in TianQin, in particular for sources around $10^6\,{\rm M_\odot}$. Only for lighter sources of the order $10^4\,{\rm M_\odot}$ and below TianQin has a SNR comparable to LISA's. We see that for sources around $10^7\,{\rm M_\odot}$ the SNR in LISA has a dip due to the confusion noise from galactic binaries. Around that dip, TianQin can contribute to increasing the SNR of the joint detection although, in general, the SNR of the joint detection follows LISA's SNR curve closely. The significant difference in the SNR also leads to a significant difference in the relative error for the mass. In TianQin the error is around $10^{-2}$ for $M<10^4{\rm M_\odot}$, has a minimum of around $4\times10^{-4}$ for $M\approx4\times10^5\,{\rm M_\odot}$, and goes up again to an order of $10^{-2}$ for heavy sources ($M\gtrsim4\times10^7\,{\rm M_\odot}$). The relative error in LISA is only of the order $10^{-2}$ for the lightest sources ($M\approx2\times10^3\,{\rm M_\odot}$), is of the order $10^{-4}$ for $5\times10^4\,{\rm M\odot}\lesssim M\lesssim2\times10^7\,{\rm M_\odot}$ with a minimum of less than $2\times10^{-4}$ for $M\approx6\times10^5\,{\rm M_\odot}$, and goes up to almost $6\times10^{-3}$ for $M=10^8\,{\rm M_\odot}$. The relative error in the joint detection follows the error in LISA closely for $M\lesssim4\times10^5\,{\rm M_\odot}$ but is significantly better for higher masses. We attribute the improvement in the higher masses to TianQin's ability to detect higher harmonics that are outside LISA's most sensitive band when the MBHB merges.

\begin{figure}[tpb] \centering \includegraphics[width=0.48\textwidth]{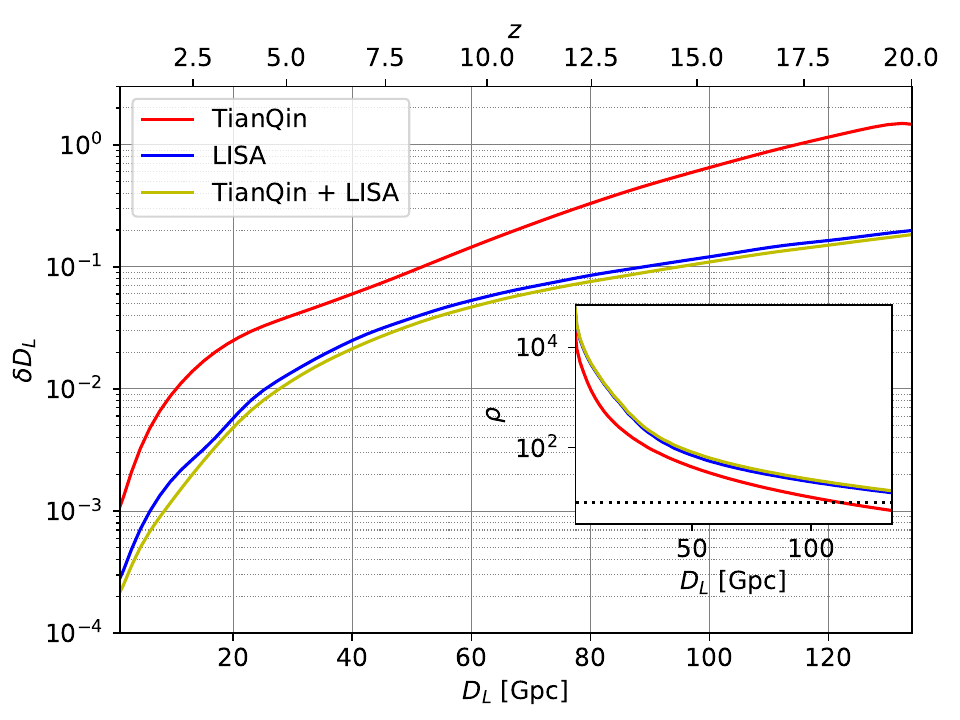}
\caption{
    The relative error in the luminosity distance $\delta D_L$ for a MBHB. The lower abscissa shows the different luminosity distances of the source $D_L$ while the upper abscissa shows the corresponding cosmological redshift $z$. The red line represents the detection accuracy for TianQin alone, the blue line for LISA alone, and the yellow line for joint detection. The inset shows the corresponding SNR of the source where the color coding is the same as in the main plot and the detection threshold of $\rho=8$ is indicated by the black dotted line.
    }
\label{fig:mbhb_ddl}
\end{figure}

\textit{Luminosity distance:} In \fref{fig:mbhb_ddl}, we see the detection accuracy for the luminosity distance of a MBHB. We see that for all distances LISA performs significantly better than TianQin, although the difference is smaller for lower redshifts. The dependence of the difference on the distance of the source arises because at higher redshifts the frequency of the GW goes down (note that we fix the mass of the source at $z=2$) and thus the signal gets further shifted to frequencies where LISA performs better than TianQin. Due to LISA's significantly better performance, joint detection performs similarly to LISA except for lower redshifts of up to around three where TianQin contributes to an improved detection error. TianQin's relative error for the luminosity distance is around $10^{-3}$ in the local universe and remains below $0.1$ up to redshifts of almost nine. However, for high redshifts above 17.5, the relative error in TianQin surpasses 1. Note that at these distances the SNR in TianQin goes below the detection threshold of 8. For LISA, and similar for joint detection, the relative error in the local universe is of the order $10^{-4}$ and even remains below $10^{-2}$ up to $z\approx5$. The SNR in LISA and the joint detection always remains above the detection threshold and the error remains below $0.2$ at a $z=20$, only surpassing a relative error of $0.1$ above redshifts of almost 13.

\begin{figure}[tpb] \centering \includegraphics[width=0.48\textwidth]{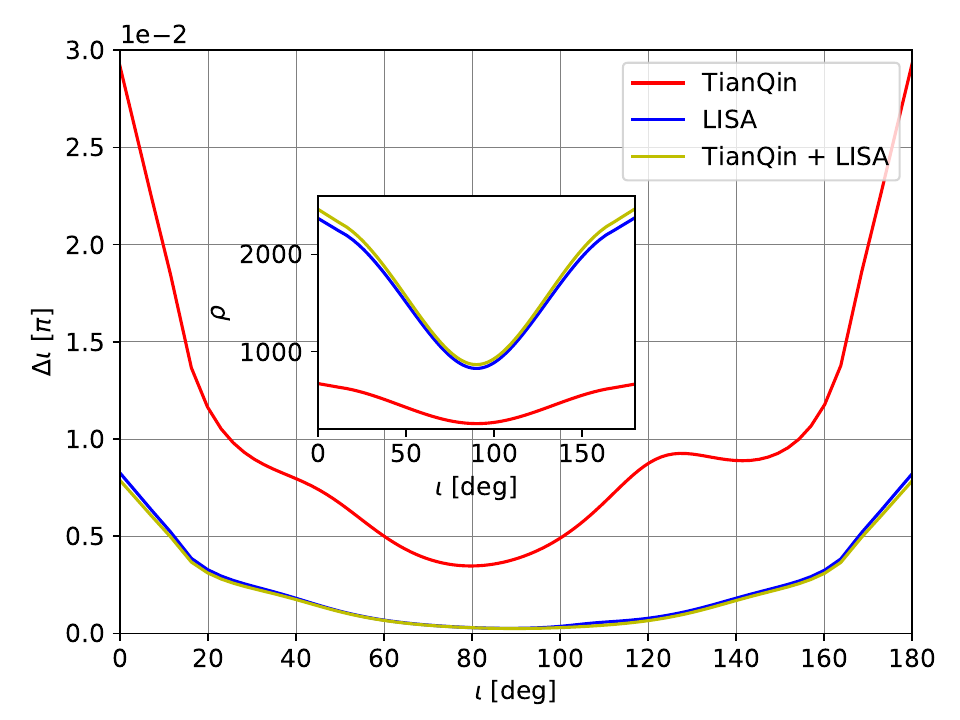}
\caption{
    The absolute error in the inclination of the source $\Delta\iota$ in fractions of $\pi$ as a function of the inclination $\iota$ for a MBHB. The error in TianQin is represented by the red line, the error in LISA by the blue line, and the error for joint detection is shown in yellow. The inset shows the corresponding SNR of the source where the color coding is the same as in the main plot.
    }
\label{fig:mbhb_diota}
\end{figure}

\textit{Inclination:} The error in the detection of the inclination of a MBHB in TianQin, LISA, and joint detection is shown in \fref{fig:mbhb_diota}. From the inset in the plot, we see that the SNR of a MBHB is symmetric for all three detection scenarios being the highest for a face-on or anti-face-on source, and the lowest for an edge-on source, which is as expected because an edge-on source only emits one polarization while for a (anti-)face-on source the two polarizations have the same strength thus resulting in a stronger signal. Moreover, we see that for all inclinations the SNR in TianQin is lower than in LISA ranging between roughly 300 and 800 for the first, and 900 and 2400 for the second. The higher SNR in LISA results in a high detection accuracy with an error of around $10^{-3}\pi$ for edge-on sources and an error between $2.5\times10^{-3}\pi$ and $7.5\times10^{-3}\pi$ for sources with high and low inclinations ($\iota\lesssim30^\circ$ and $\iota\gtrsim150^\circ$). TianQin in contrast, only has a detection accuracy of around $4\times10^{-3}\pi$ for edge-on sources and an error of the order $10^{-2}\pi$ for low/high inclination sources. Due to LISA's significantly higher SNR and better detection accuracy, the SNR and error in the joint detection resemble those of LISA closely. The inclination is constrained more accurately when the source is edge-on although the SNR is lower in this case because edge-on sources only emit one polarization resulting in a higher contribution from higher-order spherical modes which makes it easier to estimate the inclination.

\begin{figure}[tpb] \centering \includegraphics[width=0.48\textwidth]{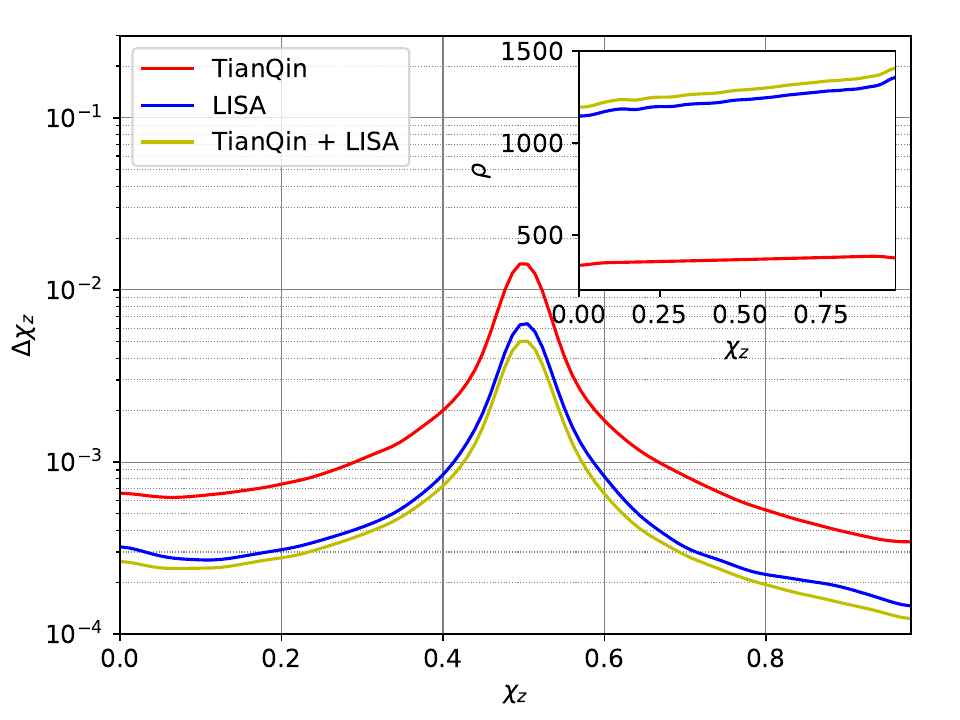}
\caption{
    The absolute error in the effective spin along the angular momentum of the binary $\Delta\chi_z$ over the magnitude of the effective spin along the angular momentum $\chi_z$ for one of the MBHs in an equal mass binary. The spin of the other BH is also assumed to be aligned with the angular momentum and its magnitude is fixed to 0.5. The error in TianQin is represented by the red line, the error in LISA by the blue line, and the error for joint detection is shown in yellow. The inset shows the corresponding SNR of the source where the color coding is the same as in the main plot.
    }
\label{fig:mbhb_dchiz}
\end{figure}

\textit{Effective spin:} \fref{fig:mbhb_dchiz} shows the absolute error of the effective spin along the angular momentum for a MBHB detected by TianQin, LISA, and in the case of joint detection. We see that for all spin values, TianQin and LISA perform at a similar level differing by a factor between two and three. Nevertheless, the joint detection only performs slightly better than LISA alone showing that TianQin contribution is minimal. The total error in TianQin is of the order $10^{-4}$ for $\chi_z\lesssim0.3$ and $\chi_z\gtrsim0.65$ being the smallest for a maximal spin of $0.98$. TianQin's error is, further, of the order of $10^{-3}$ for $0.3\lesssim\chi_z\lesssim0.65$ only going up to almost $1.5\times10^{-2}$ for $\chi_z\approx0.5$. The error in LISA if of the order $10^{-4}$ for $\chi_z\lesssim0.4$ and $\chi_z\gtrsim0.6$ and reaches its maximum of roughly $6.5\times10^{-3}$ for $\chi_z\approx0.5$ while the error in the joint detection is of the order $10^{-4}$ for $\chi_z\lesssim0.45$ and $\chi_z\gtrsim0.55$ with a maximum of around $5\times10^{-3}$ for $\chi_z\approx0.5$. As for TianQin, LISA, and joint detection have minimal error for a maximal spin of $0.98$. The minimal error at the highest spin corresponds in all three cases to a maximal SNR where, in general, we can see that the SNR increases as the spin increases; for TianQin the SNR goes from a bit more than 300 to almost 400 while for LISA and joint detection it goes from roughly 1150 to 1400. The SNR increases and the error decreases as the spin increases because a higher spin has a stronger effect on the merger process and thus the signal. The increased error around the spin magnitude of 0.5 however is induced by a degeneracy with the spin of the other MBH that also has a spin of 0.5.

\begin{figure}[tpb] \centering \includegraphics[width=0.48\textwidth]{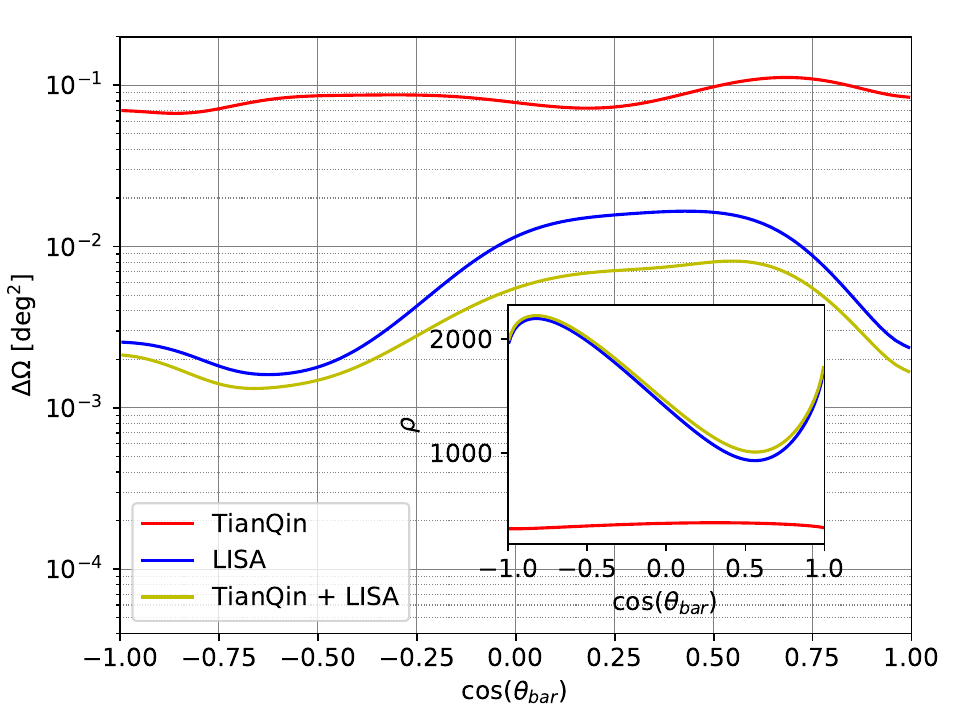}
\caption{
    The sky localization error $\Delta\Omega$ for a MBHB detected by TianQin (red), LISA (blue), and joint detection (yellow) as a function of $\cos(\theta_{\rm bar})$. The inset shows the corresponding SNR of the source where the color coding is the same as in the main plot.
    }
\label{fig:mbhb_domegat}
\end{figure}

\begin{figure}[tpb] \centering \includegraphics[width=0.48\textwidth]{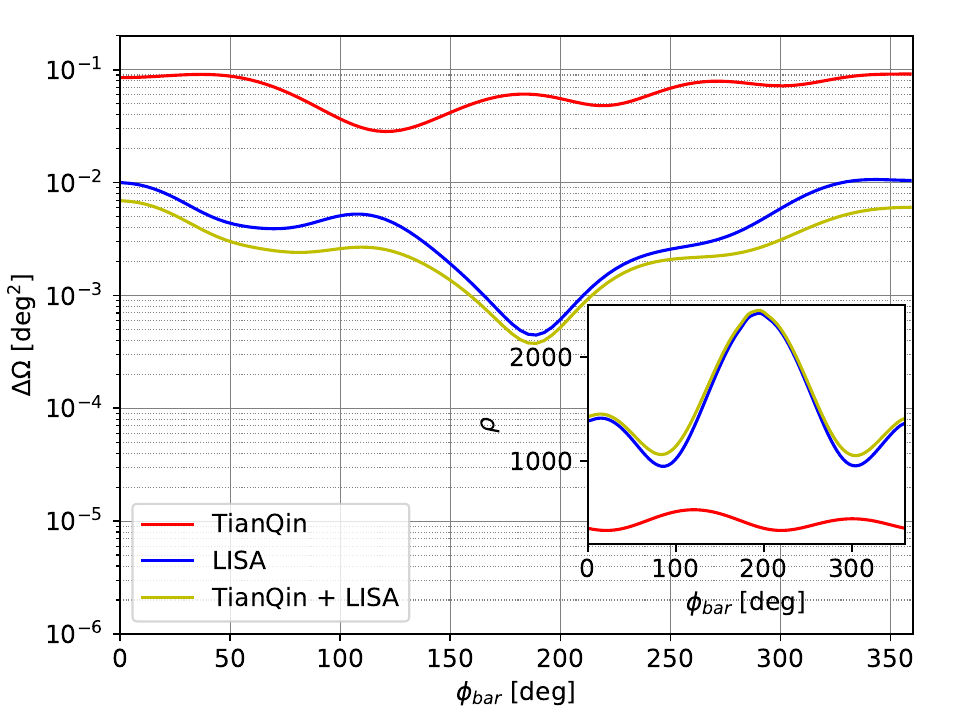}
\caption{
    The sky localization error $\Delta\Omega$ for a MBHB detected by TianQin (red), LISA (blue), and joint detection (yellow) as a function of $\phi_{\rm bar}$. The inset shows the corresponding SNR of the source where the color coding is the same as in the main plot.
    }
\label{fig:mbhb_domegap}
\end{figure}

\textit{Sky localization:} In \fref{fig:mbhb_domegat} and \fref{fig:mbhb_domegap}, we show the sky localization error for a MBHB in TianQin, LISA, and for a joint detection as a function of $\cos(\theta_{\rm bar})$ and $\phi_{\rm bar}$, respectively. We see that the SNR in TianQin does not change significantly along the latitude angle $\theta_{\rm bar}$ remaining at a level of roughly 300 and being only slightly higher for $\cos(\theta_{\rm bar})\gtrsim0$ due to TianQin's orientation towards RX J0806.3+1527. Similarly, the sky localization error in TianQin only varies a little around $10^{-1}\,{\rm deg^2}$ where the small oscillations appear because the accuracy with which TianQin can see the merger of the MBHB, which contains most of the SNR~\citep{tq_mbhb_2019a}, depends on the sky localization of the source. For LISA and the joint detection, which closely follows LISA because of its significantly higher SNR compared to TianQin, the SNR oscillates substantially having a maximum of around 2200 for $\cos(\theta_{\rm bar})\approx-0.85$ and a minimum of around 900 for $\cos(\theta_{\rm bar})\approx0.6$. Around the minimum TianQin's contribution to the SNR of joint detection becomes more significant but is still small. The sky localization error for LISA and joint detection have an almost inverse behavior, having the minimum of around $1.5\times10^{-3}\,{\rm deg^2}$ for $-0.75\lesssim\cos(\theta_{\rm bar})\lesssim-0.5$ and a maximum of roughly $10^{-2}\,{\rm deg^2}$ for $0\lesssim\cos(\theta_{\rm bar})\lesssim0.75$, where the maximum of the joint detection is less pronounced due to TianQin's contribution. Such an inverse behavior or anti-correlation is expected as it can be derived from basic properties of the FMA using that the sky localization is an extrinsic parameter~\citep{creighton_anderson_2011}. Note that the oscillations in the SNR and the sky localization error can again be attributed to the accuracy with which the merger of the MBHB is detected depending on the sky localization of the source.

For the azimuthal angle $\phi_{\rm bar}$ both detectors have a pronounced symmetry in the SNR. TianQin's SNR oscillates between roughly 300 and 600 where the maxima and minima are shifted due to its fixed orientation towards RX J0806.3+1527. LISA has a pronounced maximum of $\rho\approx2400$ around $180^\circ$, a smaller one of around 1600 for $\phi_{\rm bar}\approx0^\circ$, and two minima of $\rho\approx1000$ at $\phi_{\rm bar}\approx90^\circ$ and $\phi_{\rm bar}\approx300^\circ$. The symmetry in the SNR is partially reflected (as an anti-correlation) in the sky localization error which for TianQin and LISA has a global minimum at the location of the SNR's global maximum. The minimum for TianQin has a value of around $3\times10^{-2}\,{\rm deg^2}$ while for LISA the minimum is around $3.5\times10^{-4}\,{\rm deg^2}$. For TianQin the values outside of the minimum do not differ significantly being also of the order $10^{-2}\,{\rm deg^2}$ with maxima of almost $10^{-1}\,{\rm deg^2}$. For LISA, in contrast, the minimum is almost one order of magnitude smaller than the other values (outside $160^\circ\lesssim\phi_{\rm bar}\lesssim210^\circ$), those being of the order of magnitude $10^{-3}\,{\rm deg^2}$ and reaching $10^{-2}\,{\rm deg^2}$. The difference in the minima (maxima) of the sky localization error (SNR) for LISA appears because we fix the latitude of the source to be roughly $45^\circ$ and thus LISA is directed more towards the source for $\phi_{\rm bar}\approx180^\circ$ than for $\phi_{\rm bar}\approx0^\circ$. The SNR of the joint detection follows LISA's SNR closely with only a significant contribution from TianQin around the minima at $\phi_{\rm bar}\approx90^\circ$ and $\phi_{\rm bar}\approx300^\circ$. However, this sky localization error of the joint detection is around 1.4 times better than LISA's for most values of $\phi_{\rm bar}$, only being very similar to LISA's error close to its minima. This is because sky localization with two detectors is almost always better with two detectors, even if one detector has a significantly lower SNR.

\subsection{Comparing TianQin, LISA \& joint detection}\label{sec:mbhbc}

For MBHBs, LISA has a clear advantage compared to TianQin in terms of detection distance which also correlates with a higher SNR. The higher SNR results in LISA also having a better detection accuracy for most parameters. Only for low-mass sources with a total mass below $2\times10^5\,{\rm M_\odot}$ as well as for the source's inclination and the spin of the MBHs TianQin can perform at a similar level as LISA. Moreover, for sky localization, the contributions of TianQin to the joint detection can be significant despite performing one to two orders of magnitude worse than LISA.

\begin{figure}[tpb] \centering \includegraphics[width=0.48\textwidth]{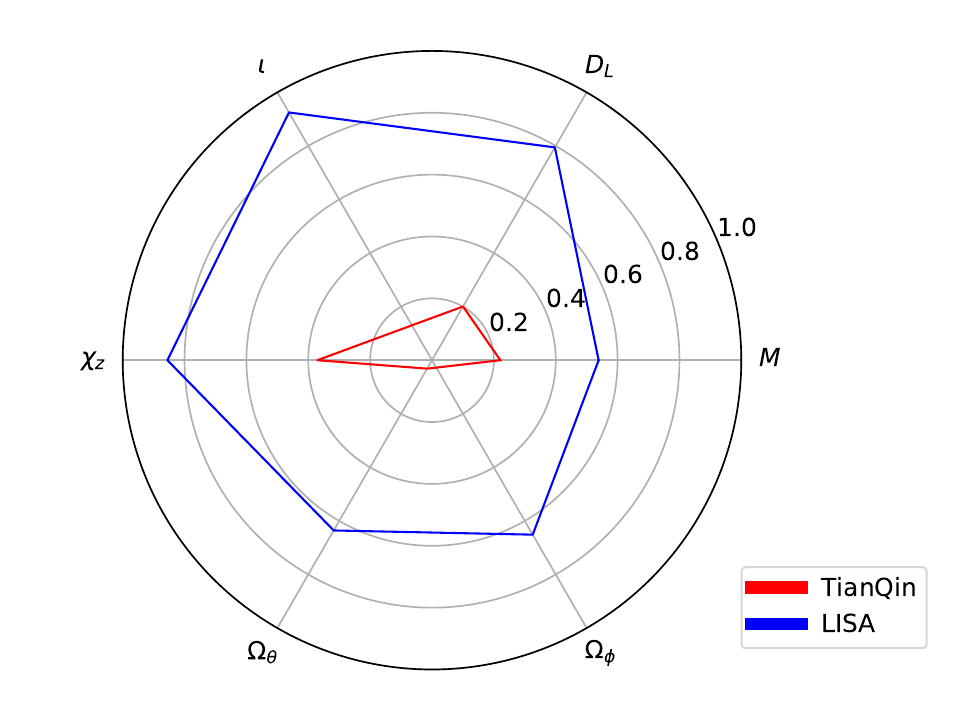}
\caption{
    The performance ratio $Q_X[\lambda]$ for TianQin and LISA, and the parameters $\lambda$ of a MBHB where we label the ratios only using the parameters. $\Omega_{\theta}$ and $\Omega_{\phi}$ are the sky localization as a function of $\theta_{\rm bar}$ and $\phi_{\rm bar}$, respectively.
    }
\label{fig:mbhb_radar}
\end{figure}

We show in \fref{fig:mbhb_radar} the performance ratio $Q$ for TianQin and LISA compared to the joint detection. For all parameters considered, on average LISA alone performs significantly better than TianQin alone. In particular, LISA dominates the result of the joint detection for the luminosity distance $D_L$, the source's inclination $\iota$, and the magnitude of the spin along the angular momentum $\chi_z$. However, TianQin's contribution to joint detection is significant for the mass due to its comparable performance for low masses and for sky localization. Therefore, a great benefit can be expected from joint detection, in particular, in terms of covering the full mass spectrum and localizing the source.

\section{Stellar-mass black holes binaries}\label{sec:sbhb}

Stellar-mass black holes (SBHs) with masses up to one hundred solar masses can be produced through the gravitational collapse of massive stars where the resulting mass depends on several parameters such as metallicity or stellar rotation~\citep{burrows_1988,belczynski_bulik_2010,oconnor_ott_2011,mapelli_zampieri_2013,spera_mapelli_2015,colpi_sesana_2017}, as primordial BHs resulting from over-densities in the early universe~\citep{bird_cholis_2016,carr_kuhnel_2016,ali-haimoud_kovetz_2017,inomata_kawasaki_2017,ando_inomata_2018,sasaki_suyama_2018}, or as a product of SBH binaries (SBHBs) merger~\citep{oleary_meiron_2016,fishbach_holz_2017,gerosa_berti_2017,rodriguez_amaro-seoane_2018,veske_marka_2020}. Before the first direct detection of GWs by the LIGO-Virgo collaboration in 2015~\citep{ligo_virgo_2016a} SBHs could only be observed through indirect detection using electromagnetic (EM) waves (mainly x-rays) emitted by a companion or accretion processes. Until now around 20 x-ray binaries have been observed, where most of them have SBHs with a mass lower than $20\,{\rm M_\odot}$~\citep{ligo_virgo_2016b}. Moreover, EM observations of SBHs suggested that there is a lower mass gap around $2-5\,{\rm M_\odot}$ between the most massive neutron stars and the lightest SBHs~\citep{ozel_freire_2016,margalit_metzger_2017,ozel_psaltis_2010,farr_sravan_2011,kreidberg_bailyn_2012,freire_ransom_2008} while pair-instability supernovae predict an upper mass gap around $50-120\,{\rm M_\odot}$~\citep{woosley_2017}. However, results from the first three observation rounds of the LIGO-Virgo collaboration revealed a large number of BHs with a mass higher than $20\,{\rm M_\odot}$ as well as candidates within the two mass gaps~\citep{GWTC1,GWTC2,GWTC3,ligo_virgo_2020a}. 

Besides the formation of SBH, another open question is how they come together to form binaries. This process can be largely divided into two channels: (i) Massive star binaries co-evolve and form SBHBs after the two stars die in a gravitational collapse~\citep{ligo_virgo_2016b,vanbeveren_2009,belczynski_dominik_2010,kruckow_tauris_2018,giacobbo_mapelli_2018,mapelli_giacobbo_2018,du-buisson_marchant_2020}, and (ii) SBHB form through dynamical processes in dense stellar environments~\citep{ligo_virgo_2016b,portegies-zwart_mcmillan_2000,gultekin_miller_2004,gultekin_miller_2006,zevin_samsing_2019,tagawa_haiman_2020,samsing_dorazio_2020,rodriguez_chatterjee_2016,rodriguez_haster_2016,liu_wang_2023}. In the first channel, we expect the SBHB to inherit the orbit and spins of their stellar progenitor and thus to have binaries with low eccentricity and the spins of the SBHs aligned to the orbital angular momentum. In the second channel, in contrast, high eccentricities and isotropic distribution of the spins are expected~\citep{ligo_virgo_2019,samsing_macleod_2014,antonini_chatterjee_2016,samsing_dorazio_2018,kremer_rodriguez_2019}.

While SBHBs merge at frequencies where ground-based GW detectors are most sensitive, their early inspiral could be observed by space-borne GW detectors like TianQin and LISA~\citep{tq_sbhb_2020,tq_sbhb_2022,sesana_2016,seto_2016,kyutoku_seto_2016,liu_shao_2020}. We expect that in the mHz band, the orbits of SBHBs have not been circularized by GWs, thus allowing us to differentiate their formation scenarios by measuring the orbital eccentricity~\citep{nishizawa_berti_2016,nishizawa_sesana_2017,breivik_rodriguez_2016}. Moreover, the detection of SBHBs in space can be used to measure the expansion of the universe and to test modified gravity with great precision~\citep{del-pozzo_sesana_2018,kyutoku_seto_2017,barausse_yunes_2016,chamberlain_yunes_2017}. In this section, we study the distance to which SBHBs will be detected by TianQin and/or LISA adopting a restricted 3-PN waveform model including eccentricity but ignoring the spins of the component SBHs~\citep{buonanno_iyer_2009,krolak_kokkotas_1995,tq_mbhb_2019a,tq_sbhb_2020}. We, further, adopt a detection threshold of $\rho=5$ which has been used for studies considering a network of GW detectors~\citep{sesana_2016,wong_kovetz_2018,tq_sbhb_2020}. The detection accuracy for the most important parameters is estimated using a FMA. A more detailed discussion on the detection of SBHBs by TianQin and LISA can be found in, e.g., \cite{tq_sbhb_2020}, \cite{tq_sbhb_2022}, \cite{sesana_2016}, \cite{moore_gerosa_2019}, and \cite{lisa_2022b}.

\subsection{Detectability}\label{sec:sbhbd}

\begin{figure}[tpb] \centering \includegraphics[width=0.48\textwidth]{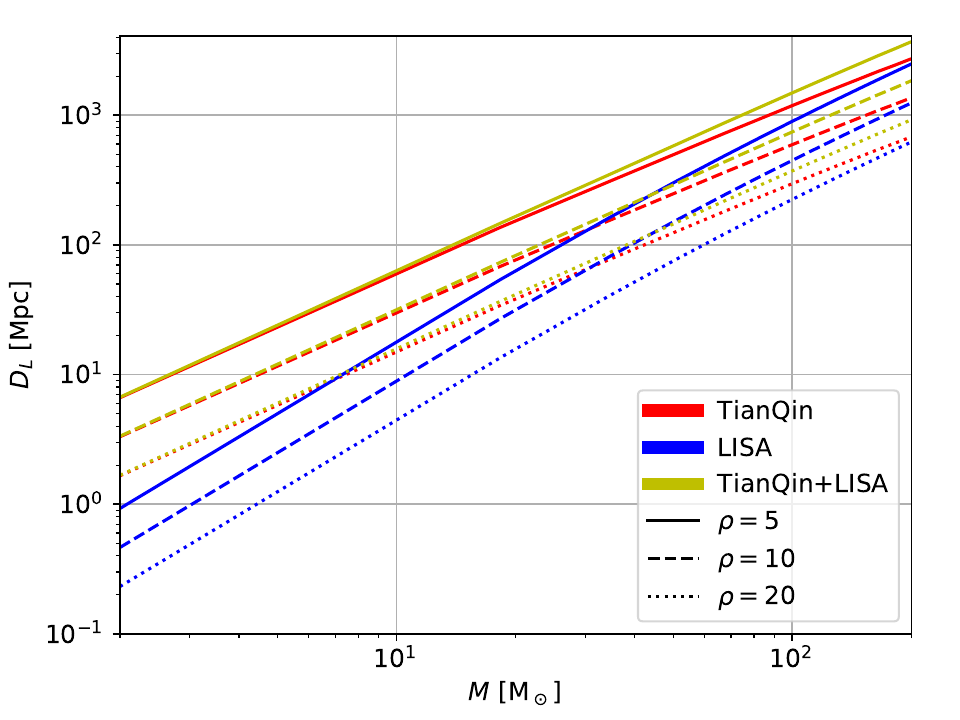}
\caption{
    The luminosity distance $D_L$ to which a SBHB can be detected with a fixed SNR $\rho$ for TianQin, LISA, and joint detection as a function of the binary's total mass $M$.
    }
\label{fig:sbhb_hor_snr}
\end{figure}

An important parameter in the study of SBHBs using GWs is the distance to which we will be able to detect them depending on their total mass. This question is discussed in this section for equal-mass binaries detected by TianQin, LISA, and joint detection. We see in \fref{fig:sbhb_hor_snr} that SBHBs will be detectable to luminosity distances of up to a few ${\rm Gpc}$ for a SNR of five if their total mass is at the higher end of the spectrum of $100\,{\rm M_\odot}$ and above. In this mass range, all three detection scenarios work similarly well with a slight improvement for joint detection. For lighter sources between $10\,{\rm M_\odot}$ and $100\,{\rm M_\odot}$ TianQin alone works significantly better than LISA alone, reaching luminosity distances between roughly $50\,{\rm Mpc}$ and $1\,{\rm Gpc}$ for a SNR of five and between $15\,{\rm Mpc}$ and $700\,{\rm Mpc}$ for $\rho = 20$. For the lighter SBHBs in this range, LISA alone will reach distances shorter by a factor of around three while for the heavier ones, the distance will be shorter by a factor of around 1.4. TianQin performs better for lighter sources because of its improved sensitivity at higher frequencies compared to LISA, which allows it to track the signal of these sources with better accuracy at times when the signal is chirping stronger. The distance reached by the joint detection is similar to the one by TianQin alone but with slight improvements for total masses of $20\,{\rm M_\odot}$ and above. For masses below $10\,{\rm M_\odot}$, TianQin alone and the joint detection will perform almost identically reaching distances between $7\,{\rm Mpc}$ and $65\,{\rm Mpc}$ while performing eight to four times better than LISA alone, respectively.

\begin{figure}[tpb] \centering \includegraphics[width=0.48\textwidth]{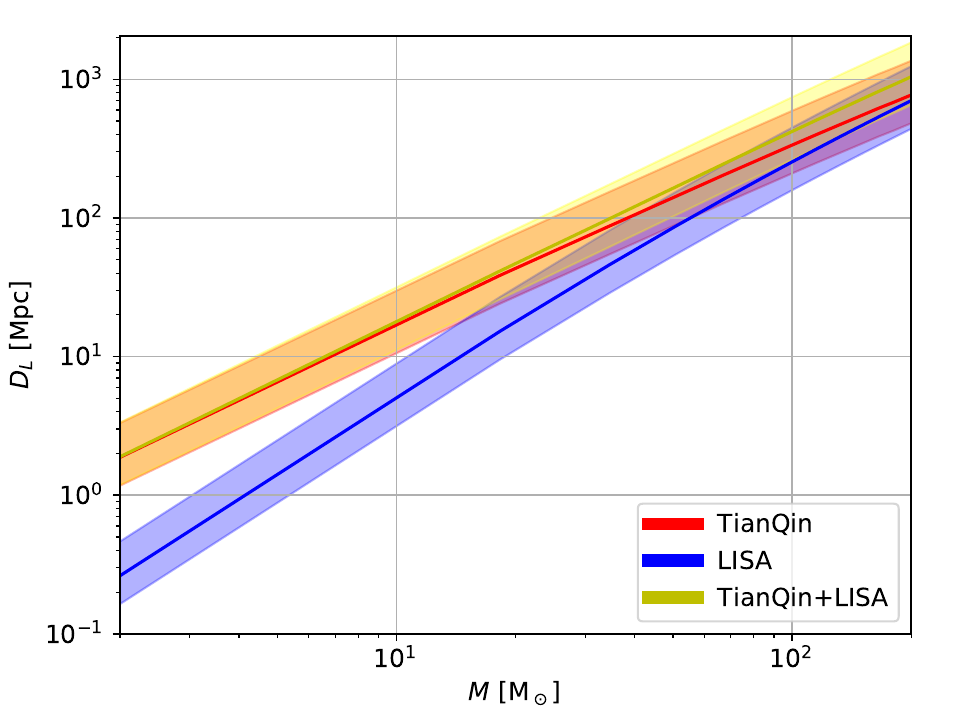}
\caption{
    The luminosity distance $D_L$ to which a SBHB with a SNR of ten can be detected for a varying inclination for TianQin, LISA, and joint detection as a function of the binary's total mass $M$. The upper edge of the shaded region corresponds to the face-on case while the lower edge corresponds to the edge-on case; the solid line corresponds to the average inclination of $60^\circ$.
    }
\label{fig:sbhb_hor_inc}
\end{figure}

\fref{fig:sbhb_hor_inc} shows the dependence of the luminosity distance to which SBHBs can be detected as a function of the total mass for different inclinations of the source and a SNR of ten for TianQin, LISA, and joint detection. We see that in all three detection scenarios, a face-on source will be detected around three times further away than an edge-on source independent of the total mass. For an average inclination of $60^\circ$, the detection distance will be around two times shorter than for the face-on case -- again independent of the detection scenario and the total mass. In particular, we see that for a typical mass of $20\,{\rm M_\odot}$ TianQin alone as well as the joint detection will see a face-on source out to a distance of around $80\,{\rm Mpc}$ while a source with an inclination of $60^\circ$ will be detectable to a distance of around $40\,{\rm Mpc}$ and an edge-on source to a distance of less than $30\,{\rm Mpc}$. The same source will be detectable by LISA alone to a distance of around $30\,{\rm Mpc}$ if it is face-on, $15\,{\rm Mpc}$ if it has an average inclination, and a bit more than $10\,{\rm Mpc}$ if it is edge-on.

\subsection{Parameter estimation}\label{sec:sbhbp}

In this subsection, we analyze how accurately different parameters of SBHB can be detected by TianQin and LISA alone as well as by their joint detection. The parameters we study are the chirp mass of the binary $\mathcal{M}_c$ (in the observer frame), the luminosity distance of the source $D_L$, the binary's eccentricity at a reference frequency of $10\,{\rm mHz}$ $e_0$, the symmetric mass ratio of the two BHs $\eta$, the sky localization error of the source $\Delta\Omega$, and the time to coalescence $t_c$. In each analysis, we only vary one parameter while fixing the other parameters to the following values $\mathcal{M}_c = 50\,{\rm M_\odot}$, $D_L = 206\,{\rm Mpc}$ (corresponding to $z = 0.05$), $e_0 = 0.05$, and $\eta = 0.125$. We, further, set the source to be at the sky location $\theta_{\rm bar} = 60^\circ$ and $\phi_{\rm bar} = 180^\circ$ where we use barycentric coordinates and set the binary to be at $2.5\,{\rm yr}$ before the merger.

\begin{figure}[tpb] \centering \includegraphics[width=0.48\textwidth]{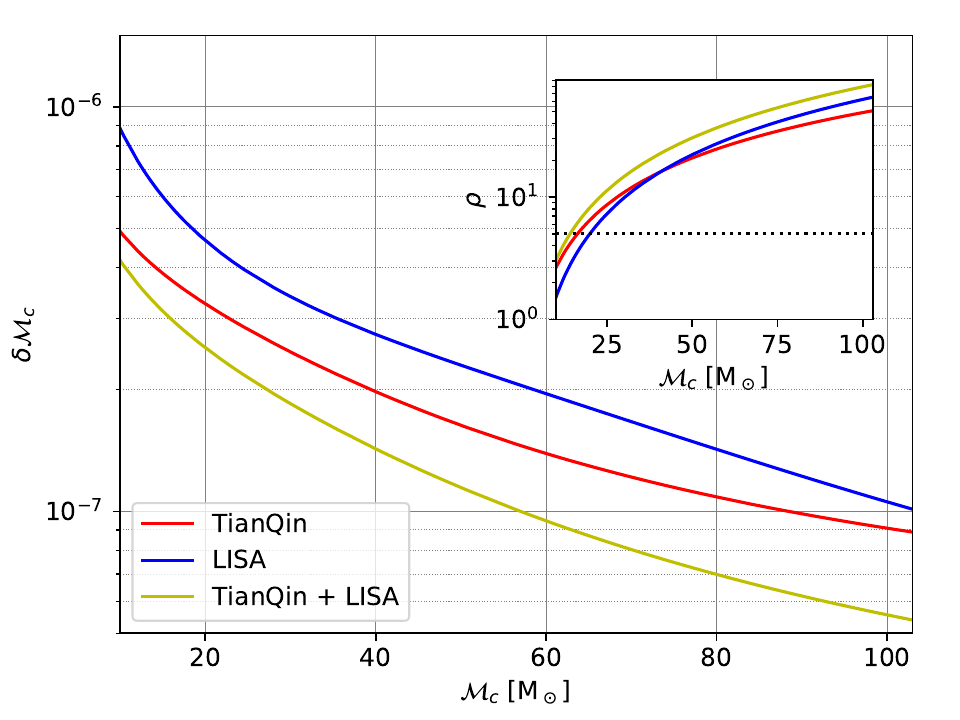}
\caption{
    The relative error in the chirp mass of a SBHB $\delta\mathcal{M}_c$ over the chirp mass of the system in the observer frame $\mathcal{M}_c$. The red line shows the detection accuracy for TianQin alone, the blue line for LISA alone, and the yellow line for joint detection. The inset shows the corresponding SNR of the source where the color coding is the same as in the main plot and the black dotted line indicates the detection threshold of $\rho=5$.
    }
\label{fig:sbhb_dmc}
\end{figure}

\textit{Chirp mass:} In \fref{fig:sbhb_dmc} we see the relative error with which the source's chirp mass in the observer frame will be measured by TianQin, LISA, and joint detection. We see that for all masses considered TianQin alone performs better than LISA alone because TianQin's better sensitivity at higher frequencies allows it to track the chirping of the signal for a longer time. TianQin will detect the chirp mass of the sources with a relative error of around $5\times10^{-7}$ for $\mathcal{M}_c \approx 10\,{\rm M_\odot}$ and of around $9\times10^{-8}$ for a chirp mass of a bit more than $100\,{\rm M_\odot}$ while LISA alone will have a relative error of $9\times10^{-7}$ and $10^{-7}$ for the same masses. That TianQin performs better than LISA can be understood from the fact that TianQin is more sensitive to higher frequencies and thus to lighter sources such as SBHBs. In particular, we can see from the SNR of the sources that TianQin has a better detection accuracy than LISA for all masses even if LISA can reach a slightly higher SNR for SBHB with a chirp mass above around $40\,{\rm M_\odot}$. Because the difference between TianQin alone and LISA alone is, however, not very big, the joint detection will perform significantly better than any single detection -- in particular, for higher masses. We see that joint detection will have a detection accuracy of $4\times10^{-7}$ for $\mathcal{M}_c \approx 10\,{\rm M_\odot}$ and that the error decreases for an increasing chirp mass to almost $5\times10^{-8}$ for $\mathcal{M}_c \gtrsim 100\,{\rm M_\odot}$.

\begin{figure}[tpb] \centering \includegraphics[width=0.48\textwidth]{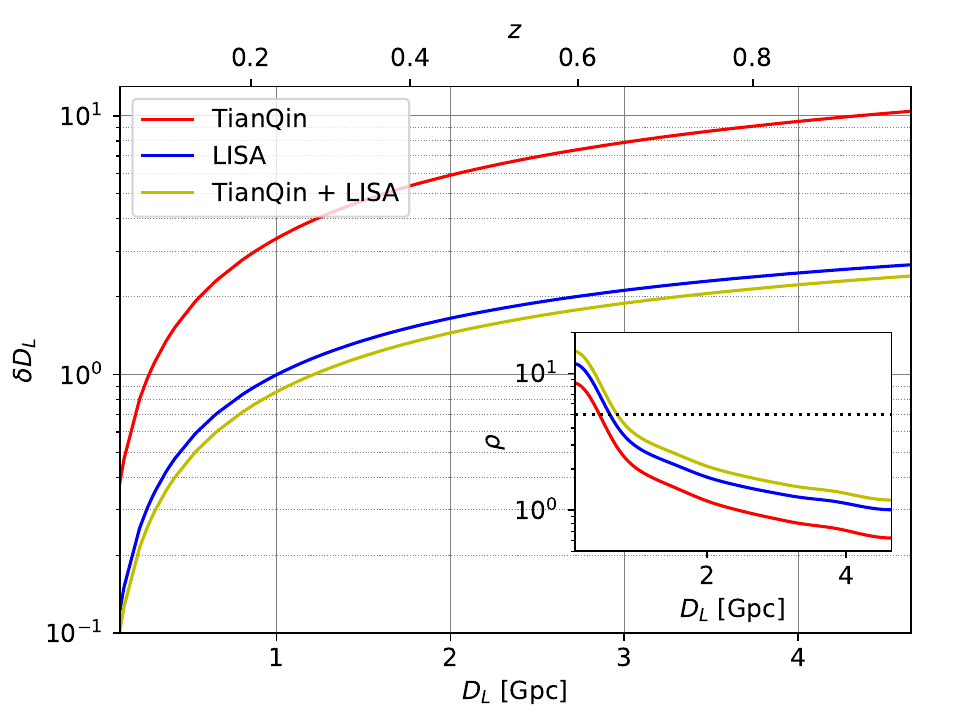}
\caption{
    The relative error in the luminosity distance $\delta D_L$ of a SBHB. The lower abscissa indicates the different luminosity distances of the source $D_L$ while the upper abscissa shows the corresponding cosmological redshift $z$. The red line shows the detection accuracy for TianQin alone, the blue line for LISA alone, and the yellow line for joint detection. The inset shows the corresponding SNR of the source where the color coding is the same as in the main plot and the detection threshold ($\rho=5$) is shown by the black dotted line.
    }
\label{fig:sbhb_ddl}
\end{figure}

\textit{Luminosity distance:} The relative error in the luminosity distance of a SBHB is shown in \fref{fig:sbhb_ddl}. For close binaries at redshifts of almost zero, the relative error is around 0.4, 0.15, and 0.1 for TianQin, LISA, and joint detection, respectively. We, further, see that for TianQin alone the relative error is only below one for redshifts of less than 0.1. For LISA alone the relative error remains below one up to $D_L = 1000\,{\rm Mpc}$ or a redshift of a bit more than 0.2 while for joint detection, the relative error remains below one for $z \approx 0.3$. The relative error increases so fast and becomes up to several times the actual distance because the SNR decreases quickly with an increasing distance, going for all detection cases below a detection threshold of $\rho=5$ at distances of less than $1000\,{\rm Mpc}$.

\begin{figure}[tpb] \centering \includegraphics[width=0.48\textwidth]{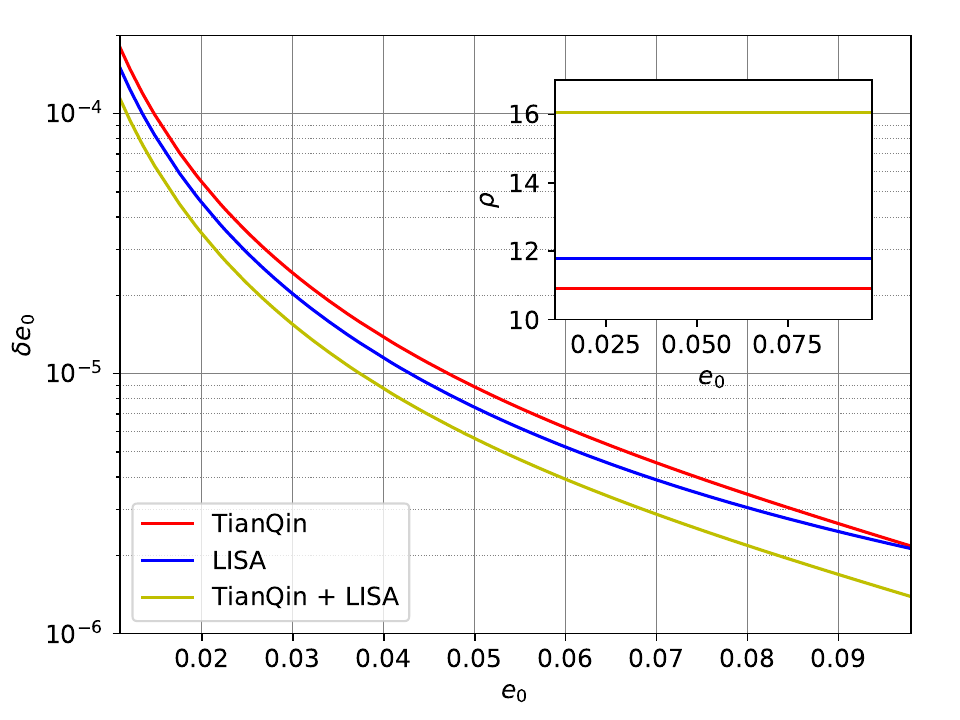}
\caption{
    The relative error in the eccentricity of a SBHB $\delta e_0$ for different eccentricities at a reference frequency of $10\,{\rm mHz}$ $e_0$. The red line shows the detection accuracy for TianQin alone, the blue line for LISA alone, and the yellow line for joint detection. The inset shows the corresponding SNR of the source with the same color coding as in the main plot.
    }
\label{fig:sbhb_de0}
\end{figure}

\textit{Eccentricity:} \fref{fig:sbhb_de0} shows that the relative error in the eccentricity of a SBHB will be detected slightly better by LISA alone than by TianQin alone correlating with a slightly higher SNR in LISA. However, in both cases, the relative error is going to be a bit more than $10^{-4}$ for $e_0 \approx 0.01$, slightly below $10^{-5}$ for $e_0 \approx 0.05$, and around $2\times10^{-6}$ for $e_0 \approx 0.1$. Joint detection will perform better than the two single detections by a factor of around 1.3 for all eccentricities. LISA can detect the (initial) eccentricity more accurately since it is mainly constrained during the early inspiral of the SBHB where LISA performs better than TianQin.

\begin{figure}[tpb] \centering \includegraphics[width=0.48\textwidth]{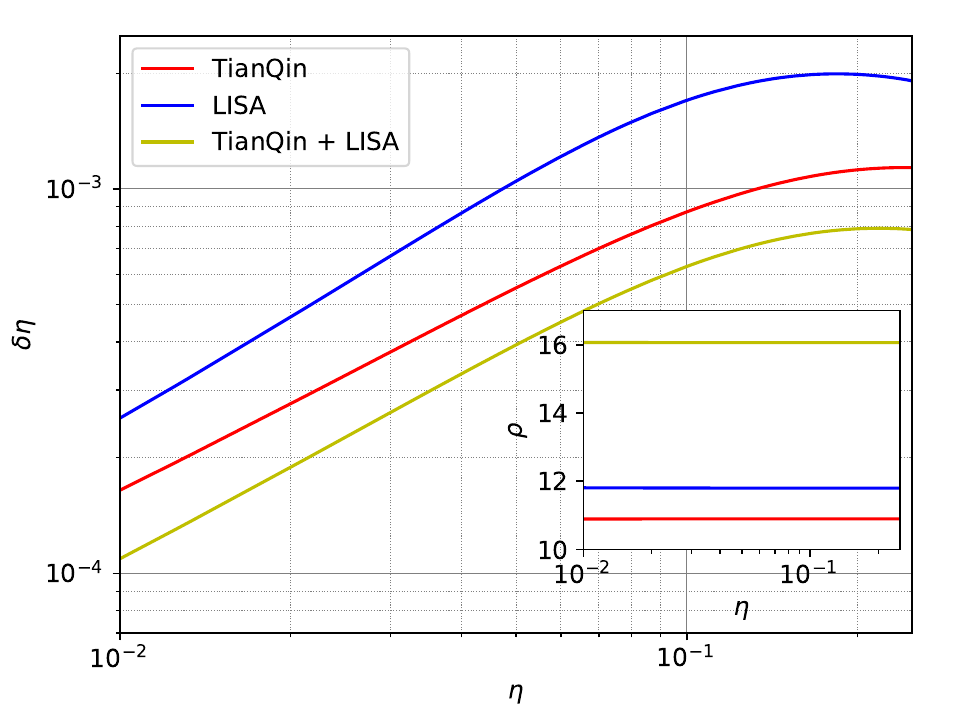}
\caption{
    The relative error in the symmetric mass ratio of a SBHB $\delta\eta$ for different symmetric mass ratios $\eta$. The red line shows the detection accuracy for TianQin alone, the blue line for LISA alone, and the yellow line for joint detection. The inset shows the corresponding SNR of the source with the same color coding as in the main plot.
    }
\label{fig:sbhb_deta}
\end{figure}

\textit{Symmetric mass ratio:} We see in \fref{fig:sbhb_deta} that the symmetric mass ratio $\eta$ will be detected slightly better by TianQin alone than by LISA alone despite the source having a slightly lower SNR in TianQin. This feature arises because the symmetric mass ratio of the SBHB is not measured directly but depends on the estimation of the source's chirp mass~\citep{maggiore_2018}. As we can see in \fref{fig:sbhb_dmc}, TianQin can measure the chirp mass of a SBHB more accurately (even for a lower SNR) and as a result the symmetric mass ratio is also constrained more tightly. The error in $\eta$ increases for an increasing symmetric mass ratio and TianQin alone will detect $\eta$ with a relative error of around $1.5\times10^{-4}$, $9\times10^{-4}$, and $10^{-3}$ for $\eta$ equal to $10^{-2}$, $10^{-1}$, and $0.25$, respectively. The relative error in LISA is around 1.6 times higher for all $\eta$ considered while the joint detection performs slightly better than the two detectors alone by a factor of around 1.4 compared to TianQin's detection accuracy.

\begin{figure}[tpb] \centering \includegraphics[width=0.48\textwidth]{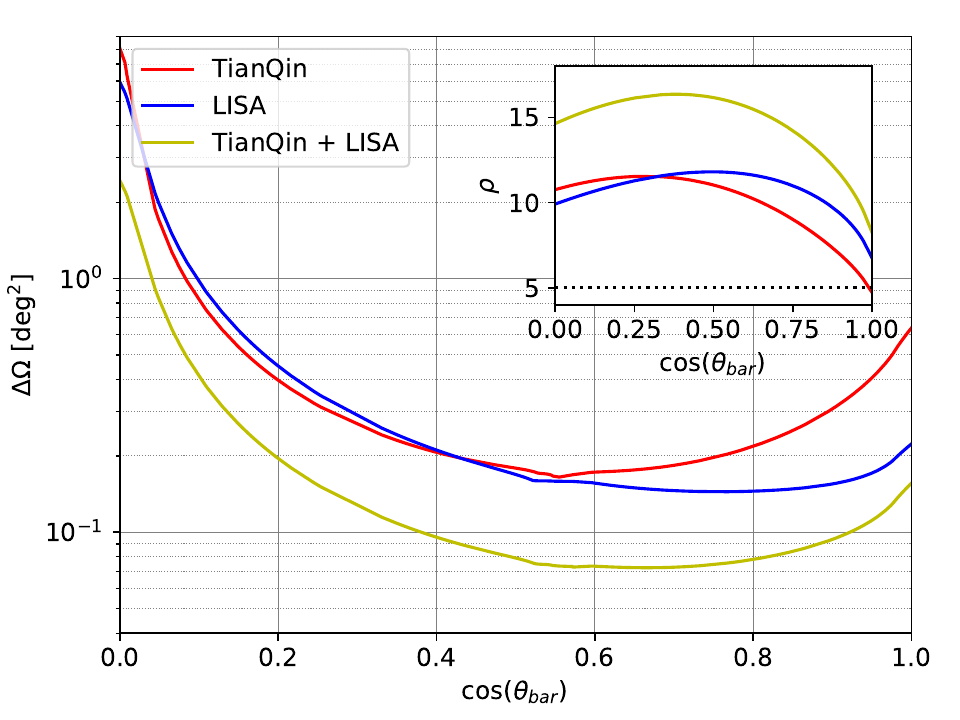}
\caption{
    The sky localization error of a SBHB $\Delta\Omega$ for different $\cos(\theta_{\rm bar})$. The red line shows the sky localization for TianQin alone, the blue line for LISA alone, and the yellow line for joint detection. The inset shows the corresponding SNR of the source where the color coding is the same as in the main plot and the detection threshold of $\rho=5$ is indicated by the black dotted line.
    }
\label{fig:sbhb_domegat}
\end{figure}

\begin{figure}[tpb] \centering \includegraphics[width=0.48\textwidth]{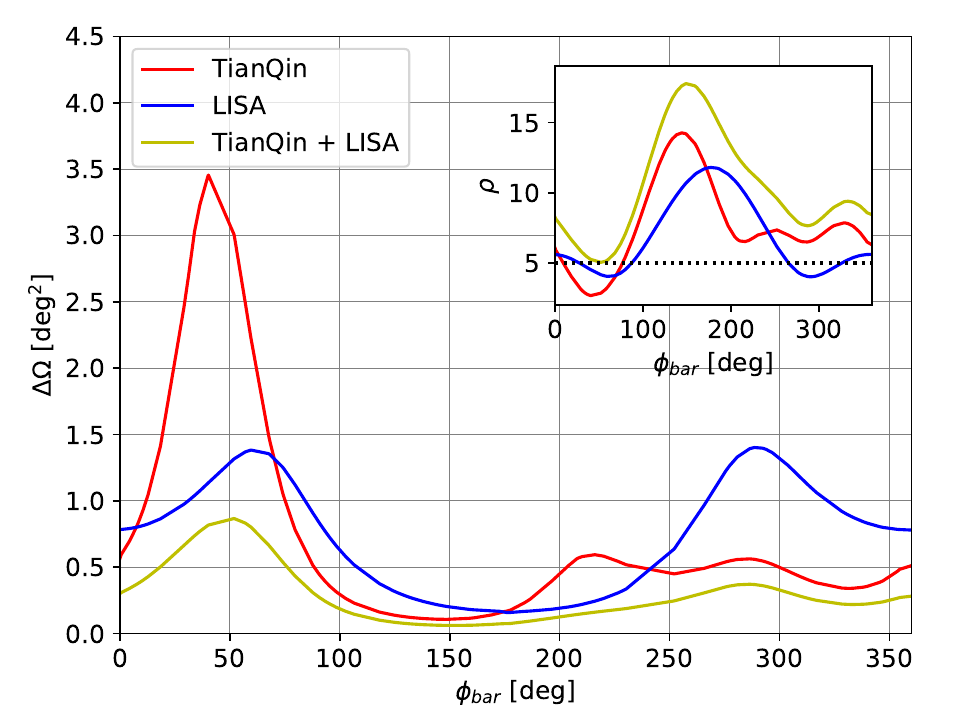}
\caption{
    The sky localization error of a SBHB $\Delta\Omega$ for different $\phi_{\rm bar}$. The red line shows the sky localization for TianQin alone, the blue line for LISA alone, and the yellow line for joint detection. The inset shows the corresponding SNR of the source with the same color coding as in the main plot and the black dotted line indicates the detection threshold ($\rho=5$).
    }
\label{fig:sbhb_domegap}
\end{figure}

\textit{Sky localization:} The sky localization error is shown in \fref{fig:sbhb_domegat} and \fref{fig:sbhb_domegap} for different $\cos(\theta_{\rm bar})$ and different $\phi_{\rm bar}$, respectively. For $\theta_{\rm bar}$ we only show the northern hemisphere since the southern hemisphere is almost symmetric. We see that for low angles between $\cos(\theta_{\rm bar}) = 0$ and $\cos(\theta_{\rm bar})\lesssim 0.5$ TianQin and LISA perform very similarly with an accuracy going from around $6\,{\rm deg^2}$ to slightly below $0.2\,{\rm deg^2}$. For angles higher than $\cos(\theta_{\rm bar}) = 0.5$ LISA alone starts performing better than TianQin alone having a sky localization error of around $0.2\,{\rm deg^2}$ for $\cos(\theta_{\rm bar}) = 1$ while TianQin has an error of around $6.5\,{\rm deg^2}$ for the same angle. That TianQin performs worse for higher angles is a result of its orientation towards RX J0806.3+1527 which lies close to the ecliptic plane, in contrast to LISA, with an orientation that changes while it orbits around the sun. The joint detection performs better than any of the two single detectors by a factor of around two to three throughout all angles considered where the improvement is more significant for angles below $60^\circ$ ($\cos(\theta_{\rm bar})=0.5$).

For the sky localization error in dependence of $\phi_{\rm bar}$, we see that it is almost symmetric in LISA reaching a minimum of around $0.2\,{\rm deg^2}$ for $\phi_{\rm bar} \approx 180^\circ$ and maxima of around $1.4\,{\rm deg^2}$ for $\phi_{\rm bar} \approx 90^\circ$ and $\phi_{\rm bar} \approx 270^\circ$. For $\phi_{\rm bar} \approx 0^\circ$ the sky localization error in LISA is of around $0.8\,{\rm deg^2}$. This symmetric behavior is expected because LISA changes its orientation throughout its orbit and makes a full rotation in one orbit. For TianQin such a symmetric behavior is not found, even when considering different orientations of the source. We attribute this behavior to TianQin having a fixed orientation throughout its orbit, a scheme of three months of data collection followed by three months where no data is collected, and the particular time when a source merges, which results in a complex function for when TianQin accumulates most of the signal. We find for TianQin alone a sky localization error between around $1\,{\rm deg^2}$ and $0.1\,{\rm deg^2}$ for $\phi_{\rm bar}$ between around $90^\circ$ and $180^\circ$ and thus slightly better than LISA. For $\phi_{\rm bar}$ between $180^\circ$ and $360^\circ$ the sky localization error oscillates around $0.5\,{\rm deg^2}$ and is thus similar or better than for LISA. However, for $\phi_{\rm bar} \approx 45^\circ$ the sky localization error in TianQin is significantly worse than in LISA going up to almost $3.5\,{\rm deg^2}$. The joint detection has also some symmetry in the sky localization error but the asymmetry in TianQin leads to a notable difference at its most extreme points. In particular, around $50^\circ$ where TianQin performs significantly worse than LISA, the joint detection also has its worst accuracy of around $0.8\,{\rm deg^2}$ while around $270^\circ$ where LISA has a maximum but TianQin performs much better, the joint detection has a sky localization error of only $0.4\,{\rm deg^2}$, and thus around three times better than LISA alone and around 1.5 better than TianQin alone. In general, the joint detection performs better than any of the two single detections having an error of less than $0.5\,{\rm deg^2}$ for all angles outside an interval between roughly $20^\circ$ and $80^\circ$.

\begin{figure}[tpb] \centering \includegraphics[width=0.48\textwidth]{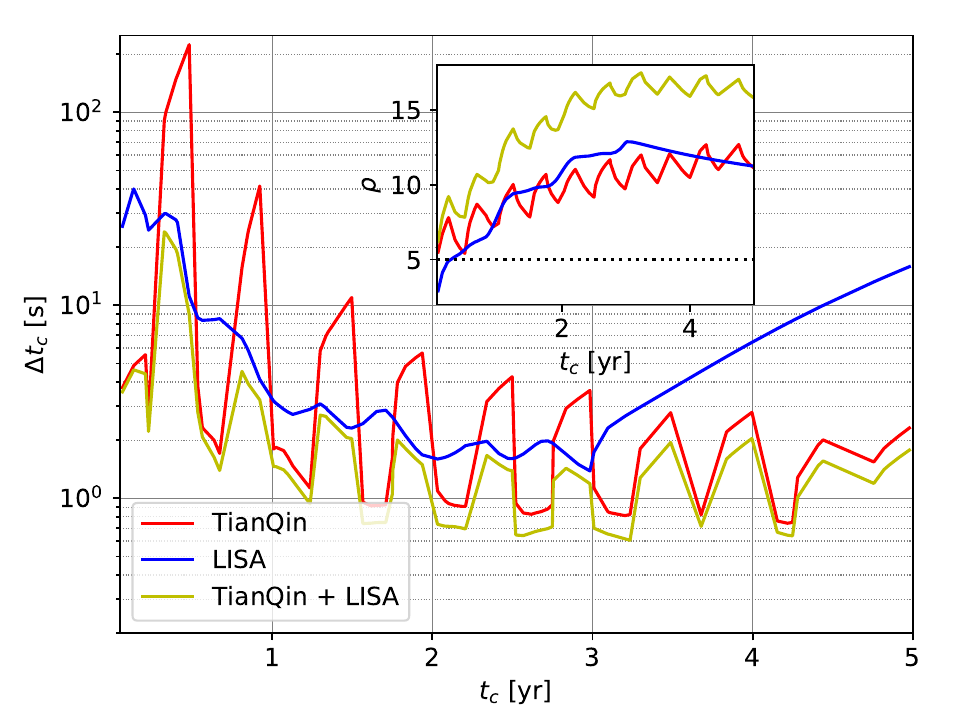}
\caption{
    The absolute error in the time to coalescence for a SBHB $\Delta t_c$ (in seconds) for different times to coalescence $t_c$ (in years). The red line shows the detection accuracy for TianQin alone, the blue line for LISA alone, and the yellow line for joint detection. The inset shows the corresponding SNR of the source where the color coding is the same as in the main plot and the detection threshold of $\rho=5$ is indicated by the black dotted line.
    }
\label{fig:sbhb_dtc}
\end{figure}

\textit{Time to coalescence:} In \fref{fig:sbhb_dtc}, we see the total error in seconds with which the time to coalescence $t_c$ of a SBHB can be determined with TianQin, LISA, and joint detection. We see that for all three detection scenarios, $t_c$ will be determined with an accuracy of at least $10\,{\rm s}$ and down to around $1\,{\rm s}$ as long as the binary is between one and four years from the merger. If the binary is less than one year from the merger then the error increases reaching up to several tens of seconds for LISA and the joint detection, and even over $100\,{\rm s}$ for TianQin since the SNR accumulated decreases significantly. For a source that is more than $3\,{\rm yr}$ from coalescence, the error in LISA increases significantly because we assumed LISA to collect $3\,{\rm yr}$ of data ($4\,{\rm yr}$ life time times 0.75 from the duty cycle) which leads the merger to not be detected and hence a decrease in SNR. For SBHBs closer than $3\,{\rm yr}$ to merger we see for LISA an oscillatory behavior due to the change of orientation of LISA in time relative to the source at the time of the merger. Note that the real detection of $t_c$ in LISA will be actually more complicated since the 25\,\% off-time actually consists of data gaps that vary in their length distributed throughout the 4\,{\rm yr} the data is collected~\citep{lisa_2022a}. For TianQin we also see an oscillatory behavior which however corresponds to TianQin being three months on and three months off. We see that for the time when TianQin is on, the detection error is lower than in LISA but it becomes bigger than the error in LISA when TianQin is off. However, for TianQin the error does not increase for an increasing time from coalescence since TianQin will collect data for the full $5\,{\rm yr}$. For joint detection, we see that the error follows quite closely the error in TianQin when TianQin is on and it becomes bounded from above by LISA's detection accuracy when TianQin is off.

\subsection{Comparing TianQin, LISA \& joint detection}\label{sec:sbhbc}

TianQin and LISA show comparable results for SBHBs, although, TianQin has an advantage towards the lower-mass end of the spectrum. SBHBs tend to have slightly higher SNR in TianQin than in LISA but the difference is, in general, only marginal. One of the parameters that show a striking difference is the time to coalescence where TianQin tends to have better accuracy but the overall performance is strongly reduced during the off-times.

\begin{figure}[tpb] \centering \includegraphics[width=0.48\textwidth]{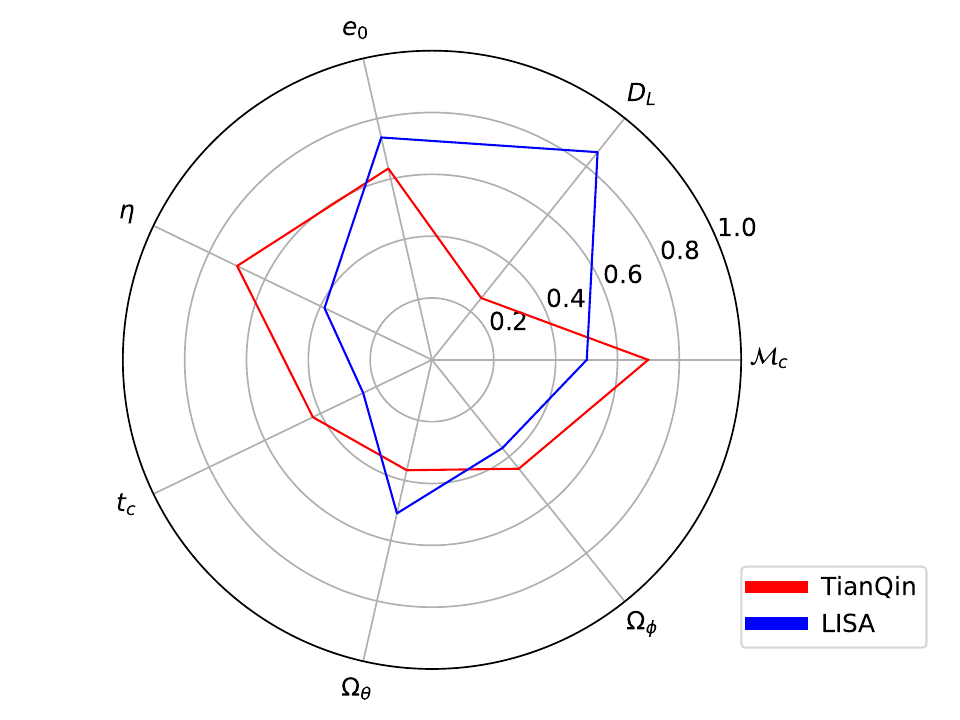}
\caption{
    The performance ratio $Q_X[\lambda]$ for TianQin and LISA, and the parameters $\lambda$ considered for SBHBs, where we label the ratios only using the parameters. $\Omega_{\theta}$ and $\Omega_{\phi}$ indicate the sky localization as a function of $\theta_{\rm bar}$ and $\phi_{\rm bar}$, respectively.
    }
\label{fig:sbhb_radar}
\end{figure}

\fref{fig:sbhb_radar} shows the performance ratio $Q$ for TianQin and LISA compared to the joint detection. We see that on average TianQin performs better than LISA for the chirp mass $\mathcal{M}_c$, the symmetric mass ratio $\eta$, and the time to coalescence $t_c$ while LISA performs better for the luminosity distance $D_L$ and the eccentricity $e_0$. For the sky localization $\Omega$ LISA performs better along $\theta_{\rm bar}$ since it covers a bigger portion of the sky but TianQin performs better along $\phi_{\rm bar}$. In general, we find that except for the luminosity distance that can be well constrained by LISA alone and lower-mass SBHBs where TianQin provides better results, joint detection performs significantly better than any of the two single detection scenarios. The greatest benefit from joint detection can thus be expected from an overall improved performance due to the high combined SNR.

\section{Double white dwarfs}\label{sec:dwd}

Double white dwarfs (DWDs) comprise the absolute majority (up to $10^8$) of all kinds of compact stellar-mass binaries in the Milky Way and hence are expected to be the most numerous GW source for space-based detectors~\citep{nelemans_yungelson_2001a,yu_jeffery_2010,lamberts_garrison-kimmel_2018,breivik_coughlin_2020}. Detecting them using GWs will significantly advance our knowledge on white dwarfs and thus potentially help shed light on different problems: (i) DWD are the end products of low-mass binary evolution thus encoding information on processes like mass transfer and common envelope phases~\citep{postnov_yungelson_2014,belczynski_kalogera_2002}, (ii) short-period ($\lesssim 1$ hour) mass-transferring DWDs -- so-called AM canum venaticorum systems -- are ideal for studying the stability of mass transfer~\citep{nelemans_yungelson_2001b,marsh_nelemans_2004,solheim_2010,tauris_2018}, and (iii) DWDs are likely to originate a variety of transient events including type-Ia supernovae~\citep{webbink_1984,iben_tutukov_1984,bildsten_shen_2007}. Furthermore, detached DWDs are an ideal laboratory to study the physics of tides, which will allow us to understand the nature of white dwarf viscosity~\citep{piro_2011,fuller_lai_2012,dallosso_rossi_2014,mckernan_ford_2016} while they also can be used to constrain deviations from general relativity~\citep{cooray_seto_2004,littenberg_yunes_2019}.

Besides studying the physics of stellar-mass compact objects (COs) and their predecessors, DWDs also allow exploring the galactic stellar population as a whole as well as the structural properties of the Milky Way~\citep{breivik_coughlin_2020,benacquista_holley-bockelmann_2006,adams_cornish_2012,korol_rossi_2019,wilhelm_korol_2021}. Moreover, a significant fraction of the population may be accompanied by a tertiary object of stellar or substellar-mass which could be studied using GWs~\citep{robson_cornish_2018,steffen_wu_2018,tamanini_danielski_2019,danielski_korol_2019}. In this section, we study the detection distance for DWDs in the Milky Way and nearby galaxies, as well as their detection accuracy using FMA for TianQin, LISA, and joint detection. Because DWDs evolve very slowly when in the band and already circularize during the common envelope phase, we model their signal using the quadrupole approximation~\citep{peters_mathews_1963,tq_dwd_2020}. Moreover, we adopt a conventional detection threshold of $\rho=7$ assuming there is no a priori observation of an EM counterpart~\citep{tq_dwd_2020,lisa_2022b}. More details on TianQin and LISA detections of DWD can be found in, e.g., \cite{tq_dwd_2020}, \cite{lisa_2017}, and \cite{lisa_2022b}.

\subsection{Detectability}\label{sec:dwdd}

\begin{figure}[tpb] \centering \includegraphics[width=0.48\textwidth]{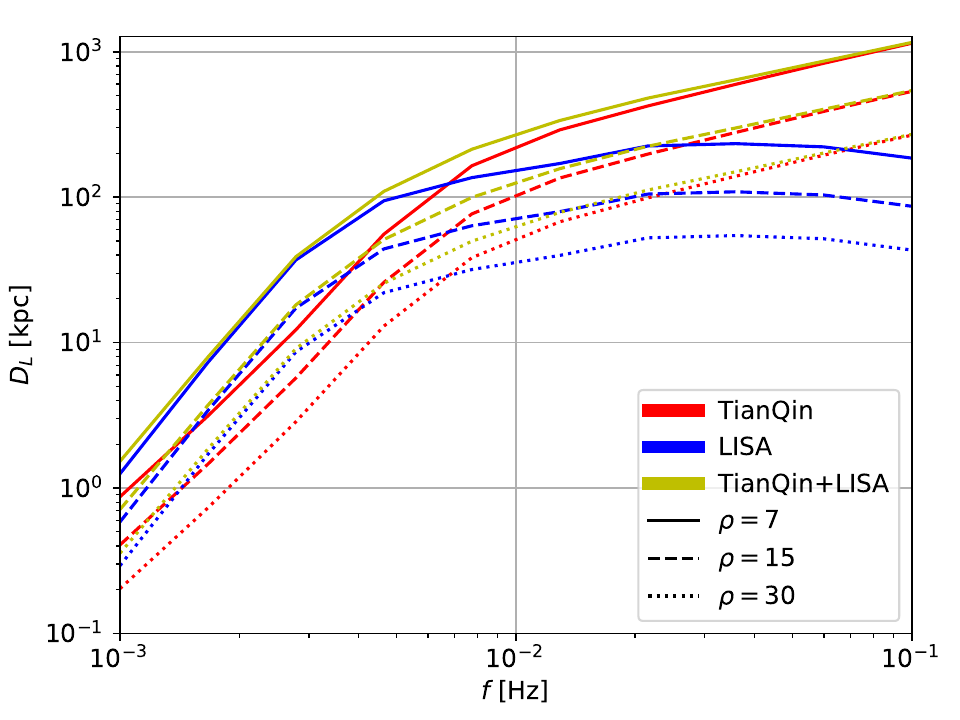}
\caption{
    The luminosity distance $D_L$ to which a DWD can be detected with a fixed SNR $\rho$ for TianQin, LISA, and joint detection as a function of the quadrupole frequency of the GW $f$.
    }
\label{fig:dwd_hor_snr}
\end{figure}

In this section, we analyze the distance to which a DWD can be detected by TianQin, LISA, and joint detection. We start again considering the case of different fixed SNRs but vary now the quadrupole frequency of the GW detected. We fix the mass of the DWD to $2.8\,{\rm M_\odot}$ as an upper limit while varying the quadrupole frequency of the wave. The quadrupole frequency is two times the orbital frequency of the binary and is thus related to the evolution of the binary's semimajor axis while the binary does not evolve much during the observation time. We see in \fref{fig:dwd_hor_snr} that for frequencies below $4\,{\rm mHz}$ LISA performs significantly better than TianQin, and joint detection follows closely the detection distance of LISA alone. For frequencies of $1\,{\rm mHz}$ LISA alone will be able to detect a DWD with a SNR $\rho=7$ to a distance of a bit more than $1\,{\rm kpc}$ while the distance increases to around $80\,{\rm kpc}$ for a frequency of $4\,{\rm mHz}$. For frequencies between $4\,{\rm mHz}$ and $20\,{\rm mHz}$ we get the most significant improvement from joint detection compared to any single detection reaching a distance between around $100\,{\rm kpc}$ and almost $500\,{\rm kpc}$ for a SNR of seven. For frequencies above $20\,{\rm mHz}$ TianQin performs much better than LISA and the joint detection follows the detection distance of the first closely. For a frequency of $20\,{\rm mHz}$ and a SNR of seven, TianQin alone will be able to detect a DWD to a distance of more than $400\,{\rm kpc}$ and up to more than $1\,{\rm Mpc}$ for a frequency of $100\,{\rm mHz}$.

\begin{figure}[tpb] \centering \includegraphics[width=0.48\textwidth]{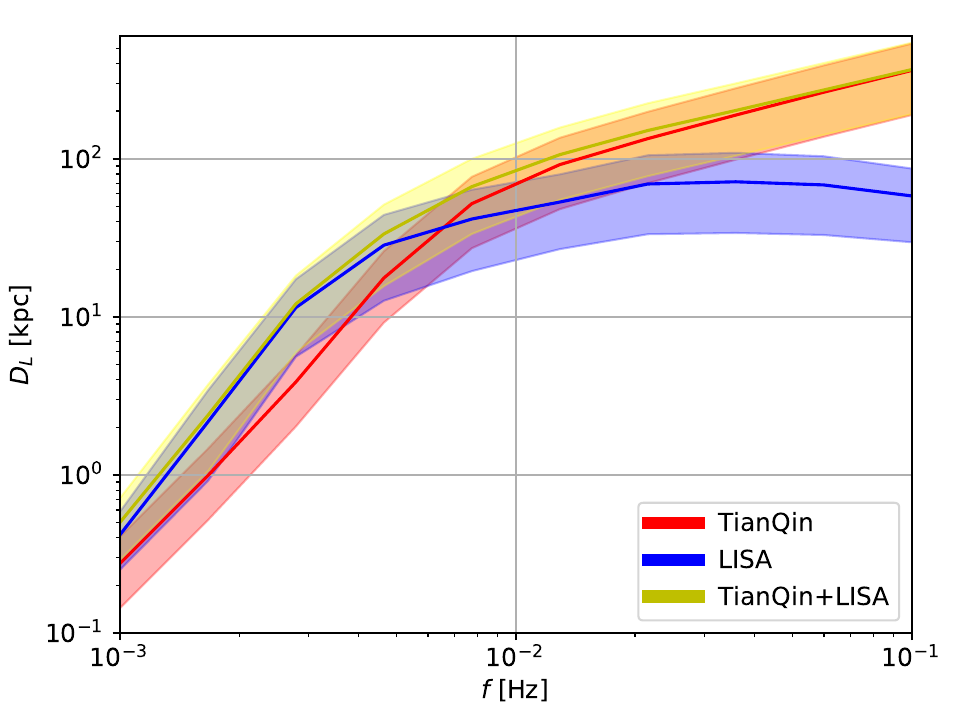}
\caption{
    The luminosity distance $D_L$ to which a DWD with a SNR of 15 can be detected for a varying inclination for TianQin, LISA, and joint detection as a function of the quadrupole frequency of the GW $f$. The upper edge of the shaded region corresponds to the face-on case while the lower edge corresponds to the edge-on case; the solid line corresponds to the average inclination of $60^\circ$.
    }
\label{fig:dwd_hor_inc}
\end{figure}

\fref{fig:dwd_hor_inc} shows how the distance to which a DWD with a fixed SNR of 15 can be detected depends on the inclination of the binary for TianQin, LISA, and joint detection. We see that again LISA alone performs better than TianQin alone for frequencies $\lesssim 4\,{\rm mHz}$ and that joint detection follows closely the behavior of the first. For frequencies $\gtrsim 20\,{\rm mHz}$, in contrast, TianQin performs better than LISA, and the joint detection follows the performance of TianQin. Only for frequencies between $4\,{\rm mHz}$ and $20\,{\rm mHz}$ do we get a considerable improvement when having joint detection. For frequencies above $20\,{\rm mHz}$ TianQin's observation distance for face-on sources is better by a factor of around 1.4 compared to a source with an average inclination and of almost three compared to an edge-on source. In the frequency range of $4\,{\rm mHz}$ to $20\,{\rm mHz}$ where the joint detection performs significantly better, we see that similar to a TianQin alone detection the face-on case can be observed out to distances that are around 1.4 times more distant than for the average case and around three times more distant than for the edge-on scenario. For LISA and a frequency of $1\,{\rm mHz}$ the detection distance for the face-on case is around 1.5 times bigger than for the average case and around 2.4 times bigger than for the edge-on case. For higher frequencies of up to $4\,{\rm mHz}$ the ratio in detection distance for the face-on case and the average inclination does not change much. However, for the edge-on case, there is a strong variation with a decrease in the detection distance around $2\,{\rm mHz}$ and an increase around $3\,{\rm mHz}$. This is probably a consequence of the galactic foreground (cf. \fref{fig:sc}), which affects the edge-on case the strongest because the signal is the weakest and thus buried the most.

\subsection{Parameter estimation}\label{sec:dwdp}

We examine in this subsection the accuracy with which different parameters of a DWD can be detected by TianQin alone, LISA alone, and by joint detection. The parameters we analyze are the binary's chirp mass $\mathcal{M}_c$, the luminosity distance of the source $D_L$, the inclination of the system relative to the line-of-sight $\iota$, the sky localization error of the source $\Delta\Omega$, and orbital period of the binary $P$. We only vary one parameter while fixing the other parameters in each analysis, where the fiducial values are $\mathcal{M}_c = 0.25\,{\rm M_\odot}$, $D_L = 8.5\,{\rm kpc}$, $\iota = 60^\circ$, and $P = 200\,{\rm s}$ (corresponding to a frequency $f = 0.01\,{\rm Hz}$). Furthermore, we set the source to be at the sky location $\theta_{\rm bar} = 95.6^\circ$ and $\phi_{\rm bar} = 267^\circ$ where we use barycentric coordinates. Note that a frequency of $0.01\,{\rm Hz}$ is only expected for a small fraction of DWDs~\citep{korol_hallakoun_2022}. However, we consider this frequency because sources at this stage of their evolution are particularly interesting as single sources since they chirp significantly during the observation time which allows an accurate study of their properties and because it is the common ``sweet spot'' of TianQin and LISA, thus allowing a fair comparison. Nevertheless, we also discuss the dependence of parameter estimation on the initial frequency at the end of this section to present a more complete picture.

\begin{figure}[tpb] \centering \includegraphics[width=0.48\textwidth]{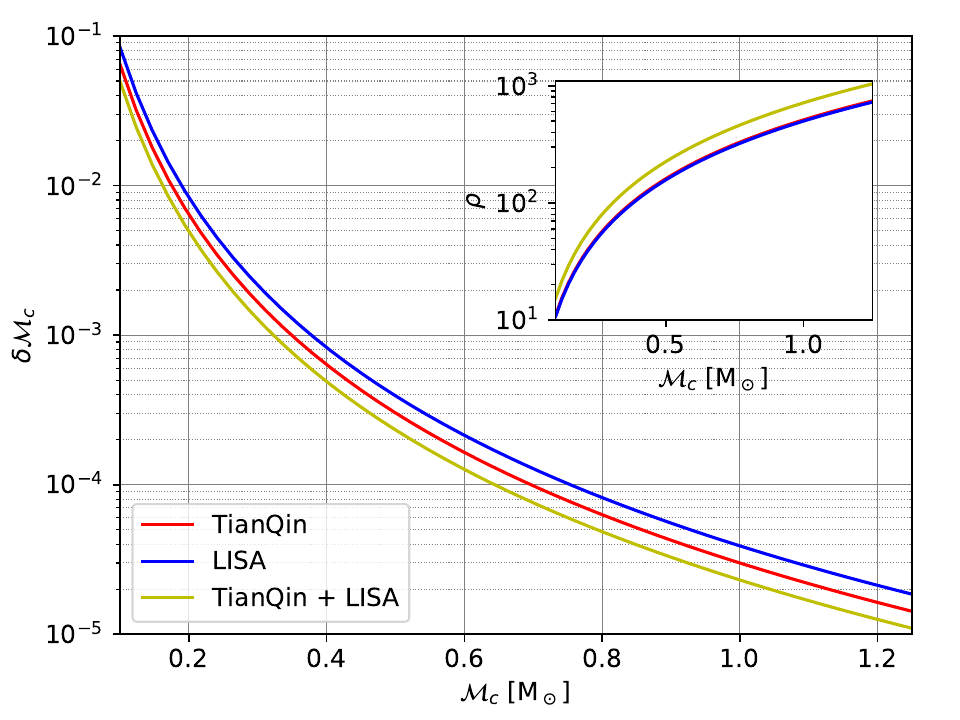}
\caption{
    The relative error in the chirp mass of a DWD $\delta\mathcal{M}_c$ as a function of the chirp mass of the system $\mathcal{M}_c$. The red line shows the detection accuracy for TianQin alone, the blue line for LISA alone, and the yellow line for joint detection. The inset shows the corresponding SNR of the source using the same color coding as in the main plot.
    }
\label{fig:dwd_dmc}
\end{figure}

\textit{Chirp mass:} In \fref{fig:dwd_dmc} we show the relative error in the chirp mass. For all masses considered, TianQin alone performs better than LISA alone by a factor of around 1.3 while the joint detection performs better than TianQin by roughly the same factor. From the inset, we see that the better accuracy for joint detection can be explained by the higher SNR, however, TianQin alone performs better than LISA alone despite having the same SNR in both detectors. TianQin performs better than LISA despite having the same SNR because TianQin is more sensitive at higher frequencies and thus can detect the chirping of the signal more accurately. We see that for all three detection scenarios and low chirp masses of below $0.2\,{\rm M_\odot}$ the relative error is at most a few percent and goes down to an order of $10^{-3}$. The relative error decreases further for higher chirp masses going down to $\sim10^{-4}$ for $\mathcal{M}_c \gtrsim 0.4\,{\rm M_\odot}$ and to $\sim10^{-5}$ for $\mathcal{M}_c \gtrsim 0.8\,{\rm M_\odot}$.

\begin{figure}[tpb] \centering \includegraphics[width=0.48\textwidth]{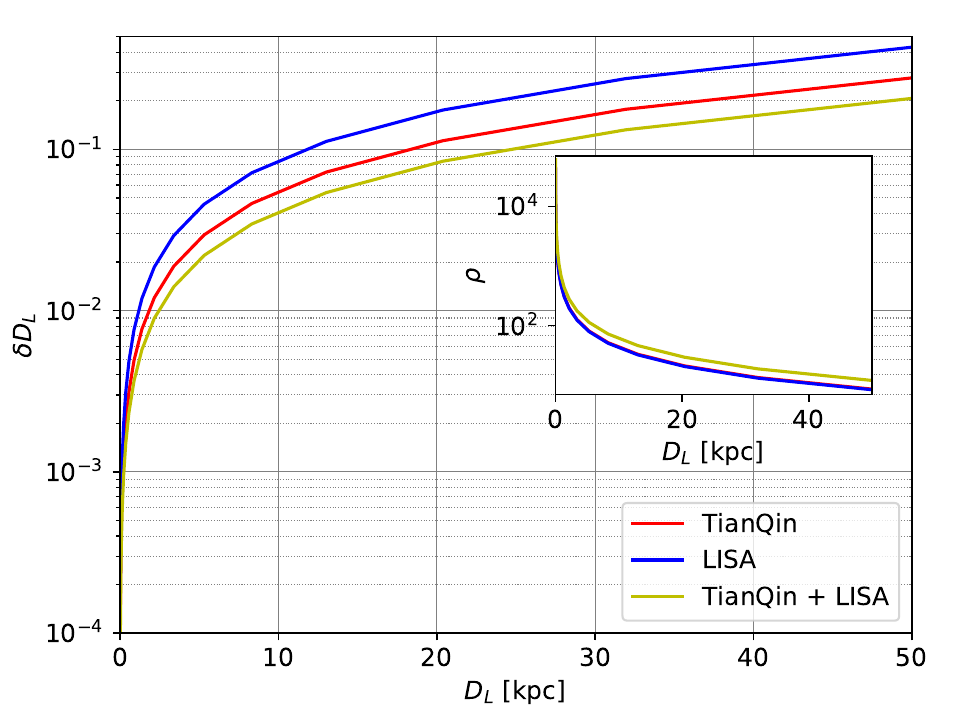}
\caption{
    The relative error in the luminosity distance for a DWD $\delta D_L$ over the luminosity distance of the system $D_L$. The red line represents the detection accuracy for TianQin alone, the blue line for LISA alone, and the yellow line for joint detection. The inset shows the corresponding SNR of the source where the color coding is the same as in the main plot.
    }
\label{fig:dwd_ddl}
\end{figure}

\textit{Luminosity distance:} For the relative error in the luminosity distance, we also find that TianQin performs better than LISA despite having the same SNR as shown in \fref{fig:dwd_ddl} and the inset, respectively. That TianQin has a smaller error in the luminosity distance than LISA is a result of TianQin's more accurate detection of the chirp mass. The luminosity distance is determined using the amplitude of the wave which also depends on the mass of the system and thus having more accurate constraints for the chirp mass reduces the uncertainty in $D_L$. However, the factor by which TianQin performs better than LISA is this time around 1.7. The joint detection once again has a higher SNR than the two single detection scenarios and performs better than TianQin alone by a factor of around 1.3. The relative error remains below 0.1 for $D_L$ smaller than around $12\,{\rm kpc}$, $18\,{\rm kpc}$, and $25\,{\rm kpc}$ for LISA, TianQin, and the joint detection, respectively. In particular, we see that for all distances considered the relative error remains below 0.5 going up to around 0.4 for LISA, 0.3 for TianQin, and 0.2 for the joint detection for $D_L=50\,{\rm kpc}$.

\begin{figure}[tpb] \centering \includegraphics[width=0.48\textwidth]{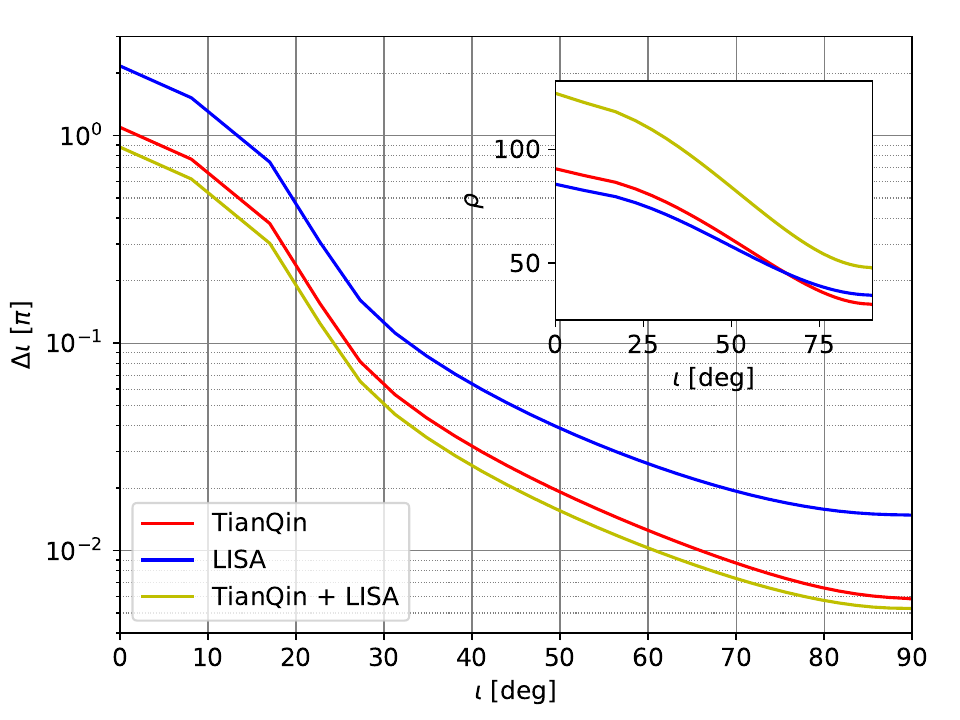}
\caption{
    The absolute error in the inclination of the system relative to the line-of-sight $\Delta\iota$ over the inclination $\iota$. The detection accuracy for TianQin alone is shown in red, for LISA alone in blue, and joint detection in yellow. The inset shows the corresponding SNR of the source using the same color coding as in the main plot.
    }
\label{fig:dwd_diota}
\end{figure}

\textit{Inclination:} \fref{fig:dwd_diota} shows the absolute error for the inclination of the source. In all three cases, the error is the biggest for the source being face-on ($\iota = 0^\circ$) and the lowest for the source being edge-on ($\iota = 90^\circ$) because the edge-on source has a higher contribution from the $\times$-polarization which helps to resolve its inclination. For all inclinations, TianQin alone has a smaller detection error than LISA alone by a factor of around two. Despite TianQin and LISA having a similar SNR for most inclinations and LISA even having a higher SNR for inclinations above roughly $60^\circ$, TianQin detects the inclination with better accuracy because the inclination and the luminosity distance of a DWD are degenerate, and TianQin can determine the luminosity distance more accurately as shown in \fref{fig:dwd_ddl}. We see that for low inclinations close to $0^\circ$ and below around $15^\circ$ for TianQin and LISA, respectively, the error is of the order $\pi$ thus making the detection of the inclination impossible. However, the error decreases quickly for higher inclinations and goes below $0.1\pi$ for $\iota\gtrsim25^\circ$ and $\iota\gtrsim33^\circ$ for TianQin and LISA, respectively. If the source is almost edge-on ($\iota\gtrsim65^\circ$), the error in LISA is below $0.02\pi$ and in TianQin it even decreases to an order of $10^{-3}\pi$. For the joint detection, we see that the SNR is significantly higher than any of the single detections but the accuracy is only slightly better than TianQin ranging between slightly less than $\pi$ for a face-on source to $\sim10^{-3}$ for an edge-on source.

\begin{figure}[tpb] \centering \includegraphics[width=0.48\textwidth]{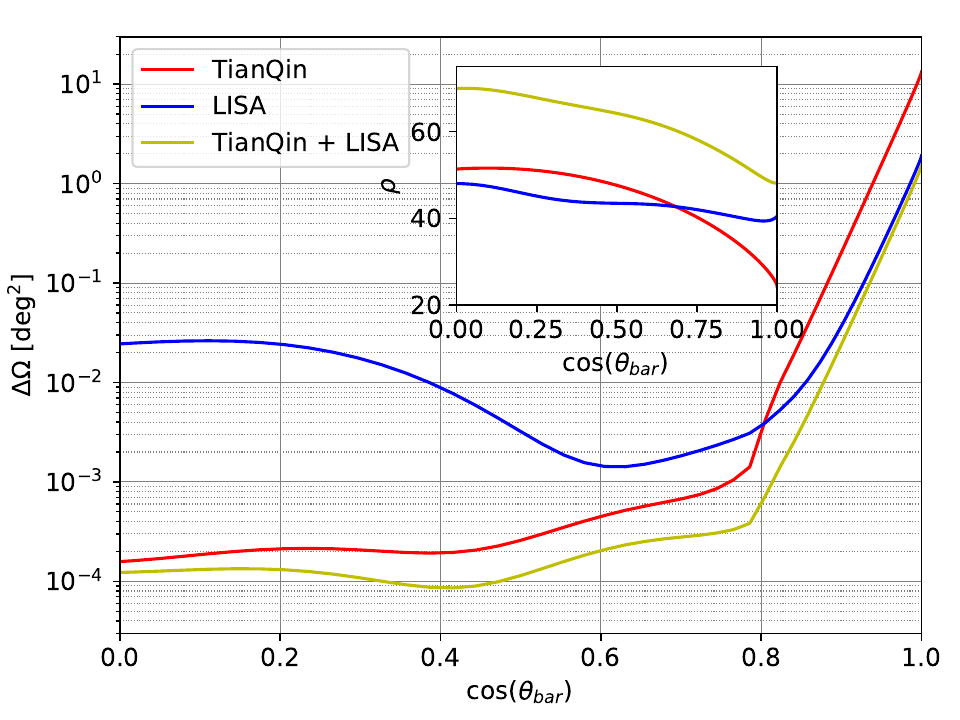}
\caption{
    The sky localization error $\Delta\Omega$ for different $\cos(\theta_{\rm bar})$. The red line shows the detection accuracy for TianQin alone, the blue line for LISA alone, and the yellow line for joint detection. The inset shows the corresponding SNR of the source where the color coding is the same as in the main plot.
    }
\label{fig:dwd_domegat}
\end{figure}

\textit{Sky localization:} The sky localization error for different $\theta_{\rm bar}$ is shown in \fref{fig:dwd_domegat} where we only present the errors for the upper half-sphere because the results for the lower half-sphere are almost symmetric. For $\cos(\theta_{\rm bar})\lesssim0.5$ the SNR in TianQin alone is slightly higher than in LISA alone while the sky localization is significantly better -- $\sim10^{-4}$ for TianQin and $\sim10^{-2}$ for LISA. This difference in performance arises because TianQin always points towards RX J0806.3+1527 which lies close to the ecliptic plane while LISA's orientation changes while it orbits around the sun. This change of orientation also makes LISA more sensitive to source at higher $\theta_{\rm bar}$ and we see that its detection error becomes smaller than TianQin's for $\cos(\theta_{\rm bar})\gtrsim0.8$ differing by almost one order of magnitude for $\cos(\theta_{\rm bar})\approx1$.  We, further, see that for joint detection, the sky localization is similar to the one of TianQin if $\cos(\theta_{\rm bar})\lesssim0.4$ and similar to the one of LISA if $\cos(\theta_{\rm bar})\gtrsim0.8$, while it performs significantly better than any of the two detections for $0.4\lesssim\cos(\theta_{\rm bar})\lesssim0.8$.

\begin{figure}[tpb] \centering \includegraphics[width=0.48\textwidth]{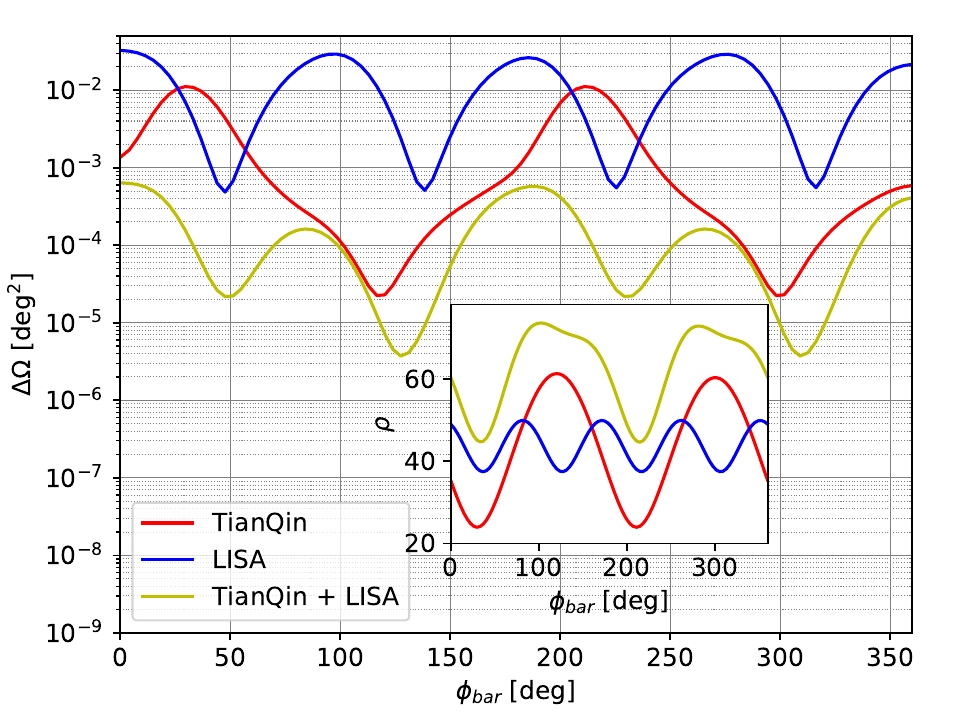}
\caption{
    The sky localization error $\Delta\Omega$ as a function of $\phi_{\rm bar}$. The detection accuracy for TianQin alone is shown in red, for LISA alone in blue, and for joint detection in yellow. The inset shows the corresponding SNR of the source using the same color coding as in the main plot.
    }
\label{fig:dwd_domegap}
\end{figure}

We show in \fref{fig:dwd_domegap}, the sky localization error as a function of $\phi_{\rm bar}$. The sky localization error for TianQin has a strong variation between $10^{-2}\,{\rm deg^2}$ in the worst case and around $2\times10^{-5}\,{\rm deg^2}$ in the best case, where the best results are obtained for $\phi_{\rm bar}\approx 120^\circ, 300^\circ$ correlating with the sky location of RX J0806.3+1527 while the worst results appear for $\phi_{\rm bar}\approx 30^\circ, 210^\circ$. For LISA the sky localization error has a smaller variation but is bigger than TianQin's for almost all $\phi_{\rm bar}$ varying around $5\times10^{-3}\,{\rm deg^2}$. Moreover, we see that LISA's sky localization error presents four minima and maxima because its satellites lie on more complex orbits. We note that the sky localization error in TianQin is significantly smaller than in LISA despite sometimes having a significantly lower SNR in TianQin than LISA because the source is set to be close to the ecliptic plane where TianQin is more sensitive. As a result of the strong difference in the SNR of TianQin and LISA, we get that the SNR of the joint detection is mostly significantly higher than in any of the two single detection scenarios. This difference also results in having a better sky localization for joint detection, in particular, for $0^\circ\lesssim\phi_{\rm bar}\lesssim70^\circ$ and $190^\circ\lesssim\phi_{\rm bar}\lesssim260^\circ$ where TianQin and LISA perform similarly.

\begin{figure*}[tpb] \centering \includegraphics[width=0.98\textwidth]{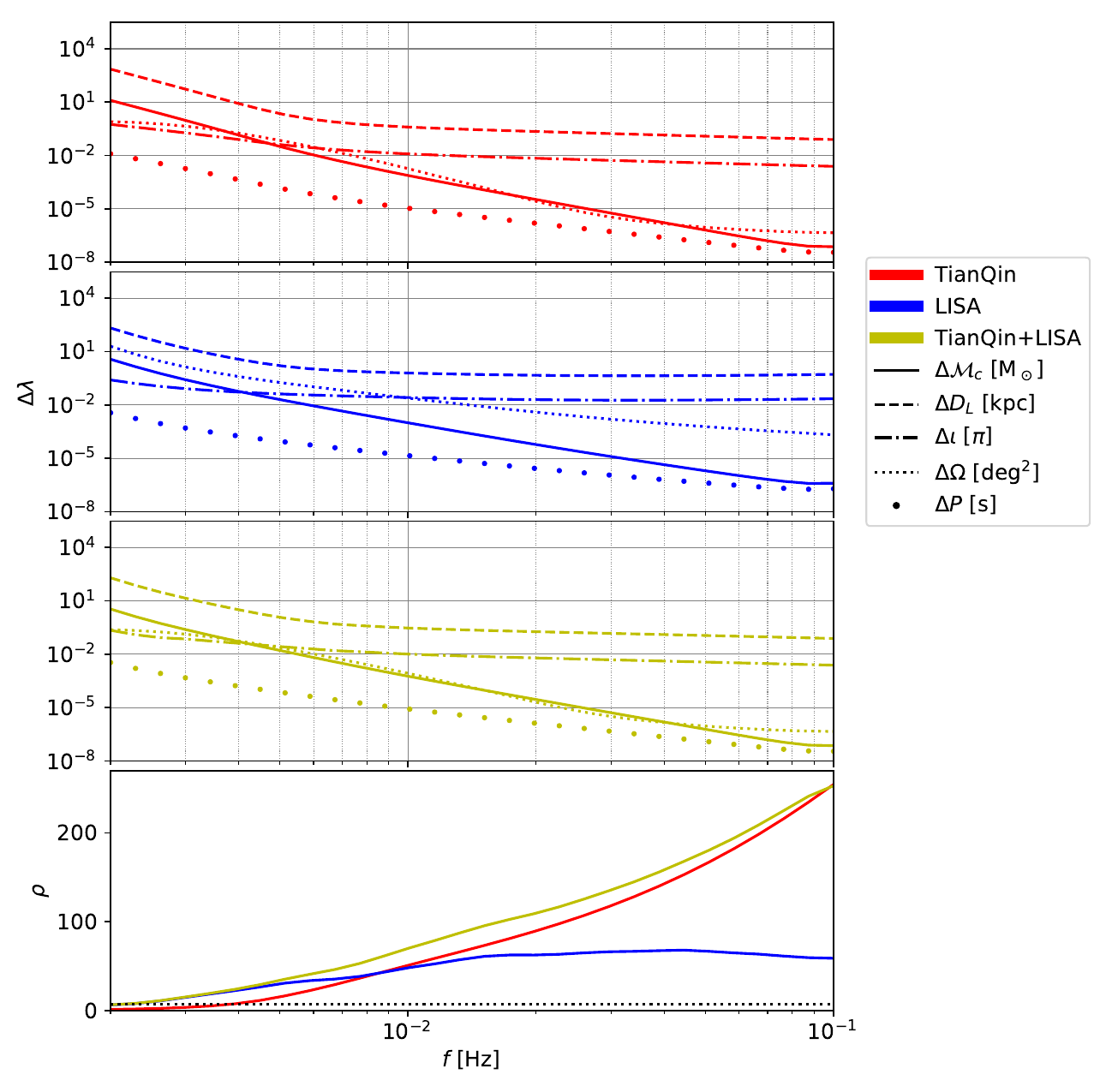}
\caption{
    The upper three plots show the absolute error in the chirp mass $\Delta\mathcal{M}_c$ (solid line), the luminosity distance $\Delta D_L$ (dashed line), the inclination of the source $\Delta\iota$ (dashed-dotted line), the sky localization $\Delta\Omega$ (dotted line), and the period of the source $\Delta P$ (points) as functions of the source's initial frequency $f$. The first plot with red lines shows the errors for TianQin alone, the second plot with blue lines for LISA alone, and the third plot with yellow lines for joint detection. The lowest plot shows the corresponding SNR of the source using the same color coding as in the other plots and the black dotted line shows the detection threshold of $\rho=7$.
    }
\label{fig:dwd_fre}
\end{figure*}

\textit{Frequency:} Due to their relatively low mass, DWDs evolve quite slowly when emitting GWs, thus having almost constant frequencies for observation times of years like for TianQin and LISA. As a result, the detection accuracy of the source's parameters strongly depends on the frequency at which the source is emitting when first detected. Therefore, we show in \fref{fig:dwd_fre} the absolute errors for all previously discussed parameters and for the orbital period as a function of the wave's frequency. We see that for all three detection scenarios, the errors of the different parameters have a similar order of magnitude where the errors tend to be smaller for LISA if $f\lesssim10\,{\rm mHz}$ and smaller for TianQin if $f\gtrsim10\,{\rm mHz}$ because of their better performance for lower and higher frequencies, respectively. The error in the chirp mass decreases in all detection scenarios as the frequency increases because sources with higher initial frequency chirp more during detection allowing a better constrain of the mass. We, further, see that the errors in the sky localization and the period of the source also constantly decrease as the frequency increases. This behavior is a direct result of the improved chirp mass estimation which allows containing other parameters more accurately. In the case of the luminosity distance and the inclination a better estimation of the chirp mass also improves their detection, however, the errors strongly decrease if $f\lesssim5\times10^{-3}\,{\rm Hz}$ but decrease at smaller rates for higher frequencies. This is because the luminosity distance and the inclination are degenerate when only detecting the dominant mode of GWs. For lower frequencies, the improvement gained from the more accurate estimation of the chirp mass allows a better estimation of these two parameters but for higher frequencies, the degeneracy starts dominating their estimation and thus the improvement significantly reduces. The SNR increases as the frequency increases for all three detection scenarios but for LISA this increase almost stagnates for frequencies above $10^{-2}\,{\rm Hz}$ and even decreases if $f\gtrsim5\times10^{-2}\,{\rm Hz}$ because of its reduced sensitivity at higher frequencies.

\subsection{Comparing TianQin, LISA \& joint detection}\label{sec:dwdc}

In the case of DWDs, the performance of TianQin and LISA mainly depends on the evolutionary stage of the binary. LISA detects sources at lower frequencies to slightly bigger distances and measures their parameters with higher accuracies. In contrast, DWDs emitting at higher frequencies, the performance of TianQin in terms of detection distance and parameter estimation is better. The advantage of TianQin towards higher frequencies might result in better overall results because higher frequency DWDs chirp more which usually allows a better parameter estimation. However, more sources are emitting at lower frequencies and thus LISA will be able to detect a higher number of sources.

\begin{figure}[tpb] \centering \includegraphics[width=0.48\textwidth]{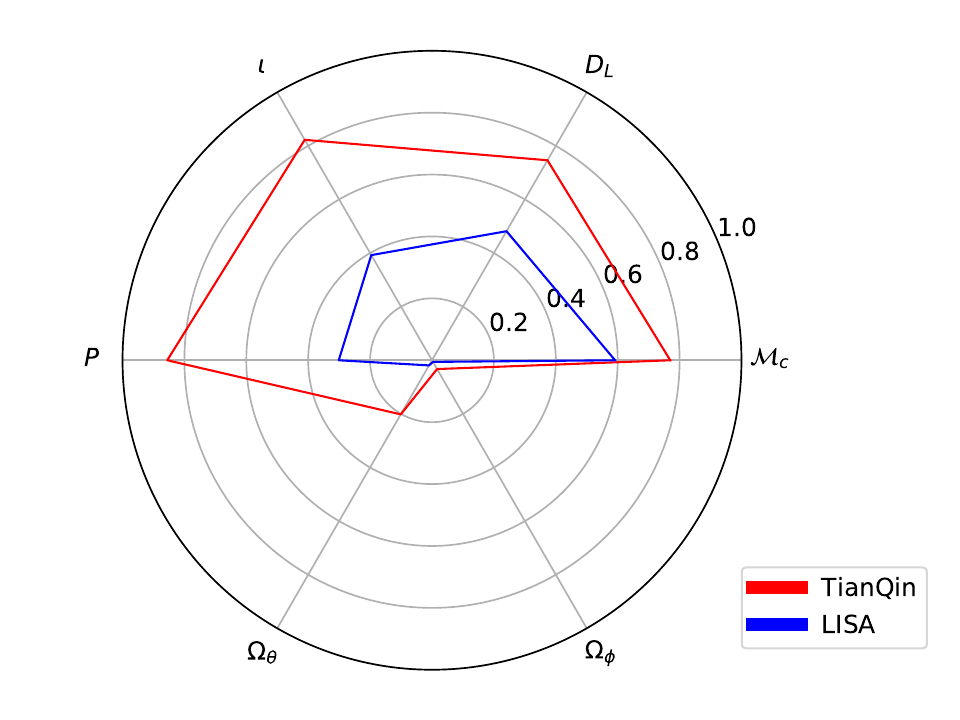}
\caption{
    The performance ratio $Q_X[\lambda]$ for the parameters $\lambda$ of a DWD, where we label the ratios only using the parameters, for TianQin and LISA. $\Omega_{\theta}$ and $\Omega_{\phi}$ indicate the sky localization as a function of the latitude $\theta_{\rm bar}$ and the azimuth angle $\phi_{\rm bar}$, respectively.
    }
\label{fig:dwd_radar}
\end{figure}

The performance ratio $Q$ for TianQin and LISA compared to the joint detection is shown in \fref{fig:dwd_radar}. For all parameters considered, on average TianQin alone performs significantly better than LISA, in particular, for the chirp mass $\mathcal{M}_c$, the luminosity distance $D_L$, the inclination of the source $\iota$, and the period of the binary $P$. However, although the sky localization in TianQin is on average much better than in LISA both single detection scenarios perform quite poorly when compared to joint detection. Therefore, the joint detection of a single source in the common ``sweet spot'' around $10^{-2}\,{\rm Hz}$ allows a better and more complete parameter reconstruction. Moreover, to obtain an understanding of most of the DWD population, joint detection is necessary due to the coverage of different frequencies by TianQin and LISA.

\section{Extreme mass ratio inspirals}\label{sec:emri}

Observations suggest the presence of MBHs surrounded by stellar clusters or cusps of a few parsecs in the center of most galaxies~\citep{kormendy_richstone_1995,magorrian_tremaine_1998,balcells_graham_2003,gebhardt_richstone_2003,ferrarese_ford_2005,alexander_2005,ferrarese_cote_2006,graham_spitler_2009,schodel_feldmeier_2014}. In such a high-density environment relaxation processes can occasionally force stars and COs onto extremely eccentric, low angular momentum orbits that result in close encounters with the central MBH. While main sequence stars usually do not survive such a close encounter, COs typically survive until they fall into the MBH~\citep{hills_1975,murphy_cohn_1991,freitag_benz_2002,gezari_halpern_2003,merritt_2015,bar-or_alexander_2016,amaro-seoane_2019,vazquez-aceves_lin_2022}. Depending on the orbital angular momentum of the CO, it can plunge directly into the MBH or be captured in an eccentric orbit, whose secular evolution decouples from the rest of the cluster and is dominated by the emission of GWs~\citep{amaro-seoane_2018a}. Such a source formed by a stellar-mass CO around a MBH is usually referred to as an extreme mass ratio inspiral (EMRI).

Detection of the GWs emitted by EMRIs will allow us to improve our understanding of astrophysics and fundamental physics in an unprecedented manner. It will allow us to obtain information about the mass distribution of MBHs~\citep{gair_tang_2010} and the stellar systems surrounding them~\citep{amaro-seoane_gair_2007}. Furthermore, EMRIs can be used to study the expansion of the universe~\citep{macleod_hogan_2008} as well as to map the space-time geometry of the MBH in great detail~\citep{gair_vallisneri_2013}. The latter will allow stringent tests of general relativity including measuring the ``bumpiness'' and the non-Kerr nature of BHs~\citep{piovano_maselli_2020,glampedakis_babak_2006}, testing the no-hair theorem, and constraining modified theories of gravity\citep{ryan_1995,ryan_1997,barack_cutler_2007,chua_hee_2018}. Last but not least, deviations from vacuum/rest sources can reveal information about the environment of the source~\citep{barausse_rezzolla_2007,barausse_rezzolla_2008,gair_flanagan_2011,yunes_kocsis_2011,barausse_cardoso_2014,barausse_cardoso_2015,derdzinski_dorazio_2021,torres-orjuela_amaro-seoane_2021}.

In this section, we study the distance to which EMRIs can be detected by TianQin, LISA, and joint detections where we use a conventional detection threshold of $\rho=20$~\citep{babak_gair_2017,tq_emri_2020}. Moreover, we use a FMA to analyze the detection accuracy for the most relevant parameters of EMRIs. We generate the waveforms using an `augmented analytic kludge model' from the `EMRI waveform software suite'~\citep{chua_moore_2017}. We use this waveform model because it can provide sufficiently accurate waveforms to make a reliable comparison between TianQin, LISA, and joint detections while being computationally efficient and covering the parameter space we want to study. For a more detailed discussion on the detection EMRIs by TianQin and LISA see, e.g., \cite{tq_emri_2020}, \cite{tq_emri_2021}, \cite{babak_gair_2017}, and \cite{lisa_2022b}.

\subsection{Detectability}\label{sec:emrid}

\begin{figure}[tpb] \centering \includegraphics[width=0.48\textwidth]{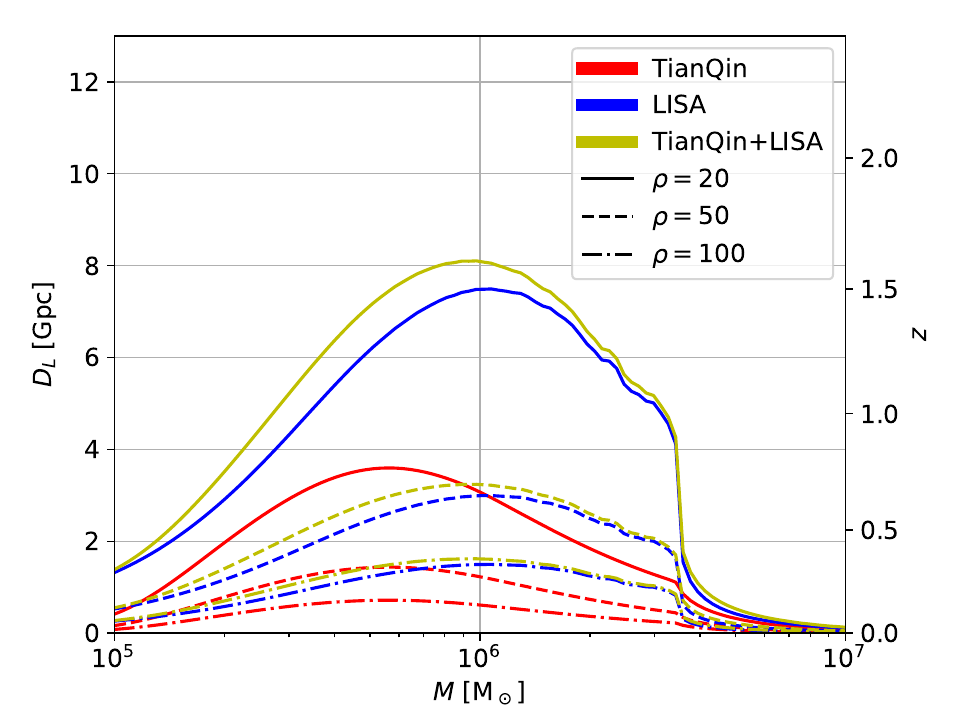}
\caption{
    The distance to which an EMRI can be detected with a fixed SNR $\rho$ for TianQin, LISA, and joint detection as a function of the mass of the MBH $M$ where we assume a mass of $10\,{\rm M_\odot}$ for the SBH. The left ordinate shows the luminosity distance $D_L$ while the right ordinate shows the corresponding redshift $z$.
    }
\label{fig:emri_hor_snr}
\end{figure}

In this subsection, we study the distance to which an EMRI can be detected depending on the mass of the MBH $M$ where for the SBH we assume a standard mass of $10\,{\rm M_\odot}$. In \fref{fig:emri_hor_snr} we see that LISA will be able to detect an EMRI with $\rho=20$ out to a distance of around $7.5\,{\rm Gpc}$ ($z\approx1.5$) if the mass of the MBH is around $10^6\,{\rm M_\odot}$. For lower masses, the detection distance goes down and decreases to around $1.5\,{\rm Gpc}$ or $z\approx0.3$ for a MBH of $10^5\,{\rm M_\odot}$. For an increasing mass, the detection distance also goes down, however, with a sudden drop around $3.5\times10^6\,{\rm M_\odot}$. We attribute this sudden drop in detection distance to the fact that for an increasing mass, the frequency of the GWs emitted goes down, and since we set up the plunge of the SBH in the MBH to always be in the band, most of the inspiral and at some point even part of the plunge move towards frequencies where LISA's sensitivity becomes quite bad. That this is the case can be seen from the oscillations at masses slightly below the sudden drop which are a result of the noise induced by the galactic foreground. Because TianQin's sensitivity also decreases towards lower frequencies the drop is also present for the detection distance in TianQin; although the drop is less pronounced because TianQin's sensitivity decreases less fast than LISA's (cf. \fref{fig:sc}). In general, TianQin will detect EMRIs at a much closer distance of up to around $3.5\,{\rm Gpc}$ but reaching this distance for a mass of the MBH of around $5\times10^5\,{\rm M_\odot}$. When going to lower masses the detection distance for TianQin also decreases and goes to $z\approx0.1$ for a MBH of the mass $10^5\,{\rm M_\odot}$. Because the difference in detection distance between TianQin and LISA strongly depends on the mass of the MBH, the joint detection performs better than any single detection. In particular for masses below $2\times10^6\,{\rm M_\odot}$ joint detection works better reaching greater distances by almost $1\,{\rm Gpc}$ than LISA alone for a SNR of 20. For masses above $2\times10^6\,{\rm M_\odot}$ as well as below around $2\times10^5\,{\rm M_\odot}$ the contribution of TianQin decreases significantly and the joint detection follows LISA's detection distance quite closely. For higher SNRs of 50 and 100, the detection distance decreases as expected: LISA alone reaches detection distances of ca. $3\,{\rm Gpc}$ and $1.5\,{\rm Gpc}$, respectively, for a mass of $10^6\,{\rm M_\odot}$, TianQin alone reaches a distance of around $1.5\,{\rm Gpc}$ and $0.75\,{\rm Gpc}$, respectively, for a mass of around $5\times10^5\,{\rm M_\odot}$, and the joint detection reaches a distance of $z\approx0.7$ and $z\approx0.35$, respectively, if the mass of the MBH ranges between around $7\times10^5\,{\rm M_\odot}$ and $10^6\,{\rm M_\odot}$.

\begin{figure}[tpb] \centering \includegraphics[width=0.48\textwidth]{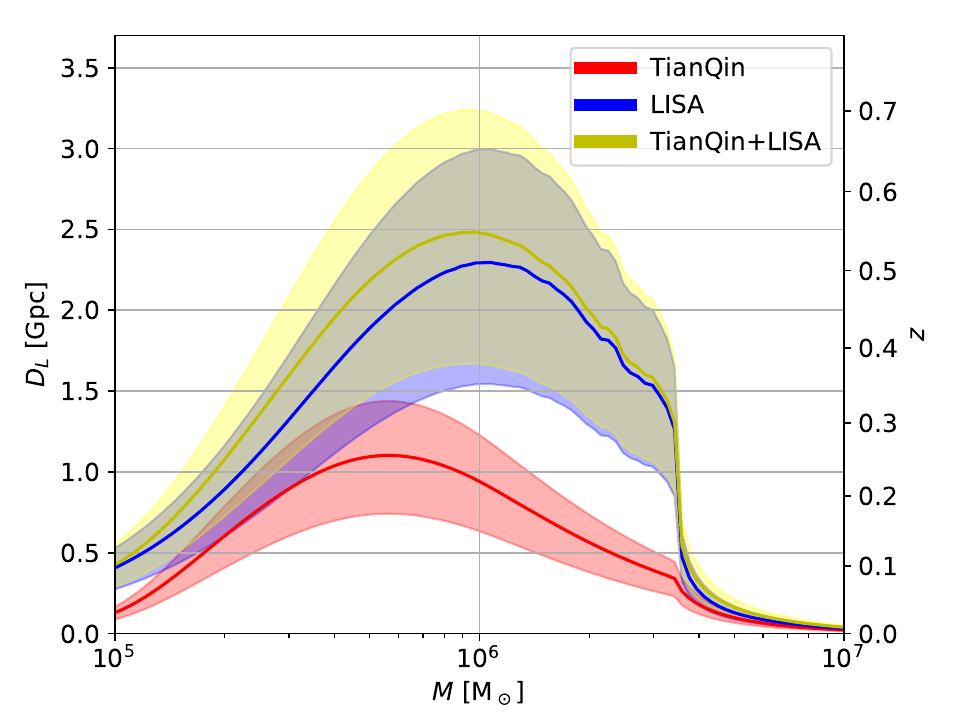}
\caption{
    The distance to which an EMRI with a SNR of 50 can be detected for a varying inclination for TianQin, LISA, and joint detection as a function of the mass of the MBH $M$ and a SBH of mass $10{\rm M_\odot}$. The upper edge of the shaded region corresponds to the face-on case while the lower edge corresponds to the edge-on case; the solid line corresponds to the average inclination of $60^\circ$. The left ordinate shows the luminosity distance $D_L$ while the right ordinate shows the corresponding redshift $z$.
    }
\label{fig:emri_hor_inc}
\end{figure}

In \fref{fig:emri_hor_inc}, we show the detection distance for an EMRI with a SNR of 50 in TianQin, LISA, and for the joint detection for different inclinations. In all three cases, the distance to which a face-on source can be detected is around 1.3 times bigger than for an average inclination of $60^\circ$ and around two times bigger than for an edge-on source. For TianQin and a mass of $M \approx 5\times10^5\,{\rm M_\odot}$ this means a face-on source will be detected to a distance of around $1.4\,{\rm Gpc}$, a source with an average inclination to around $1.1\,{\rm Gpc}$, and an edge-on source to a distance of around $0.7\,{\rm Gpc}$. For LISA and a MBH with a mass of $M \approx 10^6\,{\rm M_\odot}$ a face-on source will be detectable to a distance of around $3\,{\rm Gpc}$, a source with an inclination of $60^\circ$ to a distance of around $2.3\,{\rm Gpc}$, and an edge-on source to a distance of around $1.5\,{\rm Gpc}$. For a joint detection and the same values as for LISA, the respective distances are $3.25\,{\rm Gpc}$, $2.5\,{\rm Gpc}$, and $1.6\,{\rm Gpc}$. Sources above around $3.5\times10^6\,{\rm M_\odot}$, however, will only be detected to close distances of $z \lesssim 0.1$ almost independent of their inclination.

\subsection{Parameter estimation}\label{sec:emrip}

In this subsection, we study the accuracy with which different parameters of an EMRI can be detected by TianQin alone, LISA alone, and by joint detection. We analyze the error for the mass $M$ (in the observer frame) and the spin $s$ of the central MBH, the luminosity distance of the source $D_L$, the eccentricity of the binary at merger $e_m$, and the sky localization error of the source $\Delta\Omega$. In each analysis, we vary one parameter while keeping the other parameters fixed to the following values: $M = 10^6\,{\rm M_\odot}$, $D_L = 1\,{\rm Gpc}$, $e_m = 0.1$, and $s=0.98$. Furthermore, we set the source to be at the sky location $\theta_{\rm bar} = 60^\circ$ and $\phi_{\rm bar} = 0^\circ$ in a barycentric coordinate system. For all cases considered, we set the mass of the SBH orbiting the MBH to $10\,{\rm M_\odot}$ and the spin of the MBH to be perpendicular to the ecliptic plane.

\begin{figure}[tpb] \centering \includegraphics[width=0.48\textwidth]{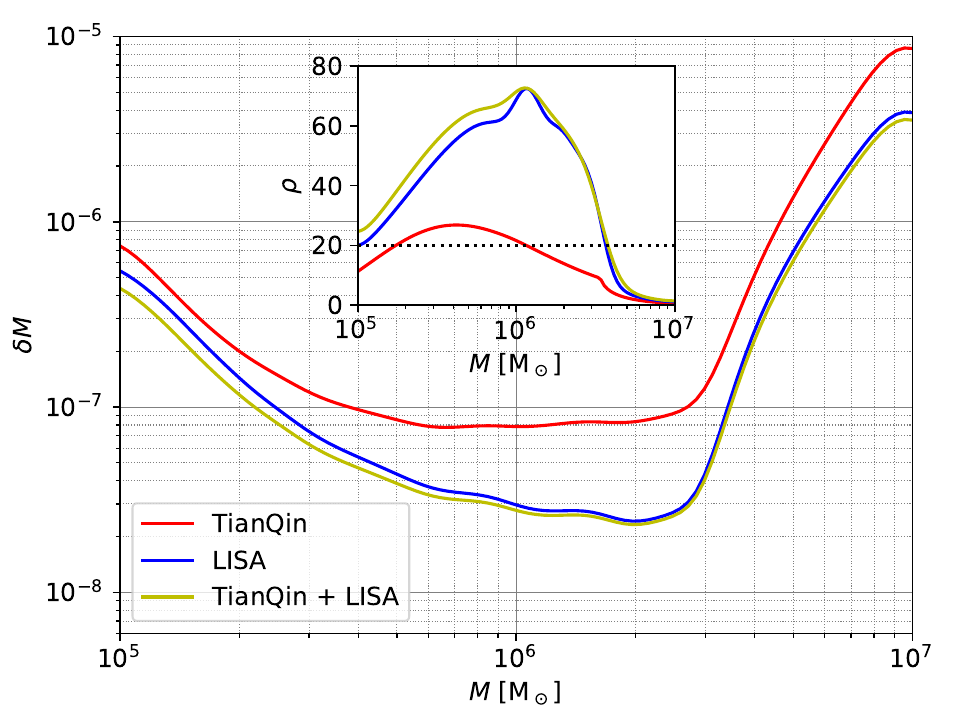}
\caption{
    The relative error in the mass (in the observer frame) of the central MBH $\delta M$ of the EMRI for different masses $M$. The red line shows the detection accuracy for TianQin alone, the blue line for LISA alone, and the yellow line for joint detection. The inset shows the corresponding SNR of the source with the same color coding as in the main plot and the detection threshold of $\rho=20$ is shown by the black dotted line.
    }
\label{fig:emri_dm}
\end{figure}

\textit{Mass of the central BH:} \fref{fig:emri_dm} shows the relative error in the mass (in the observer frame) of the central MBH of an EMRI. We see that TianQin has the best accuracy ($\approx 8\times10^{-8}$) in the mass range $6\times10^5-2\times10^6\,{\rm M_\odot}$. However, the error increases quickly for higher masses going up to almost $10^{-5}$ for $M = 10^7\,{\rm M_\odot}$ as well for lower masses going up to around $7\times10^{-7}$ for $M=10^5\,{\rm M_\odot}$. The detection error for LISA is significantly better than for TianQin by a factor of four to five in the mass range $6\times10^5-2\times10^6\,{\rm M_\odot}$ while LISA performs the best at a mass of around $2\times10^6\,{\rm M_\odot}$. For masses outside this range, LISA still performs better than TianQin but only by a factor of 1.5 to two. The higher accuracy in LISA also correlates with higher SNR which for LISA goes up to around 70 while for TianQin it remains below 30, even in the best case. Due to the significant difference between LISA and TianQin, the result of the joint detection is largely dictated by LISA. Only for lower masses $\sim10^5\,{\rm M_\odot}$ TianQin has a relevant contribution to the detection error and the SNR of an EMRI. Note that the observation times of $5\,{\rm yr}$ and $4\,{\rm yr}$ for TianQin and LISA, respectively, we use for parameter estimation in this analysis are longer than the $1\,{\rm yr}$ assumed for TianQin in \cite{tq_emri_2020} and $2\,{\rm yr}$ used for LISA in \cite{babak_gair_2017}, thus resulting in smaller errors.

\begin{figure}[tpb] \centering \includegraphics[width=0.48\textwidth]{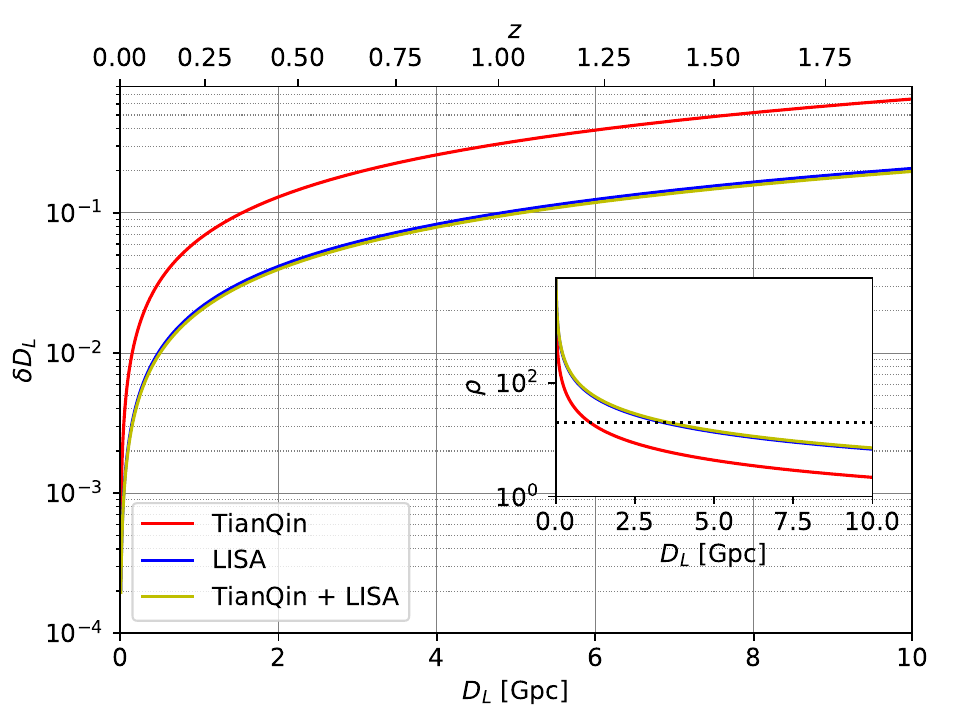}
\caption{
    The relative error in the luminosity distance of an EMRI $\delta D_L$. The lower abscissa shows the luminosity distance of the source $D_L$ and the upper abscissa its corresponding cosmological redshift $z$. The red line shows the detection accuracy for TianQin alone, the blue line for LISA alone, and the yellow line for joint detection. The inset shows the corresponding SNR of the source where we use the same color coding as in the main plot and the black dotted line indicates the detection threshold ($\rho=20$).
    }
\label{fig:emri_ddl}
\end{figure}

\textit{Luminosity distance:} In \fref{fig:emri_ddl}, we show the relative error in the luminosity distance $D_L$ for EMRIs. As for the mass, the detection accuracy in LISA is significantly better than in TianQin. For TianQin, the error remains below 0.1 for $z\lesssim0.35$ while for LISA and the joint detection -- that closely follows LISA's behavior -- the error remains below 0.1 up to redshifts of almost one. In contrast, at redshift one the error in TianQin goes up to almost 0.3. For higher redshifts, the detection error increases further, although, it should be also noted that for these distances the SNR in TianQin is well below the detection threshold of 20 while for LISA and the joint detection, it is only slightly below the detection threshold~\citep{babak_gair_2017}. For the highest distance considered ($D_L=10\,{\rm Gpc}$, $z\approx1.9$) the error in LISA/the joint detection goes up to 0.2 while for TianQin it reaches almost 0.7. We see that the luminosity distance of EMRIs can be detected with relatively good accuracy, even when the SNR is close to or below the detection threshold. We attribute this behavior to EMRIs emitting a high number of multiple modes which allow resolving the degeneracy between the luminosity distance and other parameters (e.g., total mass and inclination).

\begin{figure}[tpb] \centering \includegraphics[width=0.48\textwidth]{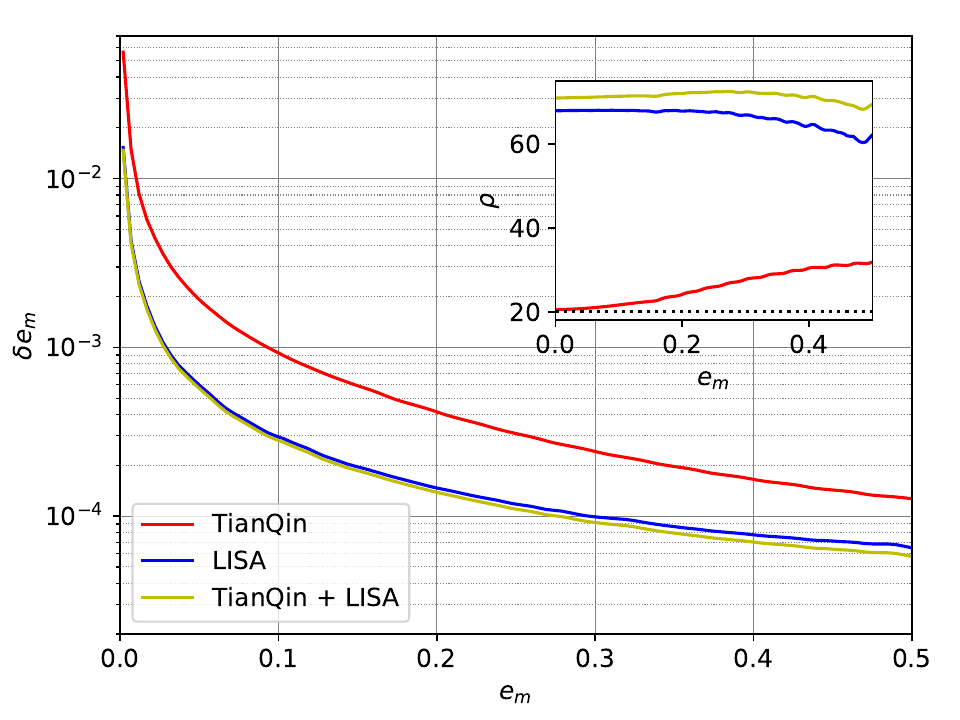}
\caption{
    The relative error in the eccentricity at the merger of an EMRI $\delta e_m$ as a function of the EMRI's eccentricity at merger $e_m$. The error for TianQin alone is shown in red, for LISA alone in blue, and the yellow line shows the error for joint detection. The inset shows the corresponding SNR of the source using the same color coding as in the main plot and the detection threshold ($\rho=20)$ is indicated by the black dotted line.
    }
\label{fig:emri_dem}
\end{figure}

\textit{Eccentricity:} The relative error in the eccentricity at the merger is shown in \fref{fig:emri_dem}. We see that the SNR in TianQin increases with higher eccentricities because higher eccentricities induce higher modes and TianQin has a better sensitivity at higher frequencies than LISA. In LISA, in contrast, the SNR decreases when the eccentricity increases because the higher modes have frequencies outside of LISA's most sensitive band. However, LISA maintains a high SNR between 60 and 70 for all eccentricities while the SNR in TianQin ranges between 20 and 30. Therefore, the error in LISA is significantly smaller than in TianQin by a factor of around four going from $1.5\times10^{-2}$ for $e_m\approx0$ to $6.5\times10^{-5}$ for $e_m=0.5$. The joint detection once again follows the accuracy in LISA closely, although, for higher eccentricities, TianQin's contribution can lead to slightly better detection accuracy. However, in general, the error in the eccentricity is for all detection scenarios and eccentricities considered, small enough that an accurate measurement of the eccentricity is possible.

\begin{figure}[tpb] \centering \includegraphics[width=0.48\textwidth]{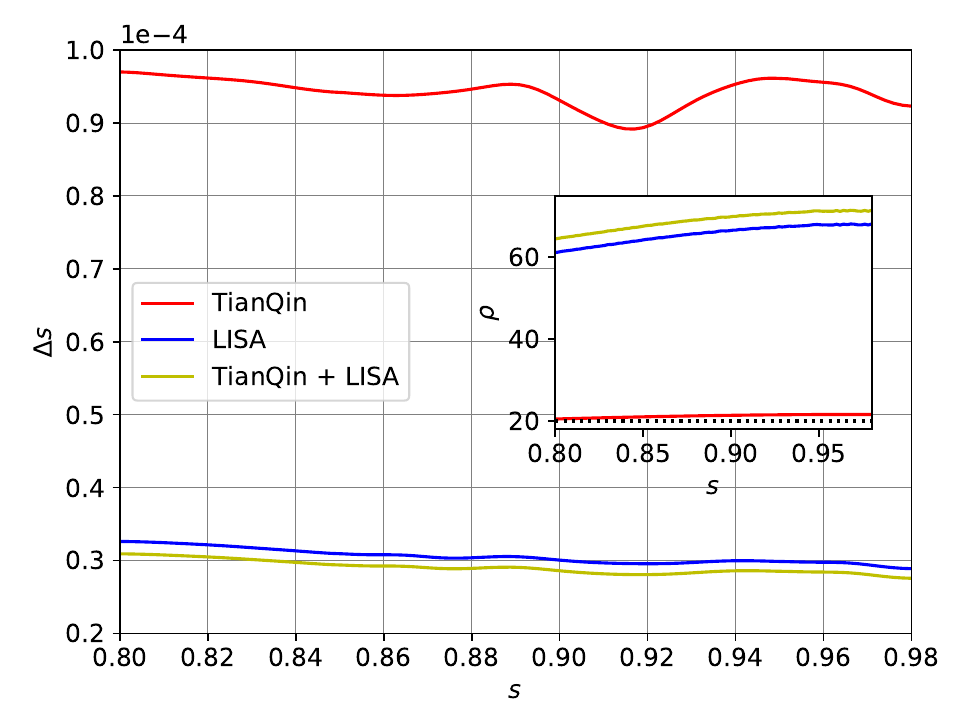}
\caption{
    The absolute error in the spin of the central MBH $s$ of the EMRI for different spin magnitudes $s$. The error for TianQin alone is shown in red, for LISA alone in blue, and joint detection in yellow. The inset shows the corresponding SNR of the source where the color coding is the same as in the main plot and the detection threshold of $\rho=20$ is shown by the black dotted line.
    }
\label{fig:emri_ds}
\end{figure}

\textit{Spin of the central BH}: \fref{fig:emri_ds} shows the detection accuracy for the magnitude of the spin of the MBH. For the three detection scenarios and all spin magnitudes considered, the error and the SNR change very little, the oscillation most likely being a result of inaccuracies in the waveform. TianQin alone, LISA alone as well as the joint detection will have errors of the order of $10^{-4}$. However, LISA and the joint detection have a SNR between 60 and 70, increasing with an increasing spin because of its increasing effect on the orbit of the SBH, while TianQin only will have a SNR of around 20, again slightly increasing with the spin. Therefore, the error in TianQin is around three times the error in the two other detection scenarios.

\begin{figure}[tpb] \centering \includegraphics[width=0.48\textwidth]{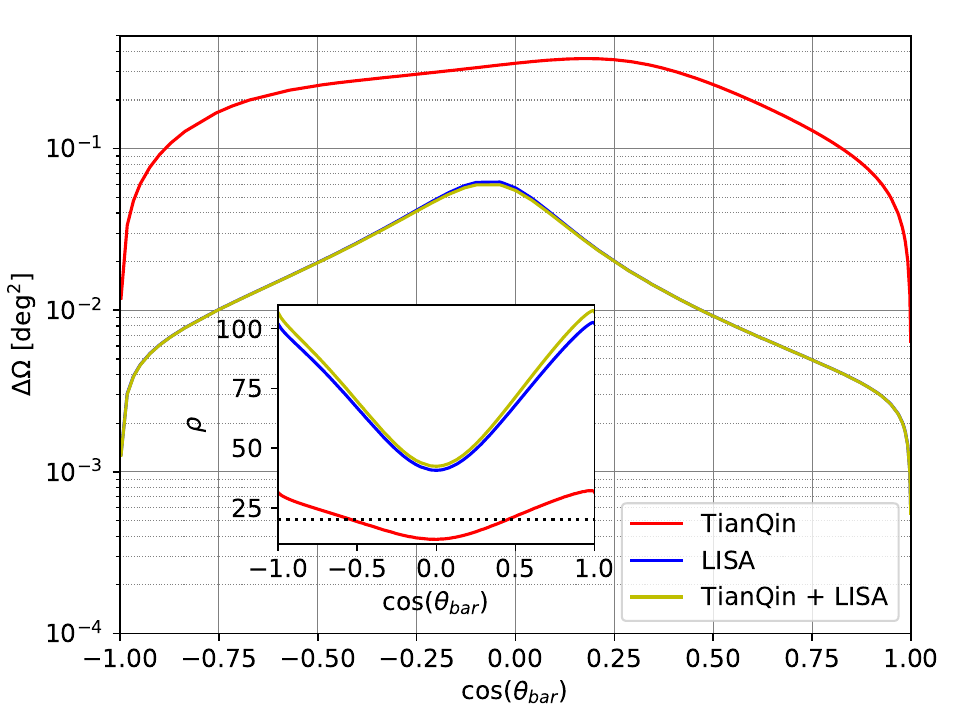}
\caption{
    The sky localization error $\Delta\Omega$ for different $\cos(\theta_{\rm bar})$. The red line shows the detection accuracy for TianQin alone, the blue line for LISA alone, and the yellow line for joint detection. The inset shows the corresponding SNR of the source where the color coding is the same as in the main plot and the black dotted line shows the detection threshold ($\rho=20$).
    }
\label{fig:emri_domegat}
\end{figure}

\begin{figure}[tpb] \centering \includegraphics[width=0.48\textwidth]{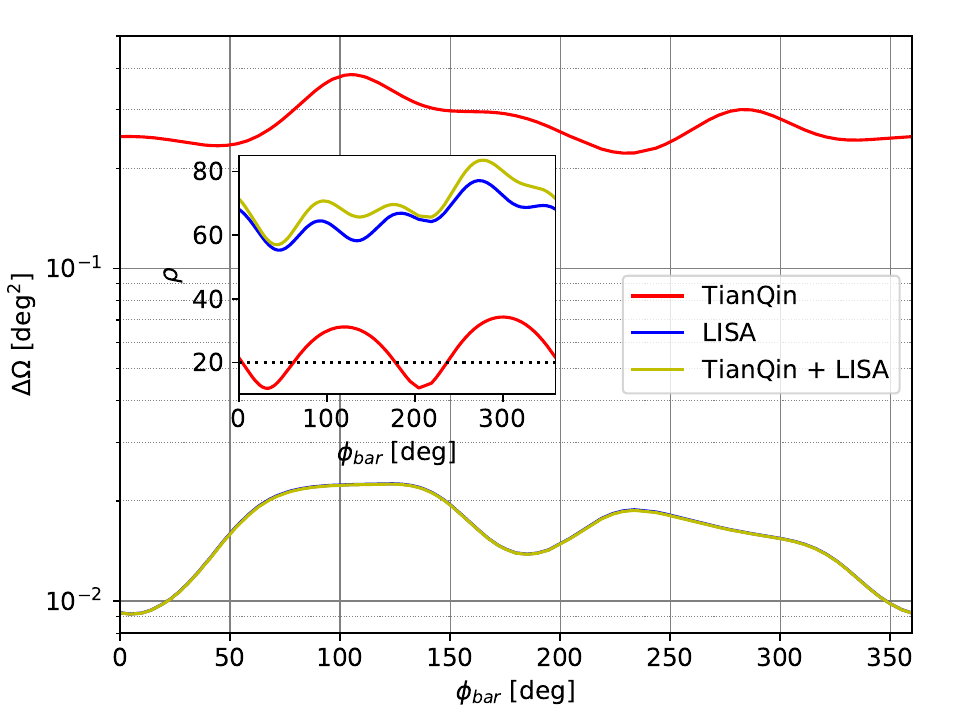}
\caption{
    The sky localization error $\Delta\Omega$ as a function of $\phi_{\rm bar}$. The red line shows the detection accuracy for TianQin alone, the blue line for LISA alone, and the yellow line for joint detection. The inset shows the corresponding SNR of the source with the same color coding as in the main plot and the detection threshold ($\rho=20$) is indicated by the black dotted line.
    }
\label{fig:emri_domegap}
\end{figure}

\textit{Sky localization:} The sky localization error for an EMRI as a function of $\cos(\theta_{\rm bar})$ and $\phi_{\rm bar}$ is shown in \fref{fig:emri_domegat} and \fref{fig:emri_domegap}, respectively. We see that in TianQin the sky localization error as a function of $\cos(\theta_{\rm bar})$ is of the order $10^{-1}\,{\rm deg^2}$ only improving to an order of $10^{-2}\,{\rm deg^2}$ for $|\cos(\theta_{\rm bar})|\gtrsim0.85$. For LISA and joint detection, which again follows the behavior in LISA closely, the detection error is mostly of the order $10^{-2}\,{\rm deg^2}$ but improves to an order of $10^{-3}\,{\rm deg^2}$ for $\cos(\theta_{\rm bar}) < -0.75$ and $\cos(\theta_{\rm bar}) \gtrsim 0.5$. Note that the detection error is bigger when close to the ecliptic plane because the spin of the MBH is set to be parallel to the normal vector of the plane and thus at this position the source is being seen closer to edge-on which results in a weaker signal. Moreover, we see that for all detection scenarios, there is an anti-correlation between the error and the SNR. As mentioned before, such an anti-correlation between the sky localization error and the SNR is expected, although it is a little surprising that EMRIs with some of the most complicated waveforms follow this simple behavior particularly closely. For the sky localization error as a function of $\phi_{\rm bar}$, we find that for TianQin it varies between $0.2\,{\rm deg^2}$ and $0.4\,{\rm deg^2}$ while for LISA and the joint detection, it ranges between $9\times10^{-3}\,{\rm deg^2}$ and $2.5\times10^{-2}\,{\rm deg^2}$. While for LISA and, accordingly, the joint detection there are distinct minima around $180^\circ$ and $360^\circ$, for TianQin the accuracy varies with no clear pattern. This stays in contrast to the SNR where TianQin has two minima at around $30^\circ$ and $210^\circ$, and two maxima around $120^\circ$ and $300^\circ$ in accordance with the position of RX J0806.3+1527 while LISA and the joint detection oscillate more irregularly.

\subsection{Comparing TianQin, LISA \& joint detection}\label{sec:emric}

For EMRIs, LISA shows a clear advantage compared to TianQin. The detection distance in LISA is significantly better than in TianQin, also resulting in higher SNRs. In contrast to MBHBs, LISA does not show an advantage towards higher masses because high-mass sources are not well detected by any of the two detectors. However, towards the lower end of the mass spectrum, TianQin can achieve detection distances and accuracies close to those of LISA.

\begin{figure}[tpb] \centering \includegraphics[width=0.48\textwidth]{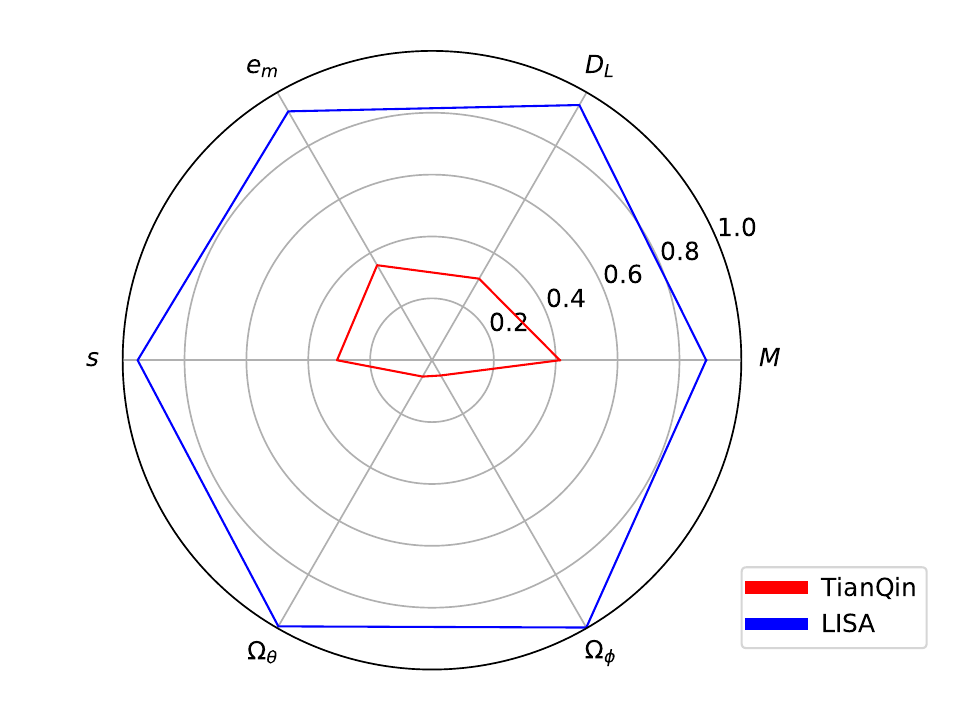}
\caption{
    The performance ratio $Q_X[\lambda]$ for the parameters $\lambda$ of an EMRI, where we label the ratios only using the parameters, for TianQin and LISA. $\Omega_{\theta}$ and $\Omega_{\phi}$ are the sky localization as a function of the polar angle $\theta_{\rm bar}$ and the azimuth angle $\phi_{\rm bar}$, respectively.
    }
\label{fig:emri_radar}
\end{figure}

The performance ratio $Q$ for TianQin and LISA compared to the joint detection is shown in \fref{fig:emri_radar}. For all parameters considered, on average LISA performs much better than TianQin. In particular, LISA dominates in terms of constraining the sky localization $\Omega$. The biggest contribution by TianQin to joint detection comes from measuring the mass on the central BH $M$ due to its improved sensitivity to lower mass sources. However, joint detection is mainly dictated by LISA due to its higher contribution to the total SNR while TianQin can slightly contribute to an improved detection.

\section{Intermediate mass ratio inspirals}\label{sec:imri}

Intermediate mass black holes (IMBHs) in the mass range $10^2-10^5\,{\rm M_\odot}$ are considered to be the ``missing link'' between SBHs and MBHs. They are thought to form in dense stellar systems such as globular clusters through the collapse of a very massive star assembled by stellar collisions~\citep{portegies-zwart_mcmillan_2002,freitag_amaro-seoane_2006,freitag_gurkan_2006,giersz_leigh_2015,mapelli_2016} or via multiple interactions/mergers between SBHs and stars~\citep{giersz_leigh_2015,di-carlo_giacobbo_2019,rizzuto_naab_2021,gonzalez_kremer_2021}. Other possible formation scenarios are the direct collapse of massive stars with extremely low metallicity~\citep{madau_rees_2001,bromm_coppi_2002,ohkubo_nomoto_2009,spera_mapelli_2017} or of gaseous clouds in the early universe~\citep{latif_schleicher_2013}, a seeding in high-redshift, metal-poor galactic halos~\citep{bellovary_volonteri_2011}, and their formation in active galactic nuclei~\citep{mckernan_ford_2012} or the circumnuclear regions of galactic discs~\citep{taniguchi_shioya_2000}. The formation of IMBHs as remnants of Population III stars has also been discussed in the literature~\citep{sakurai_yoshida_2017,wang_tanikawa_2022,liu_wang_2023}.

Some IMBH candidates have been found in galactic globular clusters~\citep{noyola_gebhardt_2010,lu_do_2013,lanzoni_mucciarelli_2013,kiziltan_baumgardt_2017}, although, it has been discussed that the presence of an IMBH in a globular cluster could also be mimicked by other processes like the presence of a dense subsystem of SBHs in the center of the cluster~\citep{van-der-marel_anderson_2010,arca-sedda_2016,askar_arca-sedda_2018,weatherford_chatterjee_2018}. Besides these indirect detections of IMBHs, the detection of GW190521 by the LIGO-Virgo-Collaboration represents the first direct detection of an IMBH as the remnant of two SBHs with masses of $85\,{\rm M_\odot}$ and $66\,{\rm M_\odot}$ merging~\citep{ligo_virgo_2020b}. However, the short duration of GW190521 in band limits the information contained thus provoking many discussions regarding the detection itself~\citep{romero-shaw_lasky_2020,fishbach_holz_2020,nitz_capano_2021,calderon-bustillo_sanchis-gual_2021b,olsen_roulet_2021,xu_hamilton_2022,gayathri_healy_2022,estelles_husa_2022} as well as its interpretation~\citep{chen_li_2019,calderon-bustillo_sanchis-gual_2021a,shibata_kiuchi_2021,torres-orjuela_chen_2022}. Therefore, probing IMBHs still represents one of the most interesting challenges in astronomy.

More direct detections of IMBHs using GWs will provide further evidence for their existence and help to clarify their properties~\citep{amaro-seoane_santamaria_2010,ligo_virgo_2017,mezcua_2017,torres-orjuela_2023}. If a compact object orbits the IMBH  -- usually denoted as ``light'' intermediate-mass ratio inspirals (IMRIs) -- it can start emitting GWs at low frequencies detectable by TianQin and LISA~\citep{will_2004,amaro-seoane_gair_2007,konstantinidis_amaro-seoane_2013,leigh_lutzgendorf_2014,haster_antonini_2016,macleod_trenti_2016,amaro-seoane_2018a,amaro-seoane_2018b,arca-sedda_amaro-seoane_2021,rizzuto_naab_2021}. The aforementioned systems are denoted as ``light'' to differentiate from the so-called ``heavy'' IMRIs that are formed by an IMBH orbiting a MBH~\citep{basu_chakrabarti_2008,arca-sedda_gualandris_2018,arca-sedda_capuzzo-dolcetta_2019,derdzinski_dorazio_2019,bonetti_rasskazov_2020,derdzinski_dorazio_2021,rose_naoz_2022}. Despite light and heavy IMRIs significantly differing in their total mass and thus in their formation, the systems where they exist, and their astrophysical implications, they are almost identical systems from the point of view of general relativity. This means they can be described by the same waveform models by rescaling their mass.

We study in this section the distance to which light and heavy IMRIs can be detected by TianQin, LISA, as well as joint detection. We, further, use a FMA to study the detection accuracy for the most relevant IMRI parameters. For light IMRIs, we adopt a detection threshold of $\rho=15$ as used in \cite{arca-sedda_amaro-seoane_2021} while for heavy IMRIs we adopt a detection threshold of $\rho=20$ from EMRIs~\citep{babak_gair_2017,tq_emri_2020}. IMRIs are among the most difficult GW sources to model because the relatively big difference in the masses of the component BHs is difficult to model in Numerical Relativity simulations while not being big enough to perform an accurate analytical approximation using the mass ratio. Here we use one of the few waveform models that can currently accurately describe IMRIs namely \texttt{BHPTNRSur1dq1e4}, a Numerical Relativity surrogate model from the `Black Hole Perturbation Toolkit'~\citep{islam_field_2022,field_galley_2014,BHPToolkit}. It can model non-spinning BHs on circular orbits with mass ratios varying from 2.5 to 10000 for durations of up to 30500\,$m_1$ (where $m_1$ is the mass of the primary BH).

For light IMRIs, we vary the mass of the primary BH in the range $6\times10^2-10^5\,{\rm M_\odot}$, thus \texttt{BHPTNRSur1dq1e4} can generate durations between around $90\,{\rm s}$ and $4.16\,{\rm hrs}$. For heavy IMRIs, we consider the mass of the primary BH to be between $10^5\,{\rm M_\odot}$ and $3\times10^6\,{\rm M_\odot}$, corresponding to durations of roughly $4.16\,{\rm hrs}$ and $125\,{\rm hrs}$, respectively. \texttt{BHPTNRSur1dq1e4} is only able to simulate the late inspiral which significantly limits the study of IMRIs which are expected to be for much longer times in band, in particular, for the lighter systems. Therefore, the results obtained here can be considered pessimistic and for longer observation times they will improve.

\subsection{Detectability}\label{sec:imrid}

\subsubsection{Light intermediate mass ratio inspirals}\label{sec:limrid}

\begin{figure}[tpb] \centering \includegraphics[width=0.48\textwidth]{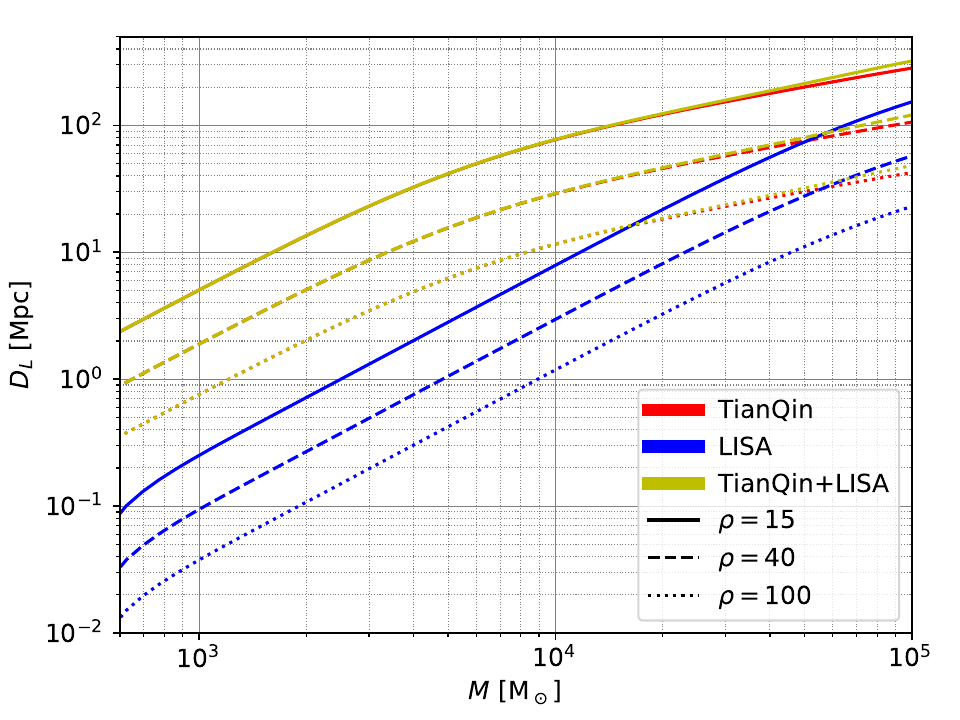}
\caption{
    The luminosity distance $D_L$ to which a light IMRI can be detected with a fixed SNR $\rho$ for TianQin, LISA, and joint detection as a function of the total mass $M$.
    }
\label{fig:limri_hor_snr}
\end{figure}

We study the luminosity distance $D_L$ at which a light IMRI with a mass ratio of 100 can be detected as a function of the binary's total mass $M$ in the observer frame in this subsection. \fref{fig:limri_hor_snr} shows at what distance TianQin, LISA, as well as joint detection, will detect light IMRIs with a SNR of 15, 40, and 100. We see that for all masses considered, TianQin will reach significantly bigger distances than LISA because these sources mainly emit GWs of higher frequencies and thus the distances of the joint detection follow those of TianQin closely. Only for masses above roughly $5\times10^4\,{\rm M_\odot}$ LISA's contribution becomes more significant and can contribute to improved joint detection. TianQin will detect sources below $1000\,{\rm M_\odot}$ out to a few ${\rm Mpc}$ for $\rho=15$ and below $1\,{\rm Mpc}$ for $\rho=100$ while LISA will detect the same sources with the same SNR to distance of the order $0.1\,{\rm Mpc}$ and $10^{-2}\,{\rm Mpc}$, respectively. The distance increases with the mass reaching almost $100\,{\rm Mpc}$ for $M\sim10^4\,{\rm M_\odot}$ and several hundred ${\rm Mpc}$ for $M\approx10^5\,{\rm M_\odot}$ in TianQin and for $\rho=15$ while for LISA and the same SNR the detection distance is around $10\,{\rm Mpc}$ and around $100\,{\rm Mpc}$ for $M\sim10^4\,{\rm M_\odot}$ and $M\approx10^5\,{\rm M_\odot}$, respectively.

\begin{figure}[tpb] \centering \includegraphics[width=0.48\textwidth]{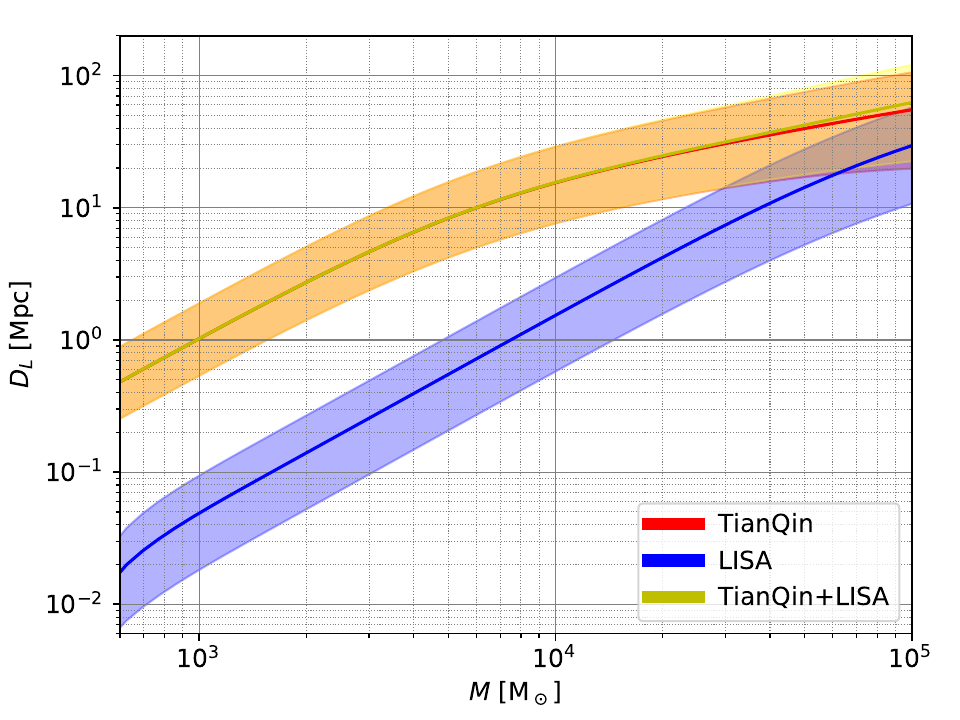}
\caption{
    The luminosity distance $D_L$ to which a light IMRI with a SNR of 40 can be detected for a varying inclination for TianQin, LISA, and joint detection as a function of the total mass $M$. The upper edge of the shaded region corresponds to the face-on case while the lower edge corresponds to the edge-on case; the solid line corresponds to the average inclination of $60^\circ$.
    }
\label{fig:limri_hor_inc}
\end{figure}

In \fref{fig:limri_hor_inc}, we show the detection distance for a SNR of 40 and different inclinations. As usual, face-on sources can be detected further away than edge-on sources differing by a factor of around 3.5 in TianQin (and the joint detection) and almost five in LISA for lighter sources with $M\sim10\,{\rm M_\odot}$. For heavier sources with masses of almost $10^5\,{\rm M_\odot}$ the difference further increases to almost five in TianQin and six in LISA. Sources with an average inclination of $60^\circ$ can be detected at distances two to three times bigger than edge-on sources in TianQin while for LISA the distance for the two inclinations differs by a factor between 2.5 and three. Thus the detection by LISA is slightly more affected by the orientation of a light IMRI than the detection by TianQin and consequently the joint detection.

\subsubsection{Heavy intermediate mass ratio inspirals}\label{sec:himrid}

\begin{figure}[tpb] \centering \includegraphics[width=0.48\textwidth]{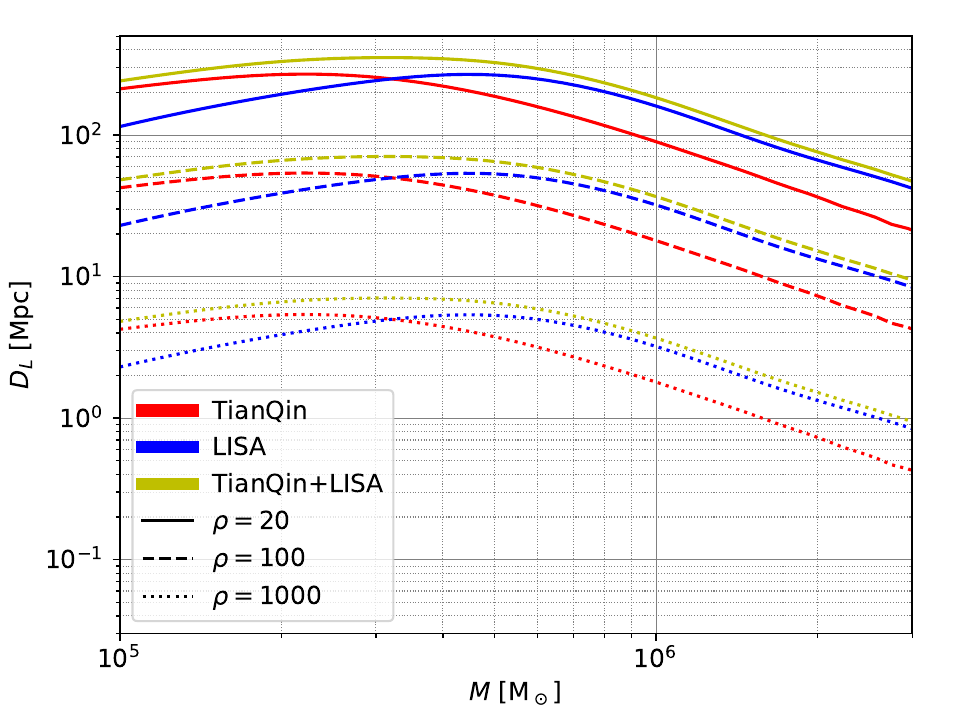}
\caption{
    The luminosity distance $D_L$ to which a heavy IMRI can be detected with a fixed SNR $\rho$ for TianQin, LISA, and joint detection as a function of the total mass $M$.
    }
\label{fig:himri_hor_snr}
\end{figure}

In this subsection, we present the distance at which a heavy IMRI ($q=100$) with a fixed SNR of 20, 100, and 1000 can be detected by TianQin alone, LISA alone, and joint detection for different total masses of the source $M$. In \fref{fig:himri_hor_snr}, we see that for $M\lesssim3\times10^5\,{\rm M_\odot}$ and $\rho=20$ TianQin alone reaches slightly bigger distances than LISA alone: around $200\,{\rm Mpc}$ for the first and $100 - 200\,{\rm Mpc}$ for the second. Due to the similar performance of TianQin and LISA in this mass range, the joint detection can reach a bigger distance between $250\,{\rm Mpc}$ and $300\,{\rm Mpc}$. For an increasing mass, the detection distance in TianQin decreases where for a total mass of $10^6\,{\rm M_\odot}$ it goes down to around $90\,{\rm Mpc}$, $20\,{\rm Mpc}$, and $2\,{\rm Mpc}$ for a SNR of 20, 100, and 1000, respectively. In LISA the detection distance first increases for an increasing mass reaching a maximum of almost $250\,{\rm Mpc}$ for $\rho=20$ and $M\approx4\times10^5\,{\rm M_\odot}$ but later it also decreases going down to around $100\,{\rm Mpc}$ for $M=10^6\,{\rm M_\odot}$ and $\rho=20$. The distance first increasing, then reaching a maximum, and decreasing again is a result of the wave's frequency decreasing as the mass increases wherefore it moves through LISA's sensitivity band. For the highest mass considered $M=3\times10^6\,{\rm M_\odot}$ the detection distance for joint detection is similar to LISA being around $45\,{\rm Mpc}$, $9\,{\rm Mpc}$, and $0.9\,{\rm Mpc}$ for a SNR of 20, 100, and 1000, respectively.

\begin{figure}[tpb] \centering \includegraphics[width=0.48\textwidth]{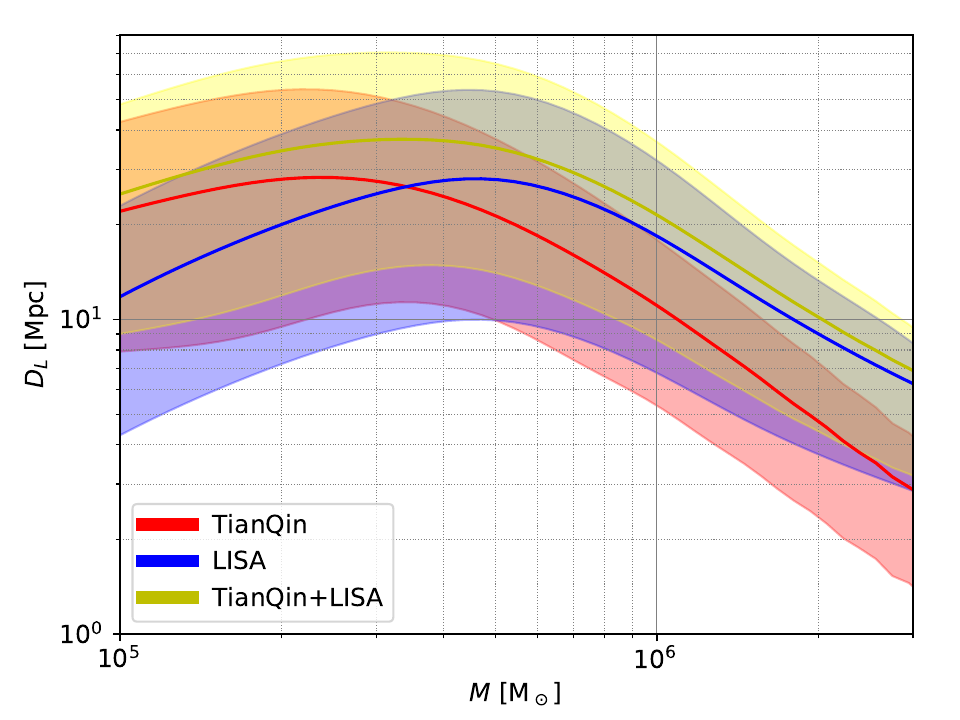}
\caption{
    The luminosity distance $D_L$ to which a heavy IMRI with a SNR of 100 can be detected for a varying inclination for TianQin, LISA, and joint detection as a function of the total mass $M$. The upper edge of the shaded region corresponds to the face-on case and the lower edge corresponds to the edge-on case, while the solid line corresponds to the average inclination of $60^\circ$.
    }
\label{fig:himri_hor_inc}
\end{figure}

The relation between the detection distance and the inclination of a heavy IMRI with a SNR of 100 is shown in \fref{fig:himri_hor_inc}. For the lowest masses $\sim10^5\,{\rm M_\odot}$ the difference between the detection distance of a face-on source and an edge-on source differs by a factor of around five for all three detection scenarios. The factor by which the distance for an edge-on source and a source with an inclination of $60^\circ$ differs is again similar for all detection scenarios being around three. For higher masses, the dependence of the detection distance on the inclination decreases. For $M=3\times10^6\,{\rm M_\odot}$ in TianQin edge-on and face-on sources differ by a factor of around three while edge-on sources and a source with $\iota=60^\circ$ by a factor of around 1.5. For LISA and joint detection, the dependence on the inclination also decreases for higher masses but is less significant. For $M=3\times10^6\,{\rm M_\odot}$ an edge-on and a face-on source differ by a factor of almost four while an edge-on source and a source with an average inclination of $60^\circ$ still differ by a factor of around three.

\subsection{Parameter estimation}\label{sec:imrip}

\subsubsection{Light intermediate mass ratio inspirals}\label{sec:limrip}

We examine in this section the accuracy with which different parameters of a light IMRI can be detected by TianQin alone, LISA alone, and by joint detection. The parameters considered are the total mass of the source $M$, its luminosity distance $D_L$, the mass ratio between the heavy and the light black hole $q$, the inclination of the system relative to the line-of-sight $\iota$ and the sky localization error $\Delta\Omega$. In each analysis we vary one parameter while fixing the other parameters using the fiducial values $M = 5000\,{\rm M_\odot}$, $D_L = 1\,{\rm Mpc}$, $q = 100$ $\iota = 60^\circ$. Moreover, we set the source to be at the sky location $\theta_{\rm bar} = 90^\circ$ and $\phi_{\rm bar} = 180^\circ$ where we use barycentric coordinates.

\begin{figure}[tpb] \centering \includegraphics[width=0.48\textwidth]{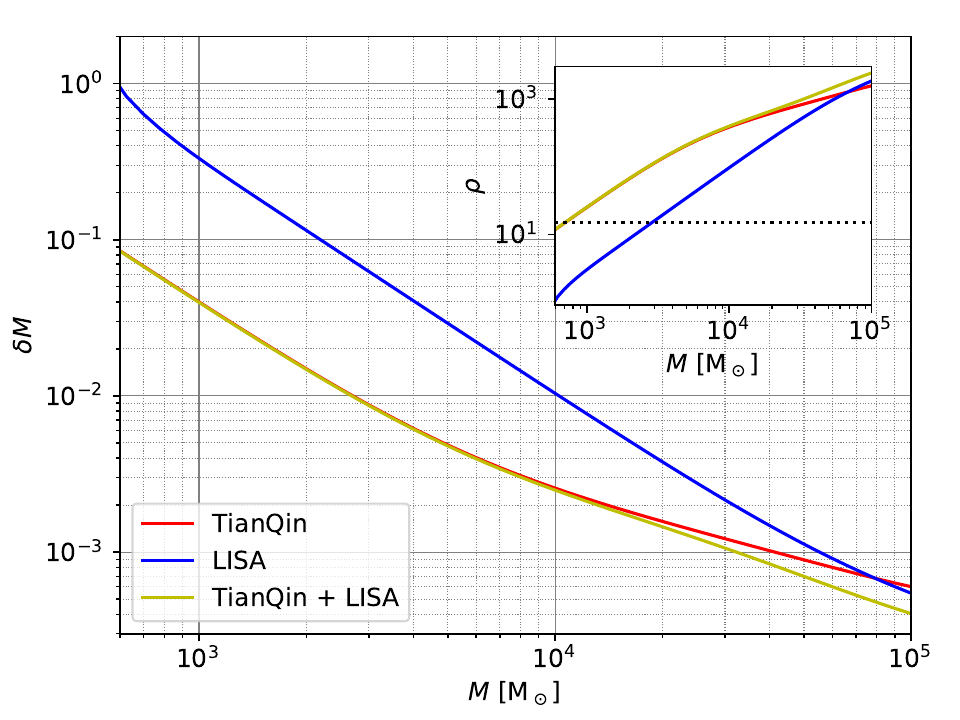}
\caption{
    The relative error in the total mass of a light IMRI $\delta M$ as a function of the IMRI's total mass $M$. The error for TianQin alone is shown in red, for LISA alone in blue, and the yellow line shows the error for joint detection. The inset shows the corresponding SNR of the source using the same color coding as in the main plot and the black dotted line indicates the detection threshold of $\rho=15$.
    }
\label{fig:limri_dm}
\end{figure}

\textit{Total mass:} \fref{fig:limri_dm} shows the relative detection error in the total mass of a light IMRI. For low-mass systems with $M\sim10^2\,{\rm M_\odot}$, the detection accuracy in TianQin alone is around 0.1 and almost an order of magnitude better than in LISA alone while the joint detection shows similar results to TianQin. The big error in LISA is related to the low SNR that remains below 10 for $M\lesssim2000\,{\rm M_\odot}$ since the source mainly emits harmonics with frequencies that LISA only can detect poorly. For increasing masses, the difference between TianQin and LISA in the detection error and the SNR decreases although for $M\approx10^4\,{\rm M_\odot}$ TianQin and the joint detection have an error of around $3\times10^{-3}$ while LISA has an error of around $10^{-2}$. Only for the highest masses of almost $10^5\,{\rm M_\odot}$ TianQin and LISA perform similarly with an error of several times $10^{-4}$ and a SNR of almost 2000 leading to a significant gain in accuracy and SNR for joint detection. 

\begin{figure}[tpb] \centering \includegraphics[width=0.48\textwidth]{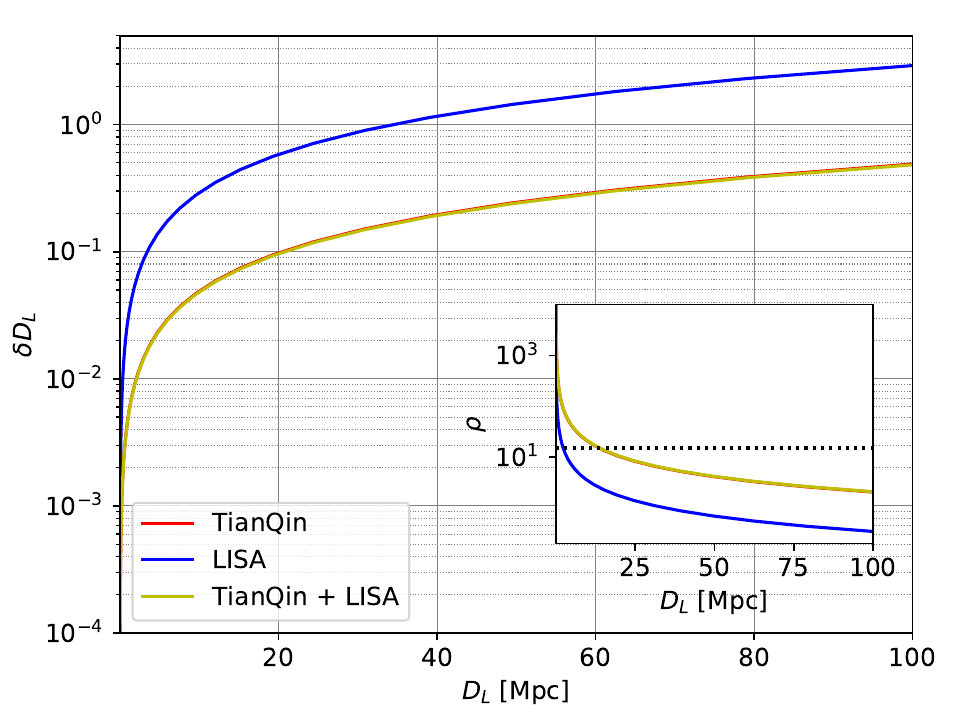}
\caption{
    The relative error in the luminosity distance $\delta D_L$ for a light IMRI as a function of the luminosity distance of the source $D_L$. The error for TianQin alone is shown in red, for LISA alone in blue, and the yellow line shows the error for joint detection. The inset shows the corresponding SNR of the source where the color coding is the same as in the main plot and the black dotted line shows the detection threshold ($\rho=15$).
    }
\label{fig:limri_ddl}
\end{figure}

\textit{Luminosity distance:} The relative error in the luminosity distance of a light IMRI is shown in \fref{fig:limri_ddl}. For all distances considered TianQin alone -- as well as the joint detection which follows TianQin's behavior closely -- perform almost an order of magnitude better than LISA alone. The relative error in TianQin remains below 0.1 up to distances of $20\,{\rm Mpc}$ while in LISA the error goes above 0.1 if $D_L\gtrsim5\,{\rm Mpc}$. These two distances are also roughly the limits for the SNR being above the detection threshold of 15. For $D_L\gtrsim20\,{\rm Mpc}$ the SNR in TianQin goes below 15 but remains close to ten while for LISA it goes even below five. Correspondingly the detection error goes up as the distance increases reaching around three and 0.4 for $D_L=100\,{\rm Mpc}$ in TianQin and LISA, respectively.

\begin{figure}[tpb] \centering \includegraphics[width=0.48\textwidth]{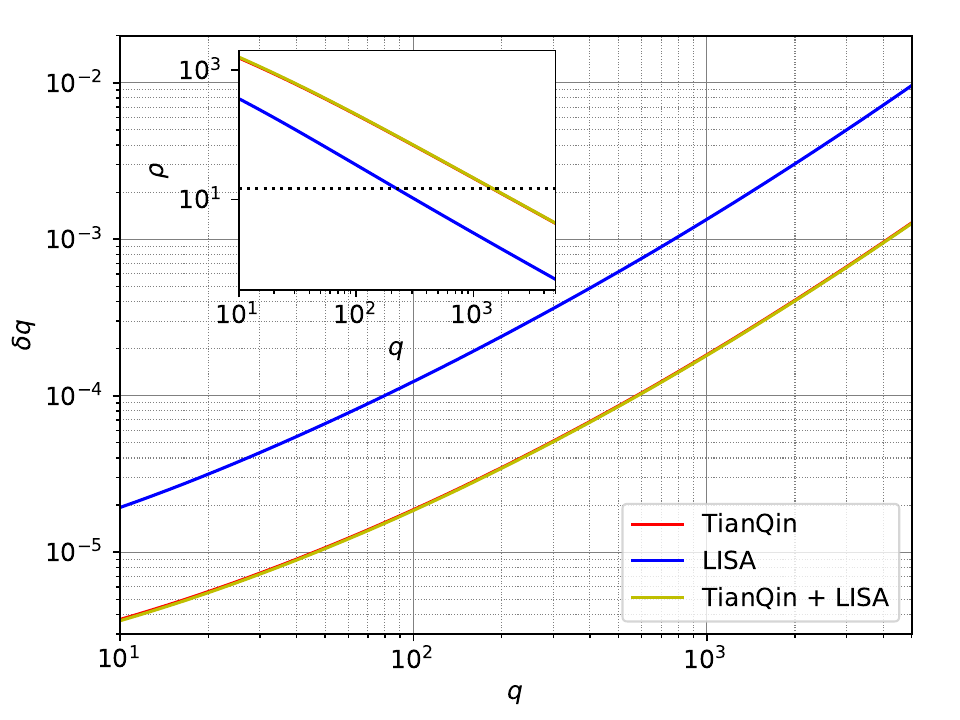}
\caption{
    The relative error in the mass ratio $\delta q$ for different mass ratios of a light IMRI $q$. The red line shows the error for TianQin alone, the blue line for LISA alone, and the yellow line for joint detection. The inset shows the corresponding SNR of the source where the color coding is the same as in the main plot and the detection threshold of $\rho=15$ is indicated by the black dotted line.
    }
\label{fig:limri_dq}
\end{figure}

\textit{Mass ratio:} In \fref{fig:limri_dq} we show the relative error for the mass ratio. We see that for all detection scenarios, the SNR goes down for an increasing mass ratio while the error goes up. TianQin and the joint detection perform better than LISA by a factor of around five having a relative error of around $4\times10^{-6}$ compared to LISA's error of $2\times10^{-5}$ for $q=10$. For such a low mass ratio the SNR is relatively high being around 1500 in TianQin and 350 in LISA. For higher mass ratios of around 100, the SNR goes down to around 300 and 50 while the error goes up to roughly $2\times10^{-5}$ and $10^{-4}$ in TianQin and LISA, respectively. For even higher mass ratios of the order 1000, the SNR further goes down being around 10 in TianQin and the joint detection, and below 5 in LISA. The error in TianQin and the joint detection is then of the order $10^{-4}$ while it is of the order $10^{-3}$ in LISA. We point out that the decreasing SNR and increasing error for increasing mass ratios are (at least partially) a consequence of the waveform used. The waveform only contains spherical modes up to $l=5$~\citep{ruiz_alcubierre_2008} while for high mass ratios, it is expected that the signal distributes between hundreds or even thousands of modes and thus we are potentially missing a significant fraction of the signal.

\begin{figure}[tpb] \centering \includegraphics[width=0.48\textwidth]{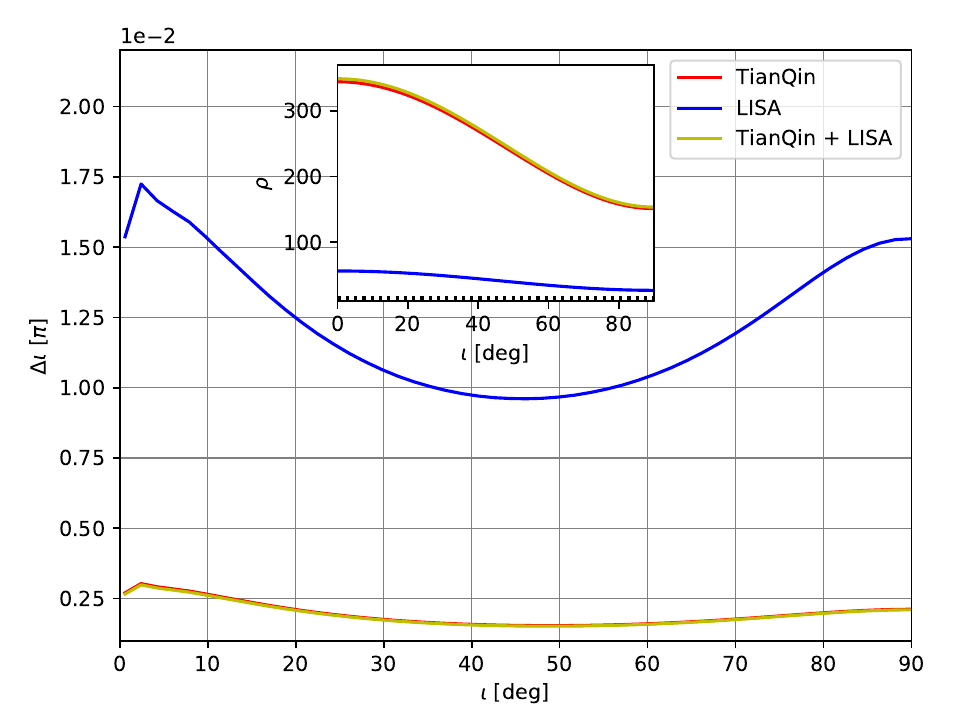}
\caption{
    The absolute error in the inclination $\Delta\iota$ of a light IMRI for different inclinations $\iota$. The red line shows the error for TianQin alone, the blue line for LISA alone, and the yellow line for joint detection. The inset shows the corresponding SNR of the source using the same color coding as in the main plot and the detection threshold ($\rho=15$) is indicated by the black dotted line.
    }
\label{fig:limri_diota}
\end{figure}

\textit{Inclination:} The absolute error in the inclination of the source is shown in \fref{fig:limri_diota}. We see that the joint detection again follows the behavior of TianQin closely with an error of around $2.5\times10^{-3}\pi$ for all inclinations considered. There is a low variation in the error but it is smaller than the change in the SNR which goes down from almost 330 for a face-on source ($\iota=0^\circ$) to around 150 for an edge-on source ($\iota=90^\circ$). In contrast, for LISA the variation in the SNR is relatively small having a value of around 50 for all inclinations while the error varies significantly between $10^{-2}\pi$ and $1.75\times10^{-2}\pi$ being the lowest for inclinations between $40^\circ$ and $50^\circ$ and the highest for a face-on source. Note that for a more realistic waveform that contains spherical modes beyond $l=5$ the detection accuracy for the inclination might further improve because the detection of higher modes is crucial to resolve the source's inclination.

\begin{figure}[tpb] \centering \includegraphics[width=0.48\textwidth]{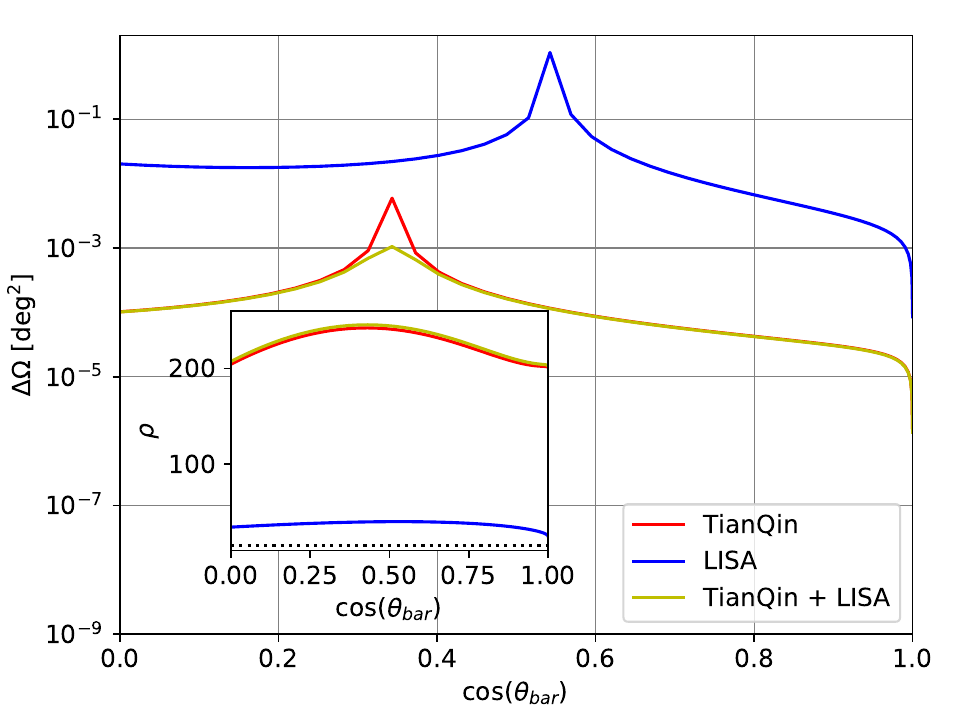}
\caption{
    The sky localization error $\Delta\Omega$ for different $\cos(\theta_{\rm bar})$ for a light IMRI. The red line shows the detection accuracy for TianQin alone, the blue line for LISA alone, and the yellow line for joint detection. The inset shows the corresponding SNR of the source where the color coding is the same as in the main plot and the black dotted line indicates the detection threshold ($\rho=15$).
    }
\label{fig:limri_domegat}
\end{figure}

\begin{figure}[tpb] \centering \includegraphics[width=0.48\textwidth]{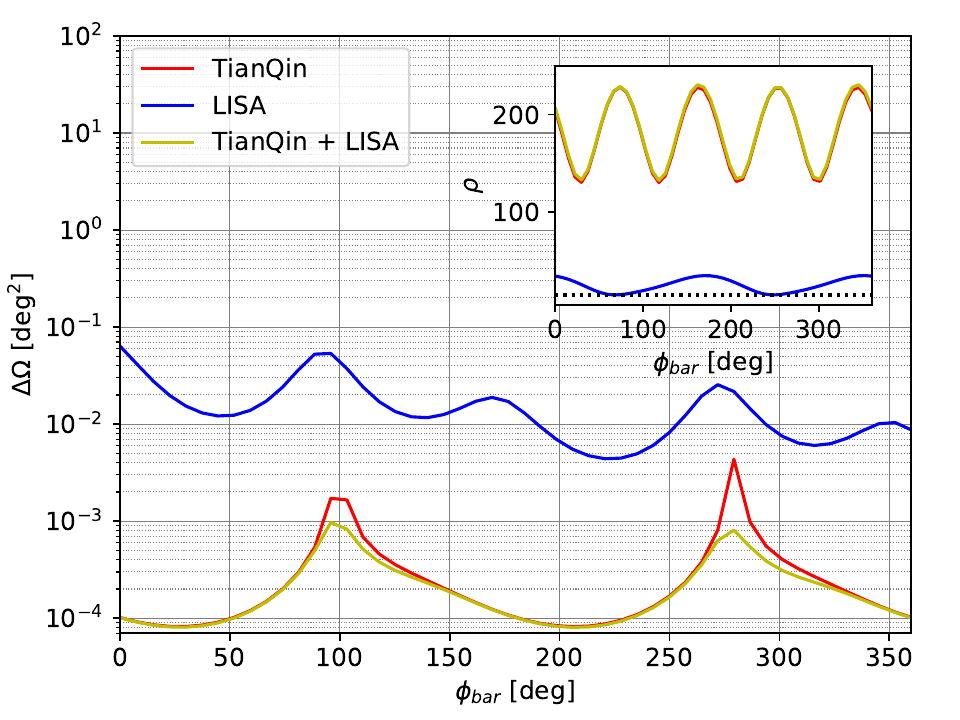}
\caption{
    The sky localization error $\Delta\Omega$ as a function of $\phi_{\rm bar}$ for a light IMRI. The red line shows the detection accuracy for TianQin alone, the blue line for LISA alone, and the yellow line for joint detection. The inset shows the corresponding SNR of the source using the same color coding as in the main plot and the detection threshold of $\rho=15$ is shown by the black dotted line.
    }
\label{fig:limri_domegap}
\end{figure}

\textit{Sky localization:} We show the sky localization error of a light IMRI in \fref{fig:limri_domegat} and \fref{fig:limri_domegap} as a function of $\cos(\theta_{\rm bar})$ and $\phi_{\rm bar}$, respectively. Note that in \fref{fig:limri_domegat}, we only show $0\leq\cos(\theta_{\rm bar})\leq1$ because $-1\leq\cos(\theta_{\rm bar})\leq0$ is almost symmetric to the first interval. In TianQin the sky localization error as a function of $\cos(\theta_{\rm bar})$ is mostly of the order $10^{-4}\,{\rm deg^2}$ going down to around $10^{-5}\,{\rm deg^2}$ for $\cos(\theta_{\rm bar})\approx1$. In LISA the sky localization error varies stronger being of the order $10^{-3}\,{\rm deg^2}$ and lower if $\cos(\theta_{\rm bar})\gtrsim0.7$, of the order $10^{-2}\,{\rm deg^2}$ for $0\leq\cos(\theta_{\rm bar})\lesssim0.45$, and of the order $10^{-1}\,{\rm deg^2}$ and bigger if $\cos(\theta_{\rm bar})\approx0.55$. The peaks in the sky localization error can be attributed to the inclination of TianQin and LISA relative to the ecliptic plane. Sources at the respective latitude angles are detected with similar accuracy for the entire (relatively short) observation time, which reduces a modulation of the signal necessary to pinpoint the location of the source. That sources at these latitude angles have a strong signal can be seen from their comparatively high SNR. The sky localization error as a function of the azimuth angle $\phi_{\rm bar}$ in TianQin of up to a few times $10^{-3}\,{\rm deg^2}$ for $\phi_{\rm bar}\approx100^\circ$ $\phi_{\rm bar}\approx280^\circ$ and of the order $10^{-3}\,{\rm deg^2}$ or slightly below for all other angles. In LISA the sky localization error is of the order $10^{-2}\,{\rm deg^2}$ if $\phi_{\rm bar}\lesssim190^\circ$ and $250^\circ\lesssim\phi_{\rm bar}\lesssim300^\circ$. For $190^\circ\lesssim\phi_{\rm bar}\lesssim250^\circ$ and $\phi_{\rm bar}\gtrsim300^\circ$the error is of the order $10^{-3}\,{\rm deg^2}$. Note that we expect a symmetry in the sky localization error for $\phi_{\rm bar}\approx0^\circ$ and $\phi_{\rm bar}\approx360^\circ$ and the small difference for LISA at these values can be attributed to numerical inaccuracies from computing the derivatives for small $\phi_{\rm bar}$ required for the FMA. The sky localization error in the joint detection follows the error in TianQin closely except that it always remains below $10^{-3}\,{\rm deg^2}$ having less pronounced peaks than TianQin alone. For the SNR we see that the joint detection again follows the behavior of TianQin alone closely varying strongly as a function of $\cos(\theta_{\rm bar})$ and even oscillating multiple times as a function of $\phi_{\rm bar}$. For LISA the SNR as a function of $\cos(\theta_{\rm bar})$ varies only a little while as a function of $\phi_{\rm bar}$ it oscillates but with a smaller amplitude and fewer peaks than for TianQin. 

\subsubsection{Heavy intermediate mass ratio inspirals}\label{sec:himrip}

In this subsection, we analyze the accuracy with which several parameters of a heavy IMRI will be detected by TianQin alone and LISA alone, as well as by joint detection. The parameters considered are the total mass of the binary $M$, the luminosity distance of the source $D_L$, the mass ratio between the primary and the secondary black hole $q$, the inclination of the system relative to the line-of-sight $\iota$ and the sky localization error $\Delta\Omega$. In each analysis we vary only one parameter and fix the other parameters to the following values $M = 10^6\,{\rm M_\odot}$, $D_L = 30\,{\rm Mpc}$, $q = 100$ $\iota = 60^\circ$. Furthermore, we set the sky location of the source to be $\theta_{\rm bar} = 90^\circ$ and $\phi_{\rm bar} = 180^\circ$ in barycentric coordinates.

\begin{figure}[tpb] \centering \includegraphics[width=0.48\textwidth]{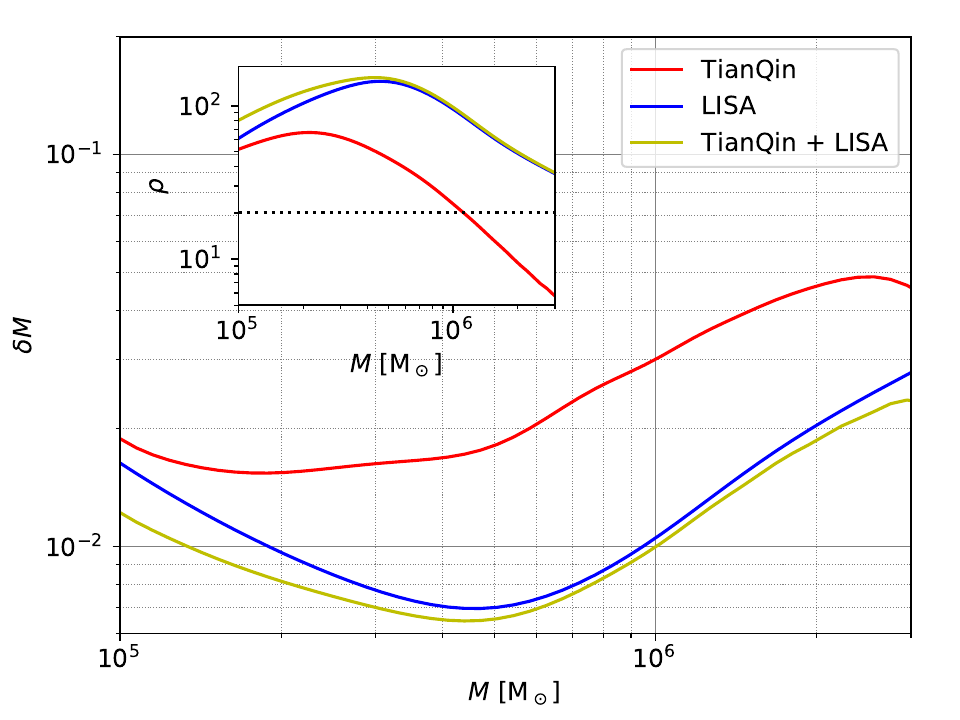}
\caption{
    The relative error in the total mass of a heavy IMRI $\delta M$ as a function of the source's total mass $M$. The error for TianQin alone is shown in red, for LISA alone in blue, and the yellow line shows the error for joint detection. The inset shows the corresponding SNR of the source where we use the same color coding as in the main plot and the black dotted line indicates the detection threshold of $\rho=20$.
    }
\label{fig:himri_dm}
\end{figure}

\textit{Total mass:} The relative error in the total mass in the observer frame of a heavy IMRI is shown in \fref{fig:himri_dm}. For low masses close to $10^5\,{\rm M_\odot}$ TianQin alone and LISA alone have a similar error between $1.5\times10^{-2}$ and $1.5\times10^{-2}$ as well as a similar SNR of around 50 and 60, respectively. Therefore, the joint detection performs slightly better than the two single detections having an error of a bit more than one times $10^{-2}$. For an increasing mass, initially, the error decreases and the SNR increases for all three detection scenarios but for TianQin the error reaches its minima of around $1.5\times10^{-2}$ and the SNR reaches its maxima of almost 70 for $M\approx2\times10^{5}\,{\rm M_\odot}$. For higher masses, the SNR in TianQin decreases rapidly and the error increases significantly to almost $5\times10^{-2}$ so that it does not affect the joint detection a lot. Instead, the joint detection follows the behavior of LISA quite closely which has its maxima in SNR of around 150 respectively its minima in the error close to $7\times10^{-3}$ for $M\approx4.5\times10^5\,{\rm M_\odot}$. For $M\gtrsim10^6\,{\rm M_\odot}$ the error increases but remains below $3\times10^{-2}$ while the SNR goes down to values between 40 and 100. We point out that TianQin's detection error increasing slower than LISA's is a result of the strong emission of higher modes that remain at frequencies where TianQin is the most sensitive even when the quadrupolar mode moves out of this region.

\begin{figure}[tbp] \centering \includegraphics[width=0.48\textwidth]{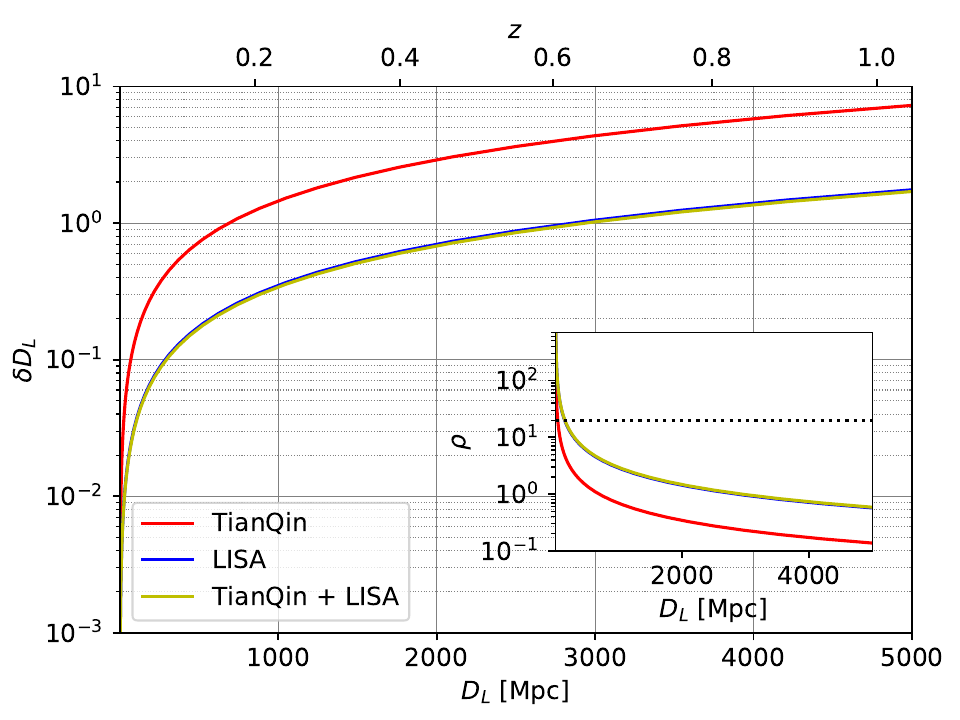}
\caption{
    The relative error in the luminosity distance $\delta D_L$ for a heavy IMRI as a function of the distance. The lower abscissa shows the luminosity distance $D_L$ while the upper abscissa shows the corresponding cosmological redshift $z$. The error for TianQin alone is shown in red, for LISA alone in blue, and the yellow line shows the error for joint detection. The inset shows the corresponding SNR of the source where the color coding is the same as in the main plot and the detection threshold ($\rho=20$) is indicated by the black dotted line.
    }
\label{fig:himri_ddl}
\end{figure}

\textit{Luminosity distance:} The error in the distance of the source is shown in \fref{fig:himri_ddl} where we see that LISA alone and the joint detection perform almost an order of magnitude better than TianQin alone. The error on TianQin is only smaller than one if $z\leq0.2$ and goes up to more than seven for a redshift of 1. Only at redshift zero where the SNR is well above the detection threshold the error in TianQin is of 0.1 or lower. For LISA alone as well as for the joint detection, the error is only smaller than 0.1 if $D_L\lesssim300\,{\rm Mpc}$ but remains below one out to redshifts of 0.6 ($\approx2700\,{\rm Mpc}$). At these distances, the SNR is of a few but for higher redshifts of around one it goes down to roughly one while the relative error goes up to almost two. Note that using a waveform that contains more spherical modes beyond $l=5$ might improve the detection accuracy for the luminosity distance. Not only because of the higher SNR that can be expected for an IMRI that limits a significant ratio of its energy at higher modes but because these modes will allow constraining the inclination of the source more accurately which is degenerate with $D_L$.

\begin{figure}[tpb] \centering \includegraphics[width=0.48\textwidth]{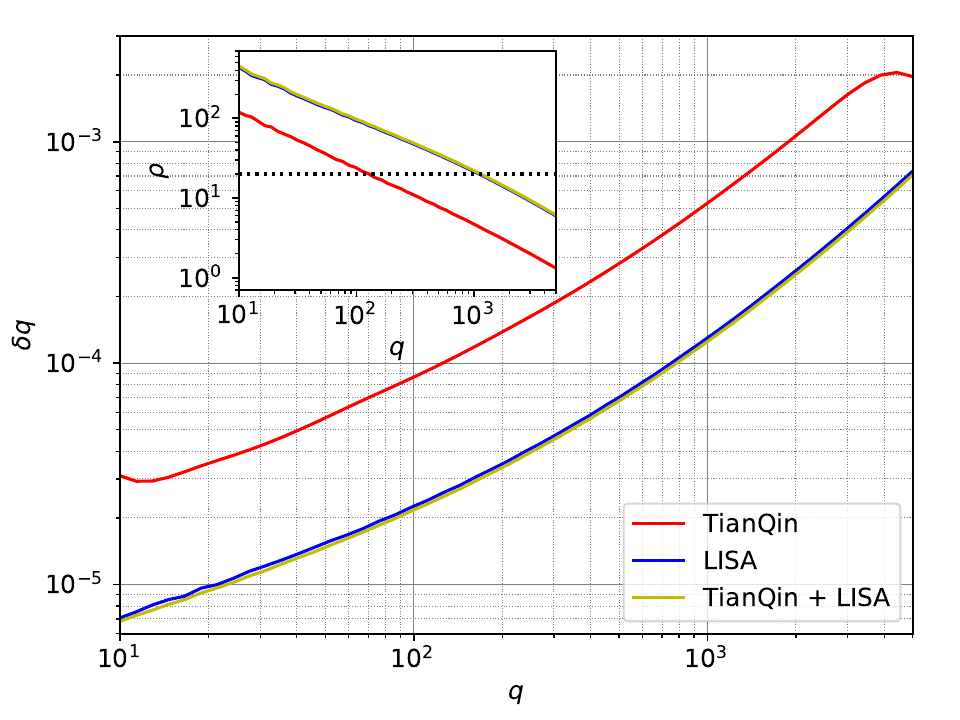}
\caption{
    The relative error in the mass ratio $\delta q$ for different mass ratios of a heavy IMRI $q$. The red line shows the error for TianQin alone, the blue line for LISA alone, and the yellow line for joint detection. The inset shows the corresponding SNR of the source where the color coding is the same as in the main plot and the black dotted line indicates the detection threshold of $\rho=20$.
    }
\label{fig:himri_dq}
\end{figure}

\textit{Mass ratio:} \fref{fig:himri_dq} shows the relative error in the mass ratio where we see that the error increases for an increasing mass ratio, although, as for light IMRIs this behavior can, at least partially, be attributed to the restricted number of spherical modes the waveform contains. For TianQin the relative error varies between $3\times10^{-5}$ and $2\times10^{-3}$ for $q=10$ and $q=5000$, respectively. We, further, see that in TianQin the SNR only remains above 20 if $q\lesssim150$. In LISA and for the joint detection the SNR remains above 20 for mass ratios of up to around 1000 where the relative error is below $1.5\times10^{-4}$. For lower mass ratios below 800, the relative error is of the order of $10^{-5}$ and even goes below $10^{-5}$ if $q\lesssim20$. For the highest mass ratios close to 5000, the error is almost $8\times10^{-4}$.

\begin{figure}[tpb] \centering \includegraphics[width=0.48\textwidth]{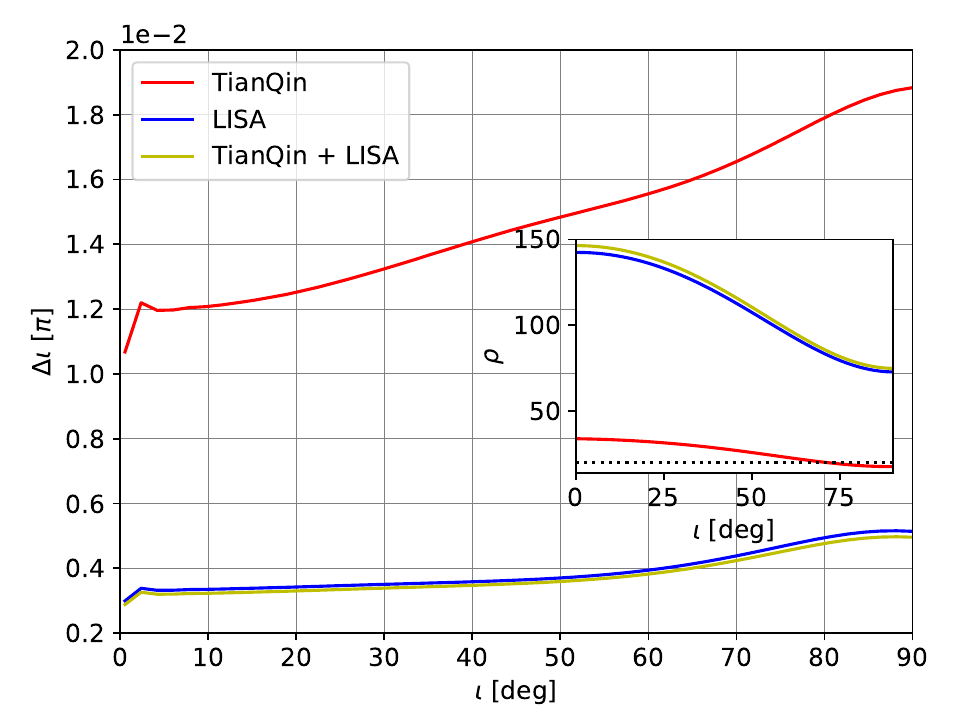}
\caption{
    The absolute error in the inclination $\Delta\iota$ of a heavy IMRI for different inclinations of the source relative to the line-of-sight $\iota$. The error for TianQin alone is shown in red, for LISA alone in blue, and the yellow line shows the error for joint detection. The inset shows the corresponding SNR of the source using the same color coding as in the main plot and the detection threshold ($\rho=20$) is shown by the black dotted line.
    }
\label{fig:himri_diota}
\end{figure}

\textit{Inclination:} In \fref{fig:himri_diota}, we show the absolute error in the inclination of the source. We see that for TianQin alone the accuracy is of the order $10^{-2}\pi$ while for LISA alone and the joint detection, the error is of the order $10^{-3}\pi$. For LISA and the joint detection, the error varies between three and five times $10^{-3}\pi$ while for TianQin the error is roughly four times higher ranging from $1.2\times10^{-2}\pi$ to $1.9\times10^{-2}\pi$. For all cases, the error is smaller when the source is face-on ($\iota=0^\circ$) due to its higher SNR. In particular, for $\iota\gtrsim30^\circ$ the SNR starts to decrease significantly -- edge-on sources having almost half of the SNR of face-on sources. We point out once again, that the waveform used only contains spherical modes up to $l=5$ which limits measuring the inclination, in particular, for edge-on sources that usually contain a high number of higher spherical modes.

\begin{figure}[tpb] \centering \includegraphics[width=0.48\textwidth]{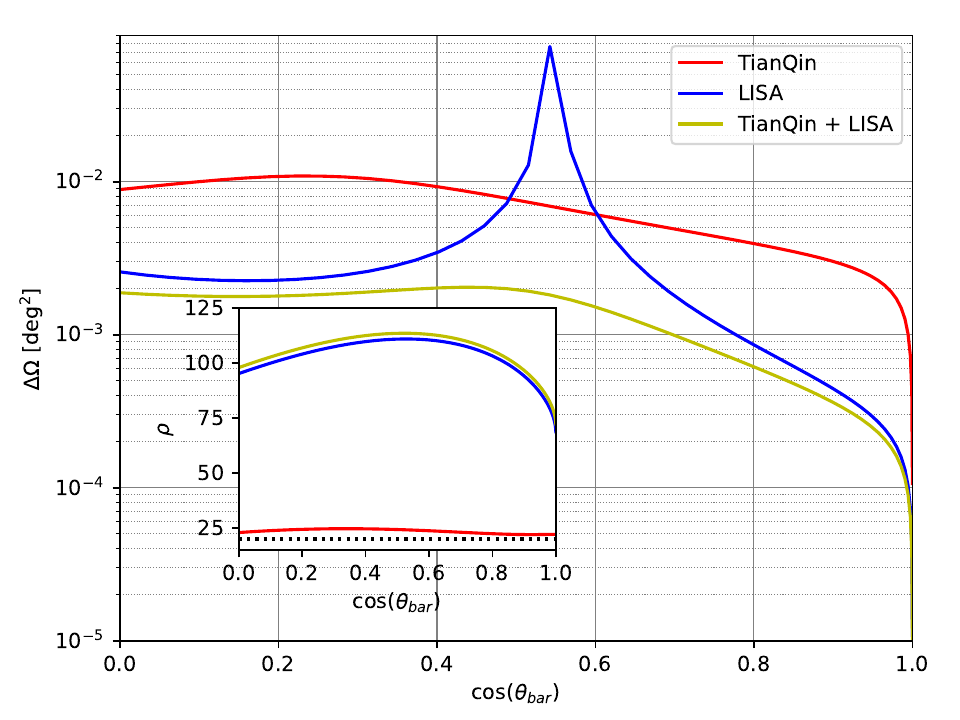}
\caption{
    The sky localization error $\Delta\Omega$ for different $\cos(\theta_{\rm bar})$ for a heavy IMRI. The red line shows the detection accuracy for TianQin alone, the blue line for LISA alone, and the yellow line for joint detection. The inset shows the corresponding SNR of the source using the same color coding as in the main plot and the black dotted line indicates the detection threshold of $\rho=20$.
    }
\label{fig:himri_domegat}
\end{figure}

\textit{Sky localization:} The sky localization error of a heavy IMRI as a function of $\cos(\theta_{\rm bar})$ is shown in \fref{fig:himri_domegat}, where we again only consider the upper half-sphere because of the symmetry of the system. For TianQin alone the error is around $10^{-2}\,{\rm deg^2}$ if $\cos(\theta_{\rm bar})$ is smaller than 0.4 while it is of the order $10^{-3}\,{\rm deg^2}$ for bigger $\cos(\theta_{\rm bar})$. For LISA alone the error is of the order $10^{-3}\,{\rm deg^2}$ if $\cos(\theta_{\rm bar})\lesssim0.5$ and $0.6\lesssim\cos(\theta_{\rm bar})\lesssim0.8$. For $\cos(\theta_{\rm bar})\gtrsim0.8$ the error goes down to an order of $10^{-4}\,{\rm deg^2}$, however, for $0.5\leq\cos(\theta_{\rm bar})\lesssim0.6$ the error in LISA is even bigger than in TianQin going up to $8\times10^{-2}\,{\rm deg^2}$. Note that LISA has the worst detection accuracy at angles where the SNR is the highest with more than 100 most likely because in this case, the source is detectable by LISA with similar accuracy at all times thus reducing a modulation of the signal necessary to pinpoint the location of the source. For TianQin, in contrast, this effect does not appear because of its fixed orientation which makes it more difficult for the source to be detected constantly well over a long observation time. The joint detection shows similar behavior to LISA due to its better accuracy for most angles. Only around the peak in the error for LISA, the error of the joint detection is restricted by TianQin so that it never goes above $2\times10^{-3}\,{\rm deg^2}$.

\begin{figure}[tpb] \centering \includegraphics[width=0.48\textwidth]{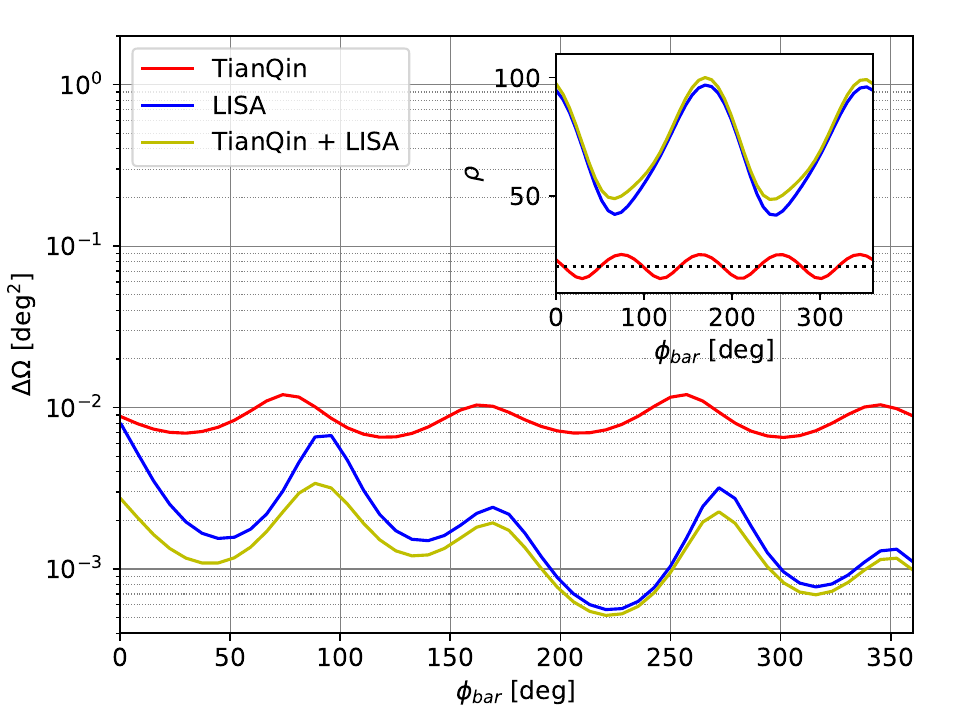}
\caption{
    The sky localization error $\Delta\Omega$ as a function of $\phi_{\rm bar}$ for a heavy IMRI. The red line shows the detection accuracy for TianQin alone, the blue line for LISA alone, and the yellow line for joint detection. The inset shows the corresponding SNR of the source where the color coding is the same as in the main plot and the detection threshold of $\rho=20$ is indicated by the black dotted line.
    }
\label{fig:himri_domegap}
\end{figure}

\fref{fig:himri_domegap} shows the sky localization error as a function of $\phi_{\rm bar}$. We see that for TianQin alone the error and the SNR only vary little oscillating between $6.5\times10^{-3}\,{\rm deg^2}$ and $1.5\times10^{-2}\,{\rm deg^2}$ for the first and between 15 and 25 for the latter. For LISA the sky localization error is mostly of the order $10^{-3}\,{\rm deg^2}$ only going below if $200^\circ\lesssim\phi_{\rm bar}\lesssim250^\circ$ or $\phi_{\rm bar}\approx320^\circ$. For $\phi_{\rm bar}\approx0^\circ$ the error in LISA goes up but this is a numerical inaccuracy from computing the derivatives for small $\phi_{\rm bar}$ required in the FMA. Instead, we expect a symmetric behavior for $\phi_{\rm bar}\gtrsim0^\circ$ and $\phi_{\rm bar}\lesssim360^\circ$ (like for TianQin) and thus the error at $\phi_{\rm bar}\gtrsim0^\circ$ should be of the order $10^{-3}\,{\rm deg^2}$. The SNR of LISA and the joint detection are similar varying between roughly 40 and 100 but the error in the joint detection is slightly lower than the errors in LISA in particular for $\phi_{\rm bar}\lesssim180^\circ$ and $\phi_{\rm bar}\approx270^\circ$.

\subsection{Comparing TianQin, LISA \& joint detection}\label{sec:imric}

For light IMRIs, TianQin performs mostly significantly better than LISA, the two only having similar results for the heaviest systems of around $10^5\,{\rm M_\odot}$. In contrast, for heavy IMRIs, LISA obtains better results for most systems again having similar accuracies in TianQin for systems with a mass of the order $10^5\,{\rm M_\odot}$. Therefore, the joint detection obtains similar results to TianQin for light IMRIs while it performs similarly to LISA for heavy IMRIs. Only at the transition mass of $10^5\,{\rm M_\odot}$ do we get significantly improved results when performing joint detection.

\begin{figure}[tpb] \centering \includegraphics[width=0.48\textwidth]{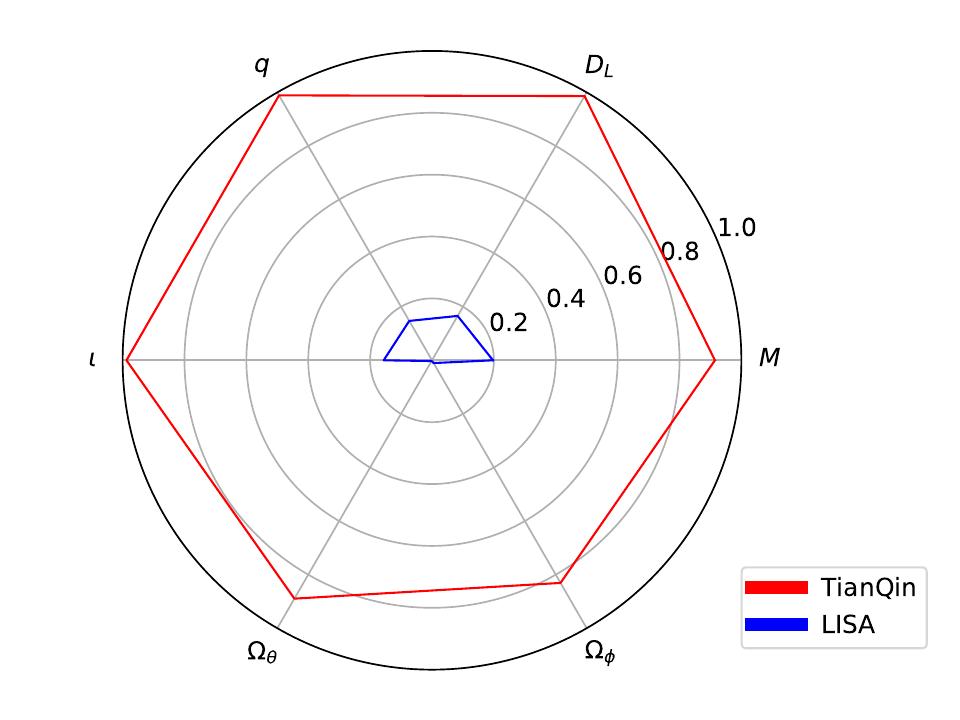}
\caption{
    The performance ratio $Q_X[\lambda]$ for TiabQin and LISA, and the parameters $\lambda$ of a light IMRI where we label the ratios only using the parameters. $\Omega_{\theta}$ and $\Omega_{\phi}$ are the sky localization as a function of $\theta_{\rm bar}$ and $\phi_{\rm bar}$, respectively.
    }
\label{fig:limri_radar}
\end{figure}

We show the performance ratio $Q$ for TianQin and LISA compared to the joint detection for light and heavy IMRIs in \fref{fig:limri_radar} and \fref{fig:limri_radar}, respectively. We see that for light IMRIs on average TianQin alone performs significantly better than LISA alone, in particular, when constraining the luminosity distance $D_L$, the mass ratio $q$, and the inclination of the source $\iota$. In contrast, for heavy IMRIs on average LISA performs better than TianQin where the best constraints are obtained again for the luminosity distance $D_L$, the mass ratio $q$, and the inclination of the source $\iota$. However, for the sky localization of heavy IMRIs, none of the two single detection scenarios performs on average as well as the joint detection. Therefore, light IMRIs will be detected significantly better by TianQin than by LISA while heavy IMRIs are better constrained by LISA, and so detecting the full mass spectrum of IMRIs requires both TianQin and LISA detection.

\begin{figure}[tpb] \centering \includegraphics[width=0.48\textwidth]{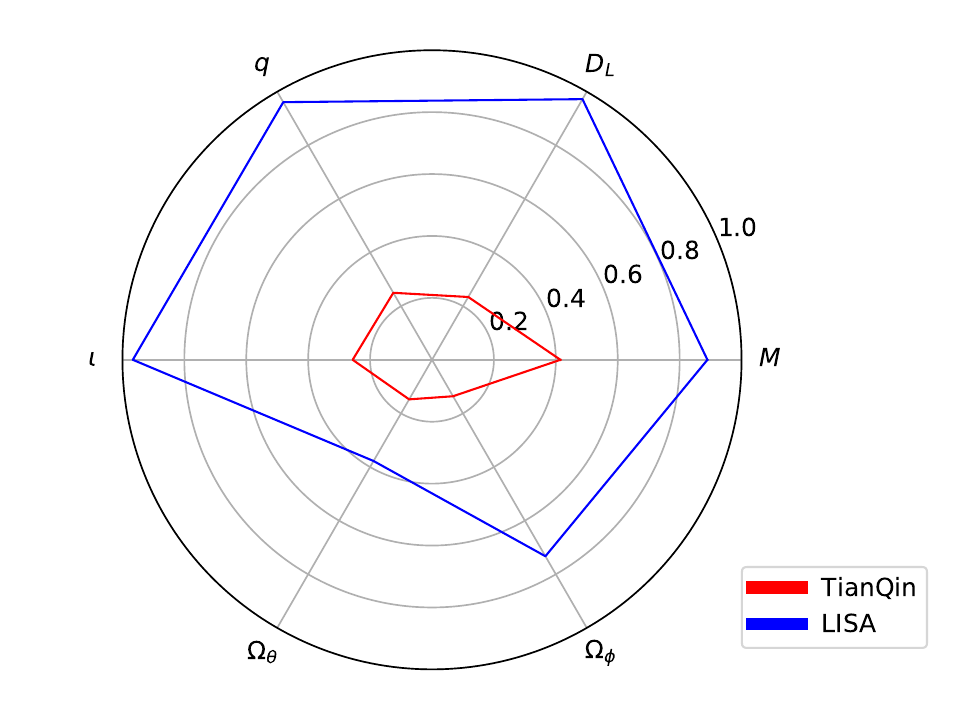}
\caption{
    The performance ratio $Q_X[\lambda]$ for the parameters $\lambda$ of a heavy IMRI, where we label the ratios only using the parameters, for TianQin and LISA. $\Omega_{\theta}$ and $\Omega_{\phi}$ indicate the sky localization as a function of $\theta_{\rm bar}$ and $\phi_{\rm bar}$, respectively.
    }
\label{fig:himri_radar}
\end{figure}

\section{Stochastic gravitational wave background}\label{sec:sgwb}

A large number of independent and unresolved GW sources can produce a stochastic gravitational wave background (SGWB)~\citep{allen_romano_1999,romano_cornish_2017,christensen_2019}. This background can be of either astrophysical origin or cosmological origin~\citep{kibble_1976,guth_pi_1982,hogan_1983,auclair_blanco-pillado_2020,caprini_figueroa_2018}. In this section, we focus on the SGWB of astrophysical origin which includes a Galactic foreground produced by Galactic DWDs~\citep{postnov_prokhorov_1998,ungarelli_vecchio_2001,adams_cornish_2014}, and extragalactic backgrounds from unresolved MBHBs~\citep{enoki_inoue_2004,antoniadis_arzoumanian_2022,arzoumanian_baker_2020,lentati_taylor_2015}, SBHBs~\citep{schneider_ferrari_2001,chen_huang_2019,dorazio_samsing_2018}, IMRIs~\citep{toscani_rossi_2020}, and EMRIs~\citep{fan_zhong_2022,berry_gair_2013,berry_gair_2013b}. The SGWB produced by Galactic sources (mainly DWDs) is comparable to the detector noise and thus often referred to as a `foreground' which has a distribution concentrated in the Galactic plane, making it anisotropic and thus more distinctive. The extragalactic background, in contrast, is assumed to be spatially homogeneous, Gaussian, stationary, and unpolarized. Detecting the SGWB for the aforementioned sources can provide important astrophysical information about the underlying populations, e.g., the mass distribution of the source, the evolution of their merger rate, and their formation mechanisms~\citep{mazumder_mitra_2014,callister_sammut_2016,maselli_marassi_2016}.

Data accumulated during the first three LIGO-Virgo observing runs (O1-O3) show no observational evidence for a SGWB. From this result the LIGO-Virgo-KAGRA Collaboration has set an upper limit on the dimensionless energy density for a frequency-independent SGWB of $\Omega_{\rm GW} \leq 5.8\times 10^{-9}$~\citep{ligo_virgo_2021}. The NANOGrav Collaboration reported in 2020 a common-spectrum process~\citep{arzoumanian_baker_2020} and recently these results have been confirmed by the Chinese Pulsar Timing Array~\citep{cpta_2023}, the Parks Pulsar Timing Array~\citep{ppta_2023} and the European Pulsar Timing Array~\citep{epta_2023}. In this section, we analyze the detectability and detection accuracy for SGWB with TianQin, LISA, and joint detection. We parametrize the energy spectrum density of the SGWB as 
\begin{equation}
    \Omega_{\rm GW} = \Omega_{\rm BG} + \Omega_{\rm FG}.
\end{equation}
$\Omega_{\rm BG}$ then represents the background formed by extra-galactic binaries that can be modeled using a power-law
\begin{equation}\label{eq:psgwb}
    \Omega_{\rm BG}(f) = \Omega_0\left(\frac{f}{f_{\rm ref}}\right)^{\alpha_0},
\end{equation}
where $\Omega_0$ is the amplitude level at the reference frequency $f_{\rm ref}$ and $\alpha_0 = 2/3$ is the spectral index for a background of binaries~\citep{farmer_phinney_2003,regimbau_2011,moore_cole_2015}. For the foreground produced by Galactic DWDs $\Omega_{\rm FG}$, we use a broken power-law
\begin{equation}\label{eq:plsgwb}
    \Omega_{\rm FG}(f) = \Omega_1\left(\frac{f}{f_{\rm ref}}\right)^{\alpha_1}\left[1 + 0.75\left(\frac{f}{f_{\rm ref}}\right)^\Delta\right]^{(\alpha_2-\alpha_1)/\Delta},
\end{equation}
where we set $\Omega_1 = 1\times10^{-6.94}$, $\alpha_1 = 3.64$, $\alpha_2 = -3.95$, and $\Delta = 0.75$~\citep{tq_sgwb_2022a}. A more detailed discussion on SGWB detection can be found in, e.g., \cite{tq_sgwb_2022a}, \cite{wang_han_2021}, \cite{tq_sgwb_2022b}, and \cite{caprini_figueroa_2018}.

\subsection{Detectability}\label{sec:sgwbd}

We calculate the SNR $\rho$ of the SGWB by employing TianQin, LISA, and joint detection. To do this, we consider various values of $\Omega_{\rm GW}$ that are compatible with the upper limit set by the LIGO-Virgo-KAGRA collaboration, as well as a reference frequency $f_{\rm ref} = 1\,{\rm mHz}$. Furthermore, we assume an operation time of one year for each detector, taking into account the duty cycles of TianQin and LISA. Note that since $\Omega_1$, $\alpha_1$, $\alpha_2$, and $\Delta$ are fixed to fit what we know about the population of DWDs in our galaxy and $\alpha_0$ is set by our assumption that we are considering binaries for the SGWB, a change of $\Omega_{\rm GW}$ basically means a change of $\Omega_0$ and thus a variation of the background $\Omega_{\rm BG}$ relative to the foreground $\Omega_{\rm FG}$.

\begin{figure}[tpb] \centering \includegraphics[width=0.48\textwidth]{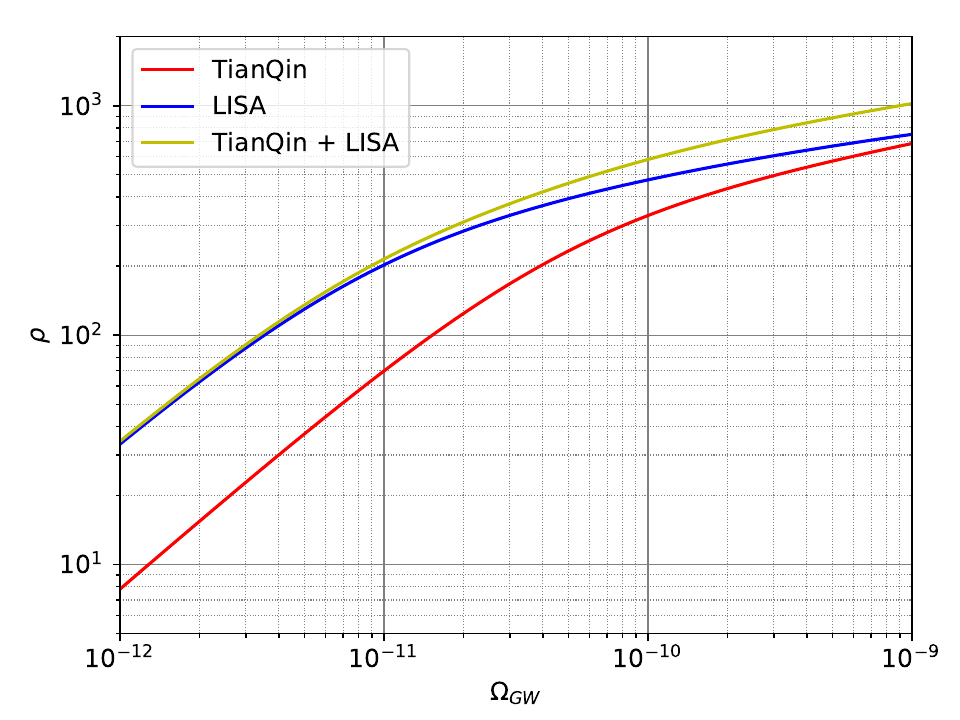}
\caption{
    The SNR $\rho$ of the SGWB in TianQin, LISA, and for joint detection as a function of $\Omega_{\rm GW}$.
    }
\label{fig:sgwb_snr}
\end{figure}

We show in \fref{fig:sgwb_snr} the SNR for the different detection scenarios. We see that $\rho$ varies between a bit less than ten and around 1000, where we get higher SNRs for increasing $\Omega_{\rm GW}$ because this corresponds to the background becoming stronger. Therefore, the SGWB should be detectable in all scenarios with a significant SNR if the energy density lies above $10^{-12}$. For $\Omega_{\rm GW} \lesssim 10^{-11}$, we see that detection by LISA alone performs similarly to joint detection with a SNR of around 30 to 200, while TianQin alone performs significantly worse by a factor of three to four. If the energy density is between $10^{-11}$ and $10^{-10}$ LISA alone still performs better than TianQin alone by a factor of 1.5 to three but the joint detection performs better than detection by any single detector reaching SNRs between around 200 and 600. For $\Omega_{\rm GW} \gtrsim 10^{-10}$ TianQin and LISA alone perform similarly reaching SNRs of around 350 to 750 and differing by a factor smaller than 1.5. In this range, joint detection performs significantly better than detection by a single detector reaching SNRs of up to 1000.

Finally, it is worth noting that the measurement of the SGWB with a single detector is done using the null channel method. On the other hand, in the joint detection scenario, both the null channel method for each component detector and the cross-correlation method for two different detectors are used (see, e.g., \cite{tq_sgwb_2022a}). Although, the two methods can achieve similar accuracy, e.g., LISA alone and joint detection for $\Omega_{\rm GW} \lesssim 10^{-11}$, the cross-correlation method is, in general, more reliable since it requires less accurate modeling of the noise in a single detector.

\subsection{Parameter estimation}\label{sec:sgwbp}

\begin{figure}[tpb] \centering \includegraphics[width=0.48\textwidth]{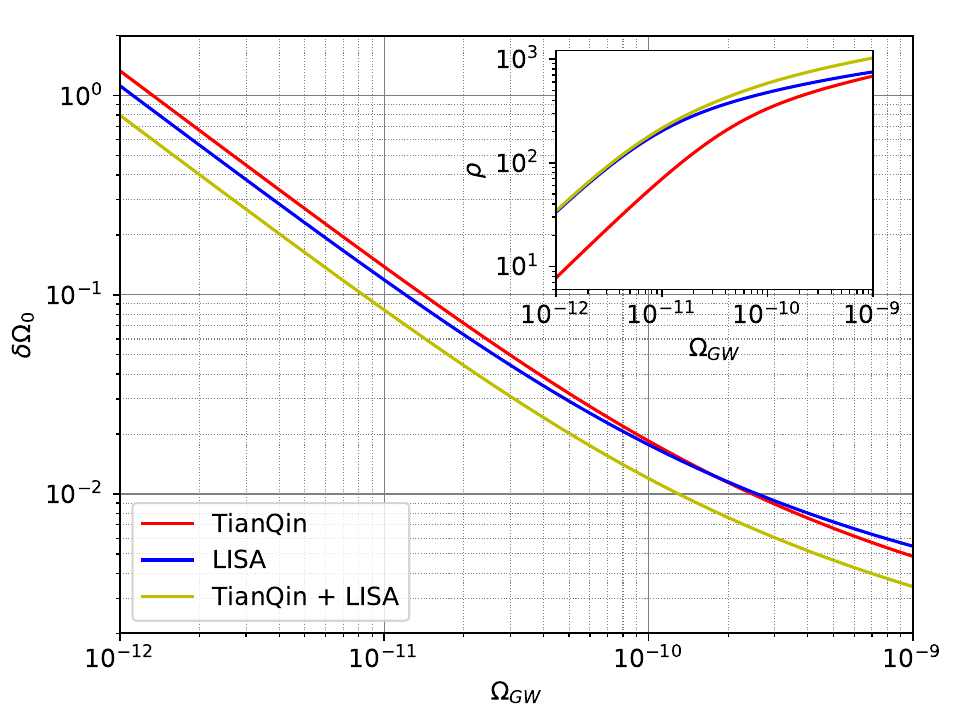}
\caption{
    The relative error for $\Omega_0$ in TianQin, LISA, and for joint detection as a function of $\Omega_{\rm GW}$. The inset shows the corresponding SNR where the color coding is the same as in the main plot.
    }
\label{fig:sgwb_domega0}
\end{figure}

The detection accuracy for the parameters of the background in \eq{eq:psgwb} and the foreground in \eq{eq:plsgwb} are estimated using a FMA, where we fix $\alpha_0$, $\Omega_1$, $\alpha_1$, $\alpha_2$, and $\Delta$ to the values indicated after the respective equations, set $f_{\rm ref} = {\rm 1\, mHz}$, and vary over $\Omega_0$ in all cases. The relative error for the energy density of the background $\Omega_0$ is shown in \fref{fig:sgwb_domega0} where we see that TianQin alone and LISA alone perform similarly with a detection accuracy of around one for $\Omega_{\rm GW}\sim10^{-12}$ corresponding to SNRs $\sim 10$. However, the relative error goes below 0.1 for $\Omega_{\rm GW} \gtrsim 1.5\times10^{-11}$ corresponding to SNRs of more than 100. For $\Omega_{\rm GW} \gtrsim 2.5\times10^{-10}$ the relative error goes below 0.01 which coincides with the range where TianQin starts performing slightly better than LISA. Through the entire range, the joint detection performs better than any single detection by a factor of at least 1.2. We further see that for all three kinds of detection, the improving detection accuracy for $\Omega_0$ correlates with an increasing SNR. However, although the SNR in TianQin is never higher than the SNR in LISA (see inset), TianQin still manages to have a better detection accuracy for $\Omega_{\rm GW} \gtrsim 2.5\times10^{-10}$.

\begin{figure}[tpb] \centering \includegraphics[width=0.48\textwidth]{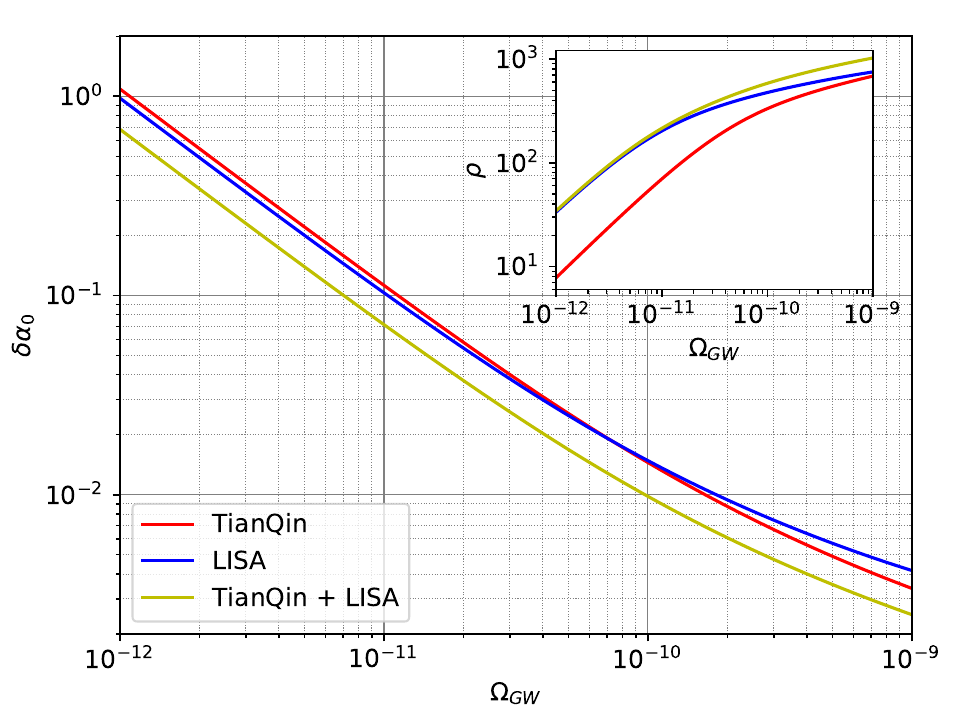}
\caption{
    The relative error for $\alpha_0$ in TianQin, LISA, and for joint detection as a function of $\Omega_{\rm GW}$. The inset shows the corresponding SNR where the color coding is the same as in the main plot.
    }
\label{fig:sgwb_dalpha0}
\end{figure}

The relative error for $\alpha_0$ is shown in \fref{fig:sgwb_dalpha0}. We see that through the entire range of $\Omega_{\rm GW}$ the joint detection again performs better than any single detection but now by a factor of around 1.4. In this case for $\Omega_{\rm GW} \le 7\times10^{-12}$ the relative error is above 0.1 but goes below 0.01 for $\Omega_{\rm GW} \gtrsim 10^{-10}$. We further see that for $\Omega_{\rm GW} \gtrsim 10^{-10}$ TianQin alone performs better than LISA alone reaching a relative error below $1.5\times10^{-2}$. Similar to the case for $\Omega_0$ the relative error for $\alpha_0$ decreases for an increasing SNR but TianQin alone performs better than LISA alone for $\Omega_{\rm GW} \gtrsim 1\times10^{-10}$ despite having a slightly lower SNR  as shown in the inset.

\begin{figure}[tpb] \centering \includegraphics[width=0.48\textwidth]{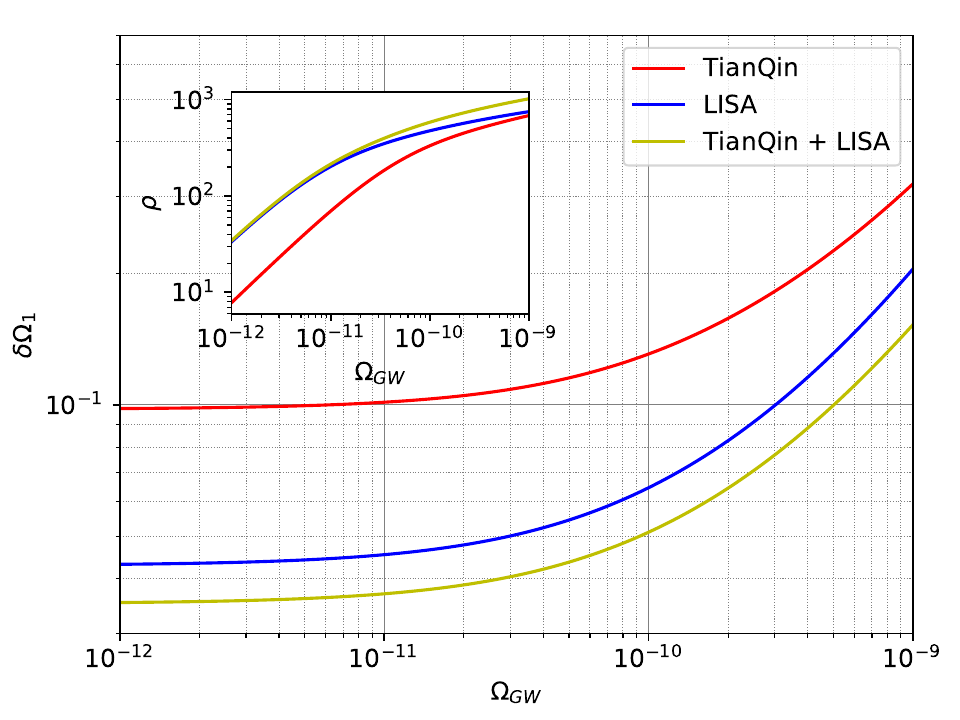}
\caption{
    The relative error for $\Omega_1$ in TianQin, LISA, and the joint detection as a function of $\Omega_{\rm GW}$. The inset shows the corresponding SNR where the color coding is the same as in the main plot.
    }
\label{fig:sgwb_domega1}
\end{figure}

\fref{fig:sgwb_domega1} shows the relative error for the energy density of the foreground $\Omega_1$. We see that, in contrast to the parameters of the background, the relative error increases when $\Omega_{\rm GW}$ increases; being it around 0.1 for TianQin alone and $\Omega_{\rm GW} \lesssim 10^{-11}$ and of the order of 0.01 for LISA alone and joint detection in the same range. For higher $\Omega_{\rm GW}$ the relative error for TianQin increases to up to 0.3 and also goes above 0.1 for LISA and joint detection if it reaches $3\times10^{-10}$ and $5\times10^{-10}$, respectively. The decrease in detection accuracy happens because the energy density of the foreground is fixed to model the known DWD population, meaning that when the total energy density is increasing it corresponds to an increase of only the background and thus the foreground becomes relatively weaker. This behavior can also be seen when comparing to the SNR shown in the inset which increases for increasing $\Omega_0$. Moreover, TianQin alone performs significantly worse than the two other detection methods because one year of observation time significantly limits TianQin's sensitivity to the foreground~\citep{tq_dwd_2020}.

\begin{figure}[tpb] \centering \includegraphics[width=0.48\textwidth]{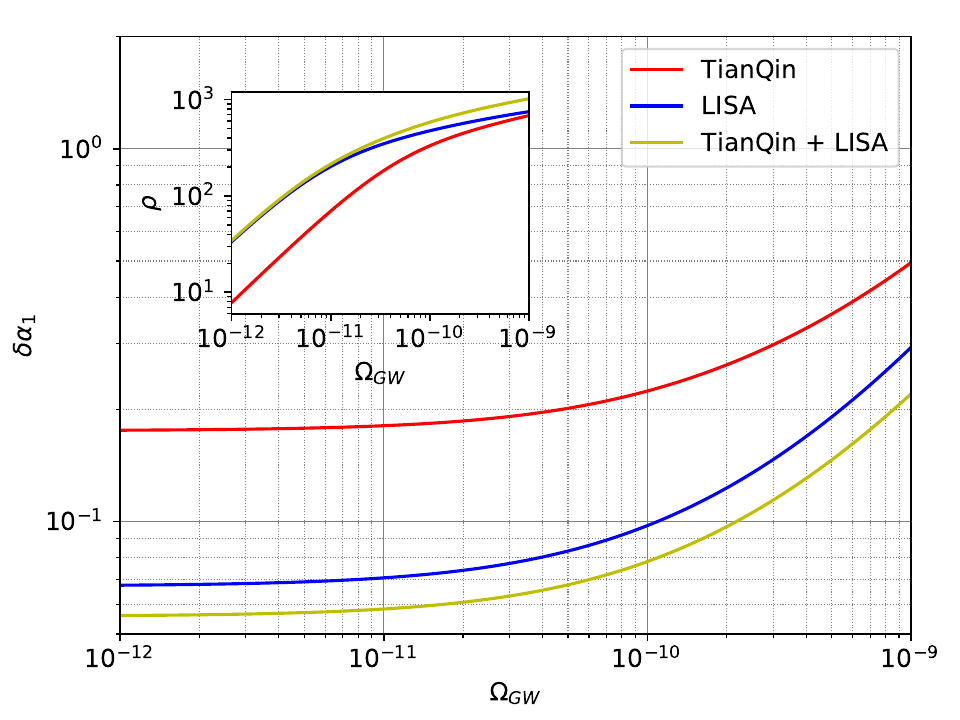}
\caption{
    The relative error for $\alpha_1$ in TianQin, LISA, and for joint detection as a function of $\Omega_{\rm GW}$. The inset shows the corresponding SNR where the color coding is the same as in the main plot.
    }
\label{fig:sgwb_dalpha1}
\end{figure}

\begin{figure}[tpb] \centering \includegraphics[width=0.48\textwidth]{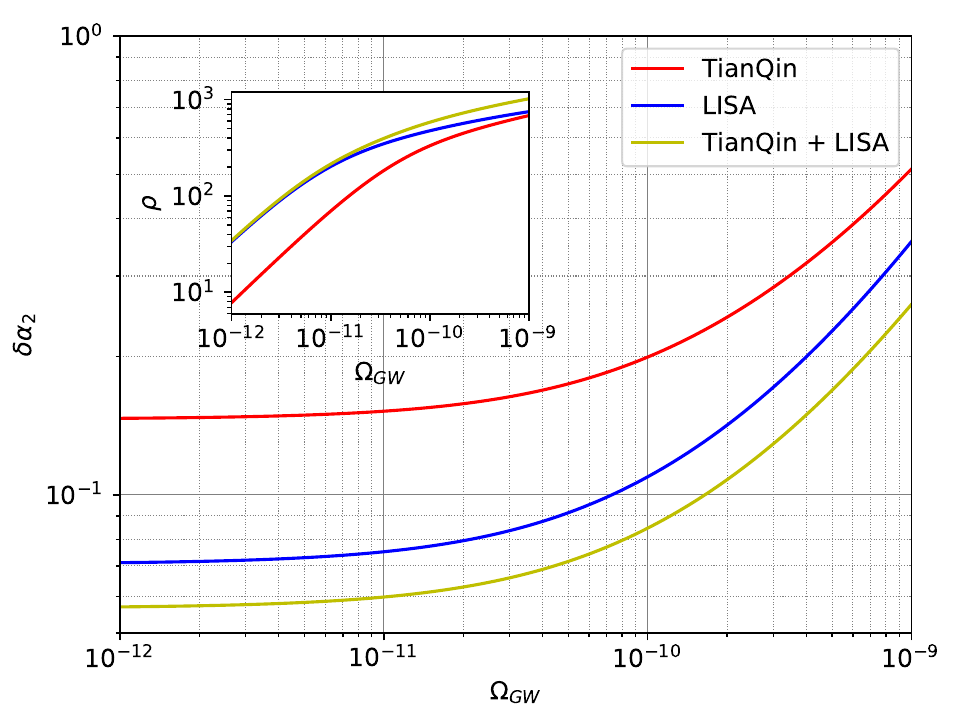}
\caption{
    The relative error for $\alpha_2$ in TianQin, LISA, and for joint detection as a function of $\Omega_{\rm GW}$. The inset shows the corresponding SNR where the color coding is the same as in the main plot.
    }
\label{fig:sgwb_dalpha2}
\end{figure}

The relative errors of $\alpha_1$ and $\alpha_2$ are shown in \fref{fig:sgwb_dalpha1} and \fref{fig:sgwb_dalpha2}, respectively. As before, the relative error for these two parameters increases when $\Omega_{\rm GW}$ increases; again TianQin alone performs significantly worse than LISA alone and a join detection. Moreover, we see again that for all detection methods, the detection accuracy decreases despite an increase in the SNR because the foreground becomes weaker relative to the background. Nevertheless, the relative error is of the order of $10^{-2}$ for LISA alone and joint detection and only goes above 0.1 if $\omega_{\rm GW} \gtrsim 10^{-10}$. For TianQin alone the error is always at the order of $10^{-1}$ going up to around 0.5.

\subsection{Comparing TianQin, LISA \& joint detection}\label{sec:sgwbc}

Using the null channel method, the SGWB has a lower SNR in TianQin than in LISA for a lower energy density but is comparable for high energy densities. As a consequence, the detection accuracy for the parameters of the background is better in LISA for lower energy densities, however, for higher energy densities TianQin can achieve more accurate measurements. For the galactic foreground, LISA obtains significantly smaller detection errors for the full range of energy densities due to its better sensitivity to lower frequencies. For the SGWB the joint detection allows cross-correlating the detection in TianQin and LISA, thus, reducing the accuracy required in modeling the noise and making the results more reliable. Moreover, the accuracies obtained are always better than in any of the two single detectors for both the background and the galactic foreground.

\begin{figure}[tpb] \centering \includegraphics[width=0.48\textwidth]{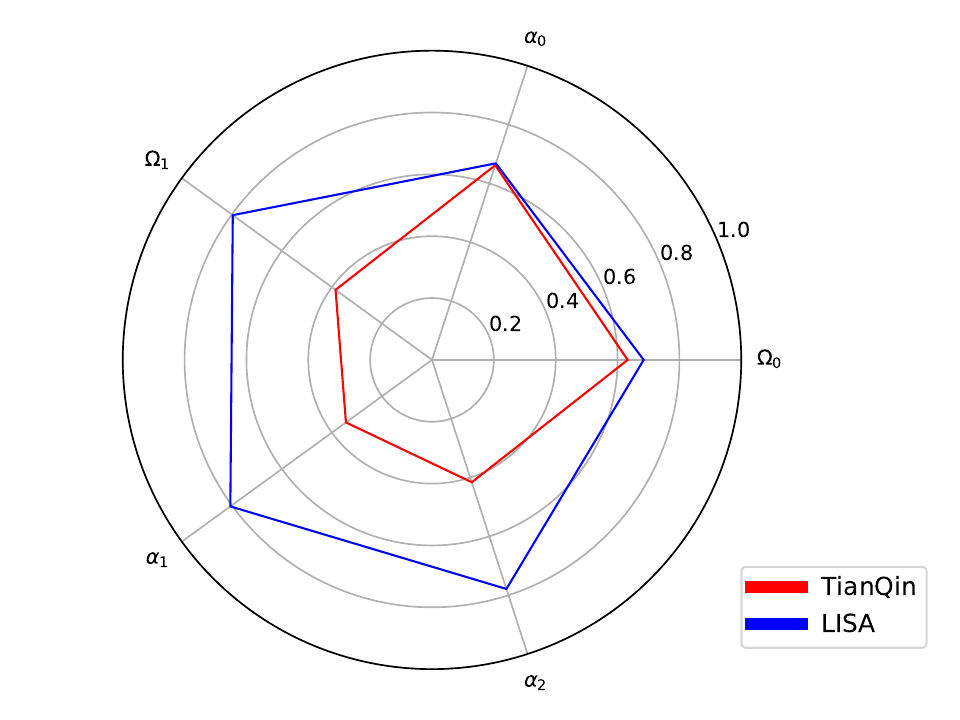}
\caption{
    The performance ratio $Q_X[\lambda]$ for TianQin and LISA and the parameters $\lambda$ of the SGWB, where we label the ratios only using the parameters.
    }
\label{fig:sgwb_radar}
\end{figure}

\fref{fig:sgwb_radar} shows the performance ratio $Q$ of the null channel method for TianQin and LISA compared to the cross-correlating joint detection. We see that for the parameter of the background $\Omega_0$ and $\alpha_0$ on average TianQin alone and LISA alone perform at a similar level although both perform significantly worse than the joint detection. For the parameters of the foreground $\Omega_1$, $\alpha_1$, and $\alpha_2$ LISA performs on average significantly better than TianQin, although, joint detection still performs better than LISA alone.

\section{Results}\label{sec:res}

Due to their similar configuration, TianQin and LISA will be able to detect the same kind of astrophysical sources. However, TianQin is more sensitive to frequencies above $10\,{\rm mHz}$ while LISA shows greater sensitivity for $f\lesssim10\,{\rm mHz}$. Therefore, TianQin tends to be more sensitive to lighter sources emitting GWs at higher frequencies while LISA tends to perform better for heavier sources that merge at lower frequencies; although, there are also exceptions to this ``rule''. Equal mass MBHBs and EMRIs will be detected by LISA to larger distances than by TianQin and most parameters will be detected with higher accuracy by LISA. However, heavy IMRIs will be detected by LISA only at bigger distances and with better accuracy for the heaviest systems with a total mass above $3\times10^5\,{\rm M_\odot}$ while TianQin performs similarly or better for the lighter ones. In contrast, TianQin will detect SBHBs at larger distances and with better accuracy but only for the lightest systems while LISA will perform similarly for the heaviest binaries with a mass close to $100\,{\rm M_\odot}$. For DWDs, the performance of the two detectors depends greatly on the evolution stage of the system. Binaries with higher frequencies above roughly $10\,{\rm mHz}$ will be detected more accurately by TianQin while LISA performs better for DWDs with lower frequencies. TianQin will reach larger distances and achieve better detection accuracies for light IMRIs although for the heaviest systems of a total mass close to $10^5\,{\rm M_\odot}$ LISA will obtain similar results.

For the SGWB we find that TianQin alone and LISA alone will perform at a similar level for the extragalactic background while LISA performs better for the galactic foreground because the DWDs that form the background are emitting at lower frequencies below $10\,{\rm mHz}$. However, here it is important to point out that the detection of the SGWB by a single detector requires accurate modeling of the noise. In contrast, joint detection using cross-correlation between two separate detectors allows us to get more reliable results. At the same time, we find that the detection accuracy for the SGWB from joint detection is better than the detection for any of the two detectors.

Joint detection by TianQin and LISA allows a significant improvement in the study of astrophysical GW sources. Since both detectors can detect the same sources, combining their detection always allows detection at larger distances and with lower detection errors. However, the biggest gain probably does not come from the joint detection of particular sources but from the possibility of covering a much larger parameter space. Getting a better picture of the astrophysical origin and evolution of most astrophysical sources requires the detection of different systems with different sets of parameters. For most sources, neither TianQin alone nor LISA alone will be able to detect their entire population but joint detection will allow a significantly improved coverage.

Despite TianQin and LISA performing the best for different sources, sometimes achieving detection distances or errors that differ by factors of more than ten, it should also be noted that the difference usually remains below one order of magnitude. Therefore, almost all sources should be detected simultaneously except for some special cases where the source is positioned at a blind spot of one of the detectors or accumulates most of its SNR while one of the detectors is not online. This aspect has been briefly discussed in this paper but should be addressed in more detail in future work. TianQin has a fixed orientation towards RX J0806.3+1527 while LISA's orientation changes along its orbit. Therefore, we find that TianQin can detect sources in certain areas of the sky with high precision through its entire cycle while performing significantly worse in other directions. LISA, in contrast, shows good detection accuracy for most sky directions but is often outperformed by TianQin for sources close to the ecliptic plane. We further find the different duty cycles of the two detectors can impact detection. TianQin will have a scheduled duty cycle of 50\,\% and the detection accuracy of certain parameters will be greatly affected by the change between on- and off-times. LISA is expected to have a duty cycle of 75\,\% but it is yet unclear when it will be off. Joint detection can greatly improve results by providing continuous coverage of the source if at least one of the detectors is operating most of the time.

\section{Conclusions}\label{sec:con}

In this paper, we study the detection of astrophysical sources with space-based laser interferometer gravitational wave detectors TianQin and LISA as well as joint detection. The sources considered are MBHBs, SBHBs, DWDs, EMRIs, light and heavy IMRIS, and the SGWB from galactic and extragalactic binaries, where for all sources we analyze to what distance they will be detected as well as the detection accuracy for different parameters. We find that, in general, TianQin will perform better for higher frequency/lower mass sources while LISA shows better sensitivity to lower frequency/higher mass sources, although the difference usually remains below one order of magnitude. Therefore, joint detection can improve the detection distance and parameter reconstruction of particular sources. Moreover, combining the detection of TianQin and LISA allows for covering a significantly larger population of astrophysical sources as well as a bigger parameter space for these sources. This greatly improved detection of sources with different properties might be one of the biggest gains from TianQin and LISA cooperation.

In the analyses performed we always vary one parameter of the source at a time while keeping all other parameters fixed. This approach allows us to understand how well TianQin, LISA, and joint detection can measure the different parameters. However, the measurement of different parameters is intertwined and sometimes even degenerate. Therefore, changing the value of the fixed parameters will, in general, affect the results obtained. We choose the values of the fixed parameters so that they are astrophysical relevant and that we obtain representative results. Nevertheless, obtaining an even more complete picture would require performing a study where all parameters of a source are varied at the same time. We do not use such an approach because we consider for each source between five and seven parameters and thus the computational expense of such an analysis would be at least six orders of magnitude bigger than the analysis performed in this paper but such a study should be aimed at in the future.

Studying and potentially coordinating joint detection by TianQin and LISA could bring further benefits to GW astronomy. The most obvious improvement probably being for the detection of the SGWB where cross-correlation of the data from both detectors greatly reduces the requirements in the modeling of the noise. Other aspects that were not extensively studied in this paper but should be considered in more detail in the future are the joint detection of sources at particular sky positions or that accumulate significant SNR in short time intervals. Due to the different orientations and duty cycles of TianQin and LISA, joint detection might provide coverage of sources and their parameters that might be missed or barely constrained by a single detector.

\section*{Acknowledgments}

We thank Lijing Shao for his helpful comments on the draft. This work was partially supported by the Guangdong Major Project of Basic and Applied Basic Research (Grant No. 2019B030302001). ATO acknowledges support from the China Postdoctoral Science Foundation (Grant No. 2022M723676). HTW is supported by the China Postdoctoral Science Foundation (Grant No. 2022TQ0011), the National Natural Science Foundation of China (Grant Nos. 12247152 \& 11975027), and the Opening Foundation of TianQin Research Center. YMH has been supported by the Natural Science Foundation of China (Grant Nos. 12173104 \& 12261131504).


\bibliography{ref}{}
\bibliographystyle{aasjournal}



\end{document}